\documentclass[11pt]{article}

\usepackage[T1]{fontenc}
\usepackage[utf8]{inputenc}
\usepackage[english]{babel}
\usepackage{lmodern}
\usepackage{geometry}
\usepackage{microtype}
\usepackage{graphicx}
\usepackage{booktabs}
\usepackage{array}
\usepackage{tabularx}
\usepackage{multirow}
\usepackage{enumitem}
\usepackage{float}
\usepackage{caption}
\usepackage{siunitx}
\usepackage{csquotes}
\usepackage{amsmath,amssymb}
\usepackage{authblk}
\usepackage{longtable}
\usepackage{pdflscape}
\usepackage{xcolor}
\usepackage{colortbl}
\usepackage{tikz}
\usetikzlibrary{arrows.meta}
\usepackage{pgfplots}
\pgfplotsset{compat=1.18}
\usepackage{pgfplotstable}
\usepackage{hyperref}

\newcolumntype{L}{>{\raggedright\arraybackslash}X}
\usepackage[
    backend=biber,
    style=apa
]{biblatex}
\definecolor{fgorange}{HTML}{C8502A}   
\definecolor{sgteal}{HTML}{2C6B5E}     
\definecolor{cvblue}{HTML}{4472C4}     
\definecolor{cellgreen}{HTML}{E8F3E6}  
\definecolor{y02green}{HTML}{2E6B30}   
\newcommand{\acct}[2]{{\color{#1}\bfseries #2}}
\newcommand{\predbar}[1]{\noindent\fcolorbox{black!30}{black!4}{%
  \parbox{\dimexpr\linewidth-2\fboxsep-2\fboxrule\relax}{\centering #1}}}
\newcommand{\verdictbar}[2]{\noindent\colorbox{#1}{%
  \parbox{\dimexpr\linewidth-2\fboxsep\relax}{\centering\color{white}\bfseries #2}}}

\title{{\bfseries Systematic Bias in Green Patent Classification: Silent Green and False Green}}

\author[1]{Hamid Bekamiri}
\author[2]{Jan Auernhammer}
\author[1]{Milad Abbasiharofteh}
\author[1]{Jesper Lindgaard Christensen}

\affil[1]{Aalborg University Business School\\
The IKE Research Group\\
Aalborg University, Denmark}

\affil[2]{Center for Design Research\\
ME Design Group\\
Stanford University, USA}

\date{}

\begin{document}
\maketitle
\begin{abstract}
Research, policy, and capital rely on the Cooperative Patent
Classification's Y02 tag to locate climate invention, yet whether Y02
measures what it is used to measure has never been tested at
corpus scale. We ask three questions: what types of error does
Y02 make, are those errors idiosyncratic, and what are their main
causes? Because neither Y02 nor a text classifier trained on it is
ground truth, our Error-as-Signal framework treats their disagreements as
evidence, not model failure: in a construct-validity test, two independent
large language models judge whether the primary function of each of
\num{517772} disputed inventions among \num{9075421} USPTO patents has a
direct climate mechanism. Y02 makes two types of error, which we term False
Green (\num{180384} inclusions) and Silent Green (\num{29465} omissions).
The errors are not idiosyncratic but form a systematic bias: False
Green concentrates in ICT, Silent Green in energy, transport, chemistry,
and industrial engineering, so raw counts overstate ICT energy
efficiency threefold. The evidence points away from applicant
strategy, misclassification did not jump when green labels became
salient, and toward bounded classification capacity. Yet complexity does
not simply make classification harder; its two forms push errors in
opposite directions, a complexity paradox. A one-standard-deviation
increase in reflection complexity (capability scarcity) is
associated with 2.45 times the odds that an error is an omission rather
than an inclusion, whereas structural complexity
(combinatorial diversity) tilts errors toward inclusion. What
matters is not how complex an invention is but whether its complexity is
legible to the taxonomy.
\end{abstract}



\section{Introduction}\label{sec:introduction}

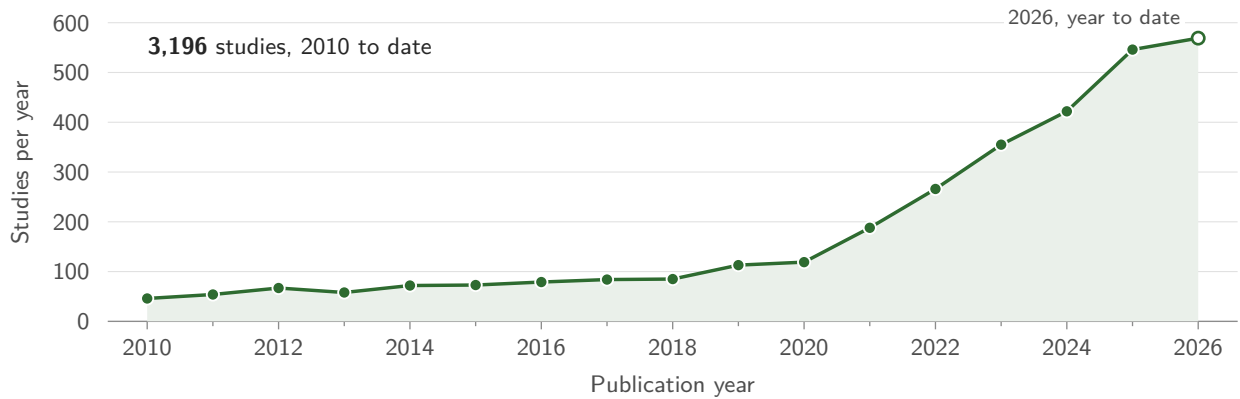
\begin{figure}[tbp]
\centering
\begin{tikzpicture}
\begin{axis}[
  width=\linewidth, height=5.8cm,
  font=\sffamily\footnotesize,
  axis lines*=left,
  x axis line style={draw=black!45, line width=0.5pt},
  y axis line style={draw=none},
  xmin=2009.4, xmax=2026.6, ymin=0, ymax=640,
  enlargelimits=false,
  xtick={2010,2012,...,2026},
  minor xtick={2011,2013,...,2025},
  ytick={0,100,...,600},
  xticklabel={\pgfmathprintnumber[assume math mode=true,1000 sep={}]{\tick}},
  yticklabel={\pgfmathprintnumber[assume math mode=true,1000 sep={,}]{\tick}},
  xtick style={draw=black!45, line width=0.5pt},
  ytick style={draw=none},
  major tick length=3pt, minor tick length=1.8pt,
  tick align=outside,
  ymajorgrids, grid style={draw=black!12, line width=0.4pt},
  ticklabel style={text=black!70},
  xlabel={Publication year}, ylabel={Studies per year},
  label style={text=black!70},
  clip=false,
]
\addplot[draw=none, fill=y02green!10, forget plot]
  table[x=year, y=n]{\yiistudies} \closedcycle;
\addplot[draw=y02green, line width=1.3pt, line join=round, line cap=round,
  no marks, forget plot]
  table[x=year, y=n]{\yiistudies};
\addplot[only marks, mark=*, mark size=2.3pt, forget plot,
  mark options={fill=y02green, draw=white, line width=0.7pt},
  skip coords between index={\yiilast}{\yiirows}]
  table[x=year, y=n]{\yiistudies};
\addplot[only marks, mark=*, mark size=2.3pt, forget plot,
  mark options={fill=white, draw=y02green, line width=1pt},
  skip coords between index={0}{\yiilast}]
  table[x=year, y=n]{\yiistudies};
\node[anchor=south east, xshift=-5pt, yshift=2pt, inner sep=1.5pt,
  fill=white, font=\sffamily\scriptsize, text=black!70]
  at (axis cs:\yiilastyear,\yiilastn) {\yiilastyear, year to date};
\node[anchor=west, align=left, inner sep=0pt, text=black!85]
  at (axis cs:2010,550)
  {\textbf{\pgfmathprintnumber[assume math mode=true,1000 sep={,}]{\yiitotal}} studies, 2010 to date};
\end{axis}
\end{tikzpicture}
\caption[Studies per year using Y02 as a green-patent measure]{Studies
per year using Y02 as a green-patent measure, 2010--2026. Counts are Google Scholar results by publication year for the query
``Y02'', retrieved September 5, 2026; the open marker is the partial count
for 2026, up to the retrieval date. The series
starts in 2010, when the European Patent Office introduced the Y02 scheme.
Google Scholar counts documents that match the query, including working
papers and theses, rather than verified uses of Y02, so the series shows
the scale and growth of the literature rather than an exact census.}
\label{fig:y02_studies}
\end{figure}

The global effort to achieve a net-zero economy represents one of the most significant attempts at directed technical change in modern times. To guide this transition, policymakers, international organizations like the OECD and WIPO, and investors rely heavily on quantitative indicators to identify, track, and support climate-related technologies. Patent statistics have long been central to measuring inventions \parencite{Griliches1990}. Within this tradition, the Cooperative Patent Classification (CPC) Y02 scheme has become the dominant administrative indicator for identifying climate-related inventions. The Y02 classification was created as a tagging scheme for climate-change mitigation technologies and has become widely used in research, policy monitoring, and ESG-related analysis \parencite{Veefkind2012}. Because these indicators shape how scholars, policymakers, and investors evaluate the direction of green technological change, their accuracy and validity are matters of first-order importance.
Since the European Patent Office introduced the Y02 scheme in 2010,
more than \num{3000} studies have used it as a green-patent measure, and
the number published each year has grown more than tenfold, from about 50
in 2010 to more than 500 in 2025 (Figure~\ref{fig:y02_studies}). The
validity of the indicator therefore conditions a large and fast-growing
body of evidence.

\pgfplotstableread{
year n
2010 46
2011 54
2012 67
2013 58
2014 72
2015 73
2016 79
2017 84
2018 85
2019 113
2020 119
2021 188
2022 266
2023 355
2024 422
2025 546
2026 569
}\yiistudies
\pgfplotstablecreatecol[create col/expr accum={\pgfmathaccuma+\thisrow{n}}{0}]{cum}{\yiistudies}
\pgfplotstablegetrowsof{\yiistudies}%
\pgfmathtruncatemacro{\yiirows}{\pgfplotsretval}%
\pgfmathtruncatemacro{\yiilast}{\yiirows-1}%
\pgfplotstablegetelem{\yiilast}{cum}\of{\yiistudies}%
\pgfmathtruncatemacro{\yiitotal}{round(\pgfplotsretval)}%
\pgfplotstablegetelem{\yiilast}{n}\of{\yiistudies}%
\pgfmathtruncatemacro{\yiilastn}{round(\pgfplotsretval)}%
\pgfplotstablegetelem{\yiilast}{year}\of{\yiistudies}%
\pgfmathtruncatemacro{\yiilastyear}{\pgfplotsretval}%

However, a growing body of evidence reveals a troubling disconnect between these widely used green patent indicators and other measures of environmental performance. Recent research documents a significant ESG innovation disconnect, where firms with high environmental ratings do not necessarily produce more green patents, and conversely, firms in traditionally capital-intensive heavy industries are often the most prolific producers of green technologies despite weaker ESG scores \parencite{Cohen2026,Lan2025}. This paradox raises a fundamental question for innovation policy.
Does the disconnect reflect firm behavior or the instruments used
to observe it, and if the instruments err, do they inflate the
disconnect or hide part of it? \textcite{Cohen2026} establish the disconnect and infer that the
environmental-rating instrument is the blind party. We take the
complementary side of the same inference and test the construct validity of the innovation-side instrument that they, like the rest of this literature, had to take as
given. Their green-patent measure follows the OECD ENV-TECH search
strategies, which rely on Y02 wherever possible \parencite{HascicMigotto2015},
so errors in Y02 pass into their comparison. Because that comparison is
between firms, what matters is relative error. If Y02 over-credited the
technologies in which traditional energy firms patent, relative to other
fields, it could have manufactured part of the disconnect; if it
under-credits them, it hides part of the disconnect, and the true gap
between ESG recognition and green invention is larger than the observed
one. At the level of technologies, we find the second pattern.

The answer matters because, to our knowledge, the construct validity of Y02 has never been tested across the full patent corpus. Existing work documents the ESG--green-patent disconnect \parencite{Cohen2026,Lan2025} or compares green-patent classification schemes in narrower samples and domains \parencite{Rainville2025,Santarlasci2023}, but does not provide a full-corpus test that attributes the source and direction of individual Y02 errors. Innovation research has long cautioned that patent-based indicators can operate as selective lenses rather than transparent measures of the underlying construct \parencite{RoachCohen2013}. The Y02 scheme's own designers likewise cautioned that inclusion in the tagging scheme does not by itself guarantee that a technology is genuinely green \parencite{Veefkind2012}, and a growing set of sample-limited studies has documented both over-inclusion, notably in information and communication technologies, and omission across green-patent classification systems \parencite{Santarlasci2023,Rainville2025}. What has been missing is a corpus-scale construct-validity test that can separate ordinary, idiosyncratic classification mistakes from structural, direction-specific measurement bias; attribute the direction of each error to an invention's primary technical function; and identify patent characteristics that systematically tilt administrative error toward over- or under-recognition.

We organize the assessment around three questions that every user
of Y02 answers implicitly when relying on it. First, what types of error
does the indicator make? A green label can fail by including an invention
without a climate function or by omitting one with it, and a model used to
check the label can err in the same two ways. Second, are these errors
idiosyncratic, noise that washes out in aggregate, or do they form a bias
that does not? Third, what are their main causes: what applicants do, or
how the label is produced? In measurement-theory terms, the
first question concerns substantive validity, whether an indicator
incorporates the properties relevant to the construct and only those; second separates threats to reliability from threats to validity
\parencite{JacobsWallach2021}; and the third asks where the invalidity
originates. The second distinction carries the practical stakes.
Random error blurs the map of green invention but leaves it centered on
the truth, and larger samples sharpen it. Bias redraws the map: it shifts the attention of researchers, the capital of investors, and the priorities of policymakers to the wrong places.
Larger samples only make the wrong map more precise.
Sections~\ref{sec:results}, \ref{sec:idiosyncratic},
and~\ref{sec:causes} answer the three questions in order.

The measurement problem can be concrete rather than abstract: Table~\ref{tab:illustrative_error_mechanisms} includes an automated retail-checkout patent describing a ``shopping environment'' that carries an administrative Y02 label even though the assessed technical function has no direct climate relevance.

Answering the first question requires a typology of errors. We treat green-patent identification as a measurement problem with
two fallible indicators and one unobserved construct. Building on
construct-validity research
\parencite{CronbachMeehl1955,AdcockCollier2001} and recent work on
measurement in computational systems \parencite{JacobsWallach2021}, we
distinguish the construct, a patent's substantive contribution to
climate-change mitigation or adaptation, from two indicators of it: the
administrative Y02 label and the prediction of a model trained on patent
text. Neither is ground truth. When they disagree, one of four errors has
occurred (Table~\ref{tab:error_attribution_framework}). Two are errors of
the label and therefore failures of construct validity: False Green, when
a non-green invention carries the tag (administrative Type~I error), and
Silent Green, when a climate-relevant invention is left untagged
(administrative Type~II error). The other two are the model's own false
positives and false negatives. Evaluation that treats Y02 as ground truth scores every disagreement as a
model error and so never sees the label's own errors. Disagreement locates
a fault; it cannot assign it.

We therefore develop an Error-as-Signal Framework for administrative measurement. The framework treats disagreement not simply as model failure, but as diagnostic evidence of possible measurement error. Although developed and applied here to green patents, the framework is indicator-agnostic: it applies wherever an authoritative administrative label can be confronted with an independent measurement of the same underlying construct. PatentSBERTa, a patent-domain language model fine-tuned to predict Y02 labels from patent text, first screens the full patent corpus for disagreement with Y02. To adjudicate these signals, we implement a construct-validity test in the form of an LLM-bagging approach: two independent LLMs (Qwen\footnote{\texttt{Qwen/Qwen3-32B}: \url{https://huggingface.co/Qwen/Qwen3-32B}.} and Gemma~4\footnote{\texttt{google/gemma-4-31B-it}: \url{https://huggingface.co/google/gemma-4-31B-it}.}) evaluate the primary technical function of each flagged invention. Their direct-mechanism consensus attributes resolved disagreements to an administrative or model error, while cross-model disagreements remain explicitly unresolved. By isolating these \num{117586} ambiguous cases, the framework distills the measurement problem from \num{9075421} granted patents down to a targeted escalation set for future research. The objective is not to claim that all classification error can be eliminated. Rather, it is to distinguish ordinary classification noise from systematic errors that bias comparisons across sectors, technologies, and firms.

Applying this framework to \num{9075421} U.S. granted patents issued between 1962 and 2024, we establish a baseline of \num{592387} administratively Y02-labelled green inventions. While initial screening with PatentSBERTa yields over 8.5 million model--administrative agreements, including \num{341357} confirmed Y02 labels, it isolates \num{517772} critical disagreements: \num{251030} downward (administratively green but predicted non-green) and \num{266742} upward (administratively non-green but predicted green). Qwen and Gemma then independently assess whether the invention's primary technical function has a direct climate-mitigation or adaptation mechanism. Direct-mechanism consensus attributes \num{180384} cases to administrative Type I error (False Green), \num{29465} to administrative Type II error (Silent Green), \num{170619} to model Type I error, and \num{19718} to model Type II error. A further \num{117586} cases remain unresolved. Correcting only the consensus-attributed administrative errors yields \num{441468} substantively green patents, or 4.86\% of the corpus. The answer to the first question is therefore that Y02 makes both types of error. The label's share of these errors is not small. Among
the \num{400186} resolved disagreements, 52.4\% are construct-validity failures of the administrative label rather than errors of the model. When Y02 says green and the
model does not, the label is the party in error in 90.1\% of resolved
cases; when the model says green and Y02 does not, the model is in error
in 85.3\%.

The answer to the second question is no: the errors are not idiosyncratic. The findings reveal a pronounced systematic bias across technological fields, with direction-specific patterns. False Green is disproportionately concentrated in asset-light and ICT-related fields, whereas Silent Green is concentrated in energy, transportation, chemistry, and industrial production. Whether a disagreement is an error of the label, and in which direction, depends far more on the specific technology than on when the patent was filed. These opposing errors partly offset one another in aggregate while redistributing measured green inventive activity across the technological landscape. This redistribution bears directly on the ESG innovation
disconnect. \textcite{Cohen2026} find that energy producers, despite lower
ESG scores, are key green innovators, especially in carbon capture. Y02's
errors work against exactly these technologies. Raw counts overstate ICT
energy efficiency about threefold but the energy family by less than a
tenth, and they understate carbon capture, the only Y02 family that grows
after correction (by 8.3\%). As a result, correction raises energy's share
of green patents from 32.8\% to 40.3\%. False Green and Silent Green do not
cancel here: both favor digital technologies over the technologies of
energy producers. At the level of technologies, correcting Y02 therefore
widens the disconnect rather than closing it
(Section~\ref{sec:discussion} discusses the firm-level test).

The third question is what causes the errors, and why they run in these directions. We set four
accounts against one another, crossing the two errors with two origins
(Section~\ref{sec:competing_accounts}). Two accounts are strategic and
concern what applicants do: greenwashing would create False Green, as
applicants use environmental language to attract the green tag, and
greenhushing would create Silent Green, as firms stay quiet about
environmental achievements. Two concern how the label is produced under
bounded classification capacity, and each is indexed by a distinct form of
technological complexity. Structural complexity, high when knowledge
exhibits a diverse combinatorial network structure, offers the classifier
many recognizable anchors and should tilt errors toward over-inclusion;
reflection complexity, high when a technology is tied to places with
scarce capabilities, offers few and should tilt errors toward omission.
All four accounts may operate at once, and each may explain some
individual errors. Our question is which of them explains the bias: the
part of the error that follows a pattern across technologies, not the
one-off mistakes that differ from case to case. The strategic accounts do not.
Misclassification does not rise discretely when green classification
became salient in 2010 and 2013, explicit green framing barely moves, and
greenhushing, though consistent with individual cases, leaves no signature
that separates it from non-strategic technical description. The capacity accounts do, in opposite directions.
Reflection complexity pushes errors toward Silent Green. A
one-standard-deviation increase is associated with 61\% higher odds that a
green invention is Silent Green rather than correctly tagged and, among
Y02's errors, with 2.45 times the odds that an error is Silent Green
rather than False Green. Structural complexity tilts errors the other way,
toward False Green. The same increase is associated with 37\% lower odds
that a green invention is Silent Green and, among Y02's errors, with 16\%
higher odds that an error is False Green rather than Silent Green. These
estimates hold filing year and technological atypicality constant.
The result is a complexity paradox. Complexity thus predicts not merely whether classification fails but how it fails, and what matters is the kind of complexity rather than
its amount. What remains is an indicator that reads environmental language
well and technical function poorly.

We contribute to three literatures. First, for research that uses
patent statistics as innovation indicators
\parencite{Griliches1990,jaffe2002patents,LernerSeru2022}, and for the tradition of validating the proxies on which management research relies \parencite{Demerjian2012}, we provide the first corpus-scale construct-validity test of the dominant green-patent indicator and show that its errors are differential rather than random
\parencite{Bound2001,Hausman2001}: they relocate measured green invention
across technological fields, so studies that use raw Y02 counts as
outcomes or regressors inherit a direction-specific bias that larger
samples cannot remove. Second, for work on learning from fallible labels
and on validating machine-learning measures \parencite{FrenayVerleysen2014,Northcutt2021,Raji2020,deKok2025}, we introduce the
Error-as-Signal framework, which treats label--model disagreement as
evidence to be adjudicated rather than as model failure, and we show that
where labels are administrative, more than half of conventionally measured
model error can be error in the label. The framework is indicator-agnostic
and applies wherever an authoritative label can be confronted with an
independent functional assessment, from ESG ratings to sustainable-finance
taxonomies. Third, for research on the ESG innovation disconnect and on
divergence among sustainability measures
\parencite{Cohen2026,Lan2025,Berg2022,Chatterji2016}, we identify a
patent-side measurement channel: the structure of an invention predicts
not only whether the administrative label errs but in which direction,
which makes climate invention embedded in engineering systematically less
visible to patent-based green metrics. Because the bias runs
against the technologies in which traditional energy firms patent,
Y02-based measurement is more likely to have understated the disconnect
documented by \textcite{Cohen2026} than to have produced it. We release a patent-level dataset
of attributed errors, judge justifications, and unresolved
cases\footnote{Repository link withheld for anonymous review.} as an
empirical platform for correcting and bounding Y02-based analyses.

The remainder of the paper is organized as follows. Section~\ref{sec:theory} develops the theoretical background, connecting the ESG innovation disconnect to construct validity, formalizing the Error-as-Signal logic of dual-measurement disagreement, and setting out the four competing accounts of administrative error. Section~\ref{sec:methodology} describes the data, the PatentSBERTa screening stage, and the construct-validity test. Sections~\ref{sec:results}, \ref{sec:idiosyncratic}, and~\ref{sec:causes} answer the three questions: Section~\ref{sec:results} identifies the types of error, Section~\ref{sec:idiosyncratic} tests whether they are idiosyncratic, and Section~\ref{sec:causes} adjudicates among four accounts of their causes. Section~\ref{sec:discussion} draws out the implications for innovation measurement, the validation of machine-learning measures, managers, investors, and policy, and Section~\ref{sec:conclusion} concludes with limitations and directions for future research. Appendix~\ref{app:atypicality} analyses technological atypicality, which every main specification holds constant.

\section{Theoretical Background}\label{sec:theory}

The global shift toward a net-zero economy is a paradigmatic example of directed technical change, in which policy and market signals are intended to steer innovation toward societal goals \parencite{acemoglu2012environment}. To assess the effectiveness of these signals, policymakers and scholars require robust, valid, and reliable indicators of inventive activity. For decades, patent statistics have served this role, providing granular, long-term data on the rate and direction of technological change \parencite{Griliches1990,jaffe2002patents}. The introduction of the CPC's ``Y'' section, and specifically the Y02 scheme for climate-change mitigation technologies, was a landmark development in this tradition, creating a dedicated policy instrument to track the outputs of this directed innovation effort \parencite{Veefkind2012}.

The theory is organized around the paper's three questions.
Sections~\ref{sec:esg_disconnect} and~\ref{sec:label_assignment} explain
why the validity of Y02 is in question and how its labels are assigned.
Section~\ref{sec:construct_validity} grounds the questions in construct
validity. Section~\ref{sec:error_as_signal} develops the typology of errors
that the first question requires and the formal distinction between
idiosyncratic error and bias that the second requires.
Section~\ref{sec:admin_mechanisms} describes the mechanisms behind each
administrative error, and Section~\ref{sec:competing_accounts} sets them
out as four competing accounts of their causes, the subject of the third
question. Section~\ref{sec:model_error_transmission} turns to the model's
own errors.

\subsection{The ESG Innovation Disconnect and the Instruments That Observe It}
\label{sec:esg_disconnect}

Despite the widespread adoption of Y02-based indicators, a significant empirical gap has emerged in the literature: the ESG innovation disconnect. \textcite{Cohen2026} show that oil, gas, and energy-producing
firms, firms with lower ESG scores that are often excluded from ESG
funds, are key innovators in the U.S. green-patent landscape: they
produce more, and higher-quality, green innovation than other firms and
are the leading innovators in some branches, notably carbon capture and
storage. Incumbent firms in carbon-intensive, heavy industries are often among the most prolific and influential producers of green patents despite weaker ESG profiles \parencite{Lan2025}.

This paradox presents a fundamental challenge to innovation theory and policy. It shapes where green inventions are believed to occur, which firms and sectors are seen as leading the transition, and how public and private resources are directed towards climate-relevant technologies. If patent-based indicators systematically overcount some fields and undercount others, observed patterns may reflect measurement bias rather than the true distribution of green inventive activity. The observed ESG innovation disconnect may therefore be shaped by two broad, non-exclusive forces. The first is a firm-side account. Firm-side mechanisms may create a substantive disconnect between environmental strategy, communication, and green innovation. Firms may fail to translate environmental commitments into corresponding R\&D activities, consistent with policy--practice decoupling \parencite{BromleyPowell2012,delmas2011drivers}. Even when environmental innovation activities are undertaken, sector-specific capabilities, organizational routines, and managerial cognition may weaken the link between those activities and their intended technological outcomes. This pattern can be interpreted as firm-level means--ends decoupling, although the innovation literature primarily identifies the underlying capability and inertia mechanisms rather than using that term directly \parencite{BromleyPowell2012,HendersonClark1990,TripsasGavetti2000}. Genuine green inventions may remain under-recognized for two distinct reasons. One reason is strategic: firms may deliberately downplay or withhold communication about their environmental achievements, consistent with recent work on greenhushing and brownwashing \parencite{Hilton2025,MontgomeryRobertsonSummers2026} and foundational research on greenhushing and strategic silence \parencite{font2017greenhushing,carlos2018strategic}. A second reason is non-strategic: an invention's climate contribution may be accurately described but embedded in specialized engineering, scientific, or legal language rather than explicitly framed in environmental terms. We refer to this mechanism as technical under-framing. Recent research documents both the distinctive linguistic complexity of patent texts and important gaps across green-patent classification systems \parencite{JiangGoetz2025,Santarlasci2023,Rainville2025}, while the original development of the Y02 taxonomy provides the foundational classification context \parencite{Veefkind2012}.

The second is measurement. Error on either side of the
relationship can manufacture a disconnect, conceal one, or distort its
shape, depending on which firms it favors. ESG ratings may imperfectly represent firms' environmental performance because providers differ in their scope, measurement methods, and weighting schemes \parencite{Berg2022,Chatterji2016}. Patent-based indicators may likewise exhibit a form of means--ends decoupling in administrative measurement: classification procedures are implemented, yet the resulting labels do not consistently achieve their intended objective of identifying substantively climate-relevant inventions \parencite{BromleyPowell2012}. Because substantive climate relevance is often difficult to infer from specialized patent language, classification systems operating under limited attention and codified taxonomies may rely on simplifying heuristics, privileging salient lexical cues and familiar technological categories over less visible functional mechanisms \parencite{JiangGoetz2025,Simon1955,TverskyKahneman1974}. The resulting gap between the substantive construct and its administrative representation can weaken measurement validity and generate systematic differences across green-patent classifications \parencite{JacobsWallach2021,Rainville2025}. More generally, patents and administrative taxonomies are imperfect indicators of innovation and technological constructs \parencite{Griliches1990,Nagaoka2010,Santarlasci2023,Veefkind2012}. Although measurement divergence among ESG ratings has been extensively documented, considerably less attention has been paid to whether the patent-based indicators on the innovation side are themselves systematically biased. Our study addresses this second measurement problem by examining whether Y02 over-recognizes some technological fields while under-recognizing others, and hence in which direction patent-side error moves the measured disconnect.

\subsection{How Green Patent Labels Are Assigned}
\label{sec:label_assignment}

Green patent indicators do not directly observe whether an invention is environmentally beneficial. Instead, they rely on administrative classification systems that translate technical patent documents into standardized technology codes. The CPC Y02 scheme was developed to identify climate-change mitigation and adaptation technologies that were previously dispersed across conventional patent classes \parencite{Veefkind2012}. Y02 has therefore become a widely used proxy for green invention in research, policy monitoring, and ESG-related analysis.

However, assigning a Y02 label is not the same as measuring substantive environmental impact. It is an administrative classification decision in which patent text is mapped onto a predefined taxonomy. This process may involve examiner judgment, classification expertise, automated search procedures, and patent-office updating systems \parencite{Santarlasci2023}. As a result, Y02 should be understood as an administrative proxy for climate relevance rather than as ground truth.

This distinction matters because the environmental function of a patent is not always explicit. Some inventions use visible green language, such as emissions reduction, renewable energy, carbon capture, or recycling. Others contribute to decarbonization through technical mechanisms such as catalyst design, thermal efficiency, fuel injection, power electronics, materials substitution, or industrial-process optimization. These inventions may be substantively green even when they do not use overt environmental vocabulary.

This creates a bounded-rationality problem in green patent classification. Patent classifiers and administrative systems operate under limited time, attention, and information-processing capacity \parencite{Simon1955,Simon1957}. Under such constraints, classification may rely on simplifying heuristics, such as salient keywords, familiar technology templates, or established Y02 categories \parencite{TverskyKahneman1974}. This does not imply intentional misclassification, but it helps explain why classification errors may be systematic. Drawing on bounded rationality as the broader theoretical foundation, we use the term bounded classification capacity to describe its manifestation in administrative classification: limits in the ability of classifiers, taxonomies, and associated procedures to map heterogeneous and technically complex inventions onto established categories. Bounded classification capacity therefore refers to a property of the classification process rather than a claim about the cognitive limitations of any individual examiner.

From this perspective, two forms of Y02 error can arise. Patents with visible green language may be over-included even when their substantive climate contribution is weak, producing administrative  Type I errors or False Green. Conversely, patents whose climate relevance is embedded in complex technical functions may be missed, producing administrative Type II errors or Silent Green inventions. We therefore treat Y02 as an indicator to be validated, not as a final measure of green invention.

Our benchmark is the construct that Y02 is \emph{used} to measure
in research, policy monitoring, and finance, an invention's substantive
contribution to climate-change mitigation or adaptation, rather than the
scheme's own inclusion rules. The distinction matters because a False
Green can arise in two ways. A tag may misapply the scheme, as when a
retail ``shopping environment'' or a pressure exchanger for hydraulic
fracturing is tagged green. Or a tag may apply the scheme correctly while
the scheme's scope exceeds the construct, as when device-level power
management enters the ICT energy-efficiency family. Both inflate
Y02-based measures for users who read the tag as evidence of climate
contribution, which is why both threaten construct validity
\parencite{CronbachMeehl1955,AdcockCollier2001}. They nonetheless call for
different remedies: reclassification in the first case, and re-scoping or
explicit down-weighting of scheme segments in the second.

\subsection{Construct Validity and the ESG Innovation Disconnect}
\label{sec:construct_validity}

This study proposes a complementary patent-side account of the ESG innovation disconnect. Rather than treating the divergence between ESG scores and green patent counts as direct evidence of a failure in corporate environmental strategy, we argue that measurement bias in Y02 may reinforce this disconnect. In particular, the Y02 classification should not be treated as a neutral or self-validating measure of green invention. It is an administrative indicator whose validity must be assessed against the substantive technological function of the patents it classifies.

This argument builds on the literature on construct validity and social-science measurement. Construct validity refers to whether an empirical indicator actually measures the theoretical concept it is intended to capture \parencite{CronbachMeehl1955}. In social-science measurement, this requires a clear distinction between the underlying concept, the operational indicator, and the empirical observations generated by that indicator \parencite{AdcockCollier2001,JacobsWallach2021}. Applied to green invention, the underlying concept is substantive climate-related technological contribution, while the empirical indicator is the administrative Y02 label. The key validity question is therefore whether Y02 accurately captures patents that contribute directly and credibly to climate mitigation or adaptation.

From this perspective, the observed disconnect is a joint product
of firm behavior and of the two instruments (ESG ratings and green-patent indicators) used to observe it, and the
direction of each instrument's error decides whether it inflates or
conceals the disconnect. If Y02 over-includes patents with weak or symbolic environmental relevance, raw green patent counts may make some firms and sectors appear more green than they are. Conversely, if Y02 under-includes technically complex decarbonization inventions, especially in engineering-intensive sectors, it may hide important green inventions from firms that are central to the net-zero transition. The two errors need not offset. Because the
disconnect is a comparison between firms, a bias that inflated every
firm's count by the same factor would leave it intact; what matters is
relative error. When over-inclusion falls on the technologies of firms
with strong ESG recognition and omission on those of traditional energy
firms, the two errors compound: raw counts overstate the green invention
of the first group relative to the second, which biases the
measured relation between ESG recognition and green invention upward and
conceals part of the disconnect. The reverse pattern would manufacture
part of it. \textcite{Cohen2026} document an ESG innovation disconnect in which traditional energy firms make substantial contributions to green innovation despite weak ESG recognition. Their green-patent measure
follows the OECD ENV-TECH search strategies, which rely on Y02 wherever
possible \parencite{HascicMigotto2015}, so Y02's error structure passes
into it. Which pattern holds is therefore an empirical question about
where Y02 errs, which Sections~\ref{sec:results}
and~\ref{sec:idiosyncratic} answer at the level of technologies.

All large-scale classification systems contain some degree of error. The central validity concern is therefore not the mere existence of mistakes, but whether those mistakes are approximately non-differential or systematically related to the characteristics of the observations being measured. Random or non-differential error primarily reduces precision and weakens estimated relationships \parencite{Copeland1977,Bound2001,Hausman2001}. Systematic error is more consequential because its probability or direction varies across sectors, technologies, firms, or linguistic forms. In that case, classification error becomes measurement bias. This distinction is especially important because an aggregate correction alone cannot reveal measurement bias: opposing errors may partly offset one another in overall counts while still redistributing measured green invention across the technological landscape. The relevant question is therefore not only how many patents are misclassified, but whether the probability and direction of error vary systematically across technological fields, a question we test directly in Section~\ref{sec:idiosyncratic}.

Measurement theory distinguishes construct reliability, roughly
analogous to statistical precision, from construct validity, roughly
analogous to unbiasedness \parencite{JacobsWallach2021}. Random
classification error mainly threatens the first, whereas systematic bias
threatens the second. Construct validity has several facets, four of which
are particularly relevant here: substantive validity, discriminant validity,
convergent validity, and consequential validity. First, substantive validity, a component of content validity, asks whether
Y02 includes patents that are genuinely climate-relevant and excludes those
that are not. Silent Green violates the first requirement because relevant
inventions are missed, whereas False Green violates the second because
non-green inventions are included. Second, discriminant validity asks whether Y02 inadvertently captures
something other than climate relevance. The concentration of False Green in
digital-context vocabulary suggests that Y02 may partly capture a
neighbouring construct, digital systems management, rather than climate
contribution. Third, convergent validity asks whether Y02 agrees with another measure of
the same underlying construct. Disagreement between Y02 and the text-based
model therefore provides a signal that one of the two measures may be wrong. Fourth, consequential validity concerns what follows from using the measure.
For Y02, its errors can redistribute measured green invention across
technologies, as documented in Section~\ref{sec:corrected_landscape}. Management research has long treated the validity of a widely used proxy as
a first-order question \parencite{Demerjian2012}. What distinguishes the
present setting is that Y02 is an administrative classification whose
failures can be examined and attributed case by case.

\subsection{Error as Signal: Dual-Measurement Disagreement}
\label{sec:error_as_signal}

We conceptualize green-patent identification as a dual-measurement problem. Let $G_i\in\{0,1\}$ denote the latent substantive climate relevance of patent $i$, $A_i\in\{0,1\}$ its administrative Y02 label, and $M_i\in\{0,1\}$ the model prediction. Because $G_i$ is not directly observed at scale, the empirical analysis uses the direct-mechanism consensus of the construct-validity test as the reference judgment for resolved cases. Administrative and model errors are respectively defined as
\[
e_i^{A}=A_i-G_i,
\qquad
e_i^{M}=M_i-G_i.
\]
A value of zero indicates a correct classification, a positive value indicates the incorrect inclusion of a non-green patent, and a negative value indicates the omission of a genuinely green patent. The distinction between random error and systematic bias can be expressed conditionally. If
\[
E[e_i^{A}\mid X_i]\approx 0,
\]
where $X_i$ represents technological domain, sector, linguistic visibility, or complexity, administrative errors behave primarily as noise with respect to those characteristics. If instead
\[
E[e_i^{A}\mid X_i]\neq 0,
\]
the administrative indicator is conditionally biased. The same logic applies to model error. Bias therefore refers not simply to non-normality or to a high number of mistakes, but to a persistent direction or unequal probability of error across relevant groups. We use the two terms accordingly throughout. An \emph{error} is a property of an individual classification: each False Green or Silent Green patent, and each model false positive or false negative, is one. \emph{Bias} is a property of the indicator: a pattern in its errors that varies with observable characteristics of the inventions it classifies, so that $E[e_i^{A}\mid X_i]\neq 0$. Any classifier of millions of documents makes errors; the claim of this paper concerns bias. The distinction also separates two consequences for research that uses Y02. Random error mainly reduces precision and weakens estimated relationships, whereas bias can shift or reverse them, and a larger sample does not remove it (Section~\ref{sec:discussion}). Model--administrative disagreement provides a scalable diagnostic signal. Define
\[
D_i=\mathbf{1}(A_i\neq M_i).
\]
When $D_i=1$, the two measurement systems cannot both be correct relative to a binary substantive truth. However, disagreement alone does not reveal which system is wrong. A downward disagreement ($A_i=1,M_i=0$) may be an administrative Type I error or a model Type II error. An upward disagreement ($A_i=0,M_i=1$) may be an administrative Type II error or a model Type I error. Substantive adjudication is therefore necessary to attribute the error source. This logic follows research on multiple imperfect observers and noisy labels, which cautions against treating any single observed label as an infallible reference standard \parencite{DawidSkene1979,FrenayVerleysen2014,Northcutt2021}. Table~\ref{tab:error_attribution_framework} summarizes the resulting attribution scheme; the empirical frequency of each outcome is reported in Section~\ref{sec:results}.

\begin{table}[H]
\centering
\caption{Four errors behind one disagreement signal}
\label{tab:error_attribution_framework}
\small
\setlength{\tabcolsep}{5pt}
\renewcommand{\arraystretch}{1.25}
\begin{tabularx}{\textwidth}{@{}>{\raggedright\arraybackslash}p{3.1cm} L L
    >{\raggedright\arraybackslash}p{1.9cm}@{}}
\toprule
& \multicolumn{2}{>{\centering\arraybackslash}p{10.2cm}}{\textbf{Construct-validity test: what does the invention do?}} & \\
\cmidrule(lr){2-3}
\textbf{Disagreement}
& \textbf{No direct climate mechanism} ($L_i=0$)
& \textbf{Direct climate mechanism} ($L_i=1$)
& \textbf{No consensus} \\
\midrule
\textbf{Downward}\newline $A_i=1$, $M_i=0$\newline
\textit{Y02 says green; the model does not}
& \cellcolor{cellgreen}\textbf{False Green}\newline
  administrative Type~I\newline
  \textit{construct validity}: Y02 errs ($e_i^{A}=+1$)
& \textbf{Model false negative}\newline
  model Type~II\newline
  \textit{model error}: the model errs ($e_i^{M}=-1$)
& Unresolved \\
\addlinespace
\textbf{Upward}\newline $A_i=0$, $M_i=1$\newline
\textit{The model says green; Y02 does not}
& \textbf{Model false positive}\newline
  model Type~I\newline
  \textit{model error}: the model errs ($e_i^{M}=+1$)
& \cellcolor{cellgreen}\textbf{Silent Green}\newline
  administrative Type~II\newline
  \textit{construct validity}: Y02 errs ($e_i^{A}=-1$)
& Unresolved \\
\bottomrule
\end{tabularx}

\vspace{0.4em}
\begin{minipage}{0.97\textwidth}
\footnotesize
\textit{Notes:} Shaded cells lie on the construct-validity
diagonal: the test contradicts the administrative label, and only these
cells correct the indicator. Attributed cells on the anti-diagonal are
model errors: the test contradicts the model, and they correct the model,
not the label. Evaluation that treats Y02 as ground truth scores all four
attributed cells as model error. $A_i$ is the administrative Y02 label and
$M_i$ the PatentSBERTa prediction (1 = green). $L_i$ is the Qwen--Gemma
direct-mechanism consensus: 1 indicates agreement that the invention's
primary function contains a direct climate-mitigation or adaptation
mechanism, 0 agreement that it does not; incidental environmental benefits
are coded 0. Cases without cross-model consensus are unresolved and retain
their administrative status. Confidence is analysed separately and does
not alter the consensus label. Counts for each cell are reported in
Figure~\ref{fig:error_as_signal}.
\end{minipage}
\end{table}

The framework changes the role of prediction error. Under conventional evaluation, a mismatch with an observed label is counted as model failure. Under imperfect administrative measurement, the same mismatch becomes an error signal: it identifies a case in which the validity of either the administrative label or the model prediction should be examined. Once validated, the direction, confidence, sectoral concentration, and linguistic characteristics of the errors reveal whether the measurement system is affected by lexical over-inclusion, functional opacity, taxonomic under-coverage, model overgeneralization, or domain shift.

\subsection{Administrative Error Mechanisms: False Green and Silent Green}
\label{sec:admin_mechanisms}

False Green and Silent Green refer specifically to consensus-attributed administrative errors. They should not be inferred directly from model--administrative disagreement. A downward disagreement becomes False Green only when the construct-validity test supports a non-green judgment; an upward disagreement becomes Silent Green only when the test supports a direct climate mechanism. In the large-scale analysis, this attribution is made only under cross-model LLM consensus.

\subsubsection{Administrative Type I Error: False Green}

False Green refers to the misclassification of a patent as green when it offers no substantive climate-mitigation or adaptation benefit. These false positives often arise from lexical ambiguity, where broad terms such as environment, efficiency, hybrid, or power management are used in non-climate contexts, for example in expressions such as software environment, hybrid data architecture, or generic power management.

The literature suggests two broad explanations for this form of over-inclusion. The first is a firm-level symbolic explanation. From an institutional perspective, firms may respond to regulatory, investor, and market expectations by adopting visible green signals that enhance legitimacy, even when these signals are only weakly connected to substantive environmental change \parencite{meyer1977institutionalized,dimaggio1983iron}. This logic is consistent with the greenwashing literature, which emphasizes institutional, market, and disclosure-related drivers of symbolic environmental claims \parencite{delmas2011drivers,marquis2016scrutiny}. Recent patent-based studies provide related evidence that green finance, ESG pressure, and environmental regulation can increase visible green innovation outputs without necessarily improving the substantive quality of those inventions, suggesting a form of patent-based greenwashing or symbolic green innovation \parencite{shi2023green,zhang2022environmental,peng2024corporate}.

A second explanation is measurement-based. Y02 errors do not necessarily imply intentional firm-level greenwashing. Instead, they may arise because the indicator used to measure green invention is itself imperfect. The construct-validity literature emphasizes that observed indicators should not be treated as ground truth unless they are validated against the underlying construct they are intended to measure \parencite{CronbachMeehl1955,AdcockCollier2001}. Applied to Y02, this means that administrative green labels should be evaluated against the substantive technical function of the invention rather than assumed to capture true climate relevance. Recent green-patent classification studies support this concern by showing that standard green-patent indicators can generate both false positives and false negatives \parencite{Santarlasci2023}.

This study builds on the measurement-based view while remaining attentive to the firm-level symbolic explanation. We distinguish between firm-induced False Green, where firms strategically produce or frame patents as green signals, and indicator-induced False Green, where the measurement system over-includes patents with green-sounding but substantively non-climate technical language. The same observed outcome, an inflated green-patent count, can therefore arise from strategic symbolic behaviour or from structural bias in the classification infrastructure. We test this distinction in Section~\ref{sec:greenwashing_event} by examining whether explicit environmental framing rose after green labels became salient.

\subsubsection{Administrative Type II Error: Silent Green}

Silent Green inventions are genuinely climate-relevant patents that are not tagged as green because their environmental contribution is embedded in technical functionality rather than explicitly described with environmental keywords \parencite{Block2024}. These errors are especially likely in engineering, chemistry, energy systems, transport, and heavy industry, where decarbonization benefits are expressed through process optimization, materials design, thermal efficiency, emissions-control mechanisms, power electronics, catalysts, or energy-conversion systems rather than overt climate-oriented vocabulary \parencite{Cohen2026}. Silent Green represents Y02 under-inclusion and can systematically disadvantage engineering-intensive technologies central to decarbonization \parencite{Cohen2026,Lan2025}.

A bounded-rationality perspective helps explain why this under-inclusion may be systematic rather than random. Patent examiners and administrative classification systems operate under limited time, attention, and information-processing capacity \parencite{Simon1955,Simon1957}. Classification may therefore rely partly on simplifying heuristics, including visible environmental keywords, familiar Y02 templates, and established technology-category mappings. A catalyst, thermal-control system, battery component, engine-control mechanism, or industrial process may contribute directly to decarbonization while being described primarily in engineering or chemical language. Technically complex green inventions can consequently remain invisible to an administrative indicator that is more sensitive to explicit environmental language than to functional climate relevance. Section~\ref{sec:complexity_silent_green} tests this account directly by asking whether the probability and the direction of administrative error vary with the complexity of the invention's technological knowledge base.

Greenhushing or technical under-framing may provide a secondary firm-level mechanism. Greenhushing refers to the deliberate under-communication of genuine sustainability practices, while related work on strategic silence shows that organizations may withhold positive environmental information to avoid reputational scrutiny or accusations of hypocrisy \parencite{font2017greenhushing,carlos2018strategic}. In the patent context, firms or patent attorneys that do not explicitly articulate the environmental contribution of an invention may reduce the probability that examiners or automated systems recognize its climate relevance.

However, Silent Green does not require intentional under-communication. It can arise even when applicants accurately describe an invention because the environmental contribution is embedded in complex technical functionality that the Y02 classification system fails to recognize. Our framework therefore treats greenhushing as a possible contributing mechanism, while emphasizing structural measurement bias under bounded classification capacity.

The distinction between False Green and Silent Green matters because the two errors create different policy risks. Administrative Type I errors can generate a false sense of progress and reward low-impact or merely green-sounding activity. Administrative Type II errors can conceal strategically important inventions and may contribute to their under-recognition in research and investment decisions \parencite{Cohen2026}. Silent Green is therefore not simply missing data; it is a systematic form of under-recognition.

We use technological legibility to describe the classification
problem that bounded capacity creates: the extent to which an invention's
substantive function can be readily mapped onto established technological
categories. Two measures of the technological knowledge base capture
distinct sources of legibility. Reflection complexity, building on the
method-of-reflections approach to complex technological knowledge
\parencite{BallandRigby2017}, is high when a technology is associated with
places that possess diversified and relatively scarce technological
capabilities; we refer to it as capability scarcity. Structural complexity
captures complexity in the combinatorial network structure of
technological knowledge \parencite{Broekel2019}; we refer to it as
combinatorial diversity. We do not assume ex ante that these dimensions have identical effects; their distinct constructions allow us to test which form of complexity is associated with administrative omission.

This distinction suggests two different routes to technological legibility. Reflection complexity captures technologies that rely on rare and sophisticated capabilities. These technologies may be economically advanced but harder for administrative classification systems to recognize correctly. Structural complexity captures a different property: diversity in the combinatorial structure of technological knowledge. Technologies with more recognizable connections to existing technological categories may be easier for the classification system to identify, reducing the risk of Silent Green. At the same time, these multiple classification cues may create more opportunities for erroneous inclusion, making errors relatively more likely to take the form of False Green. We interpret these patterns as consistent with under- and over-matching, rather than as evidence of a proven causal mechanism.

Complexity is not the only source of illegibility. An invention may
also recombine technological knowledge in unfamiliar ways, a dimension
that research on recombinant search separates from the difficulty
generated by the underlying technological structure
\parencite{Fleming2001}. We hold this technological atypicality constant
in every main specification and analyse its own, nonlinear relationship
with omission in Appendix~\ref{app:atypicality}; atypicality and
reflection complexity are only weakly correlated.

The bounded-capacity account therefore yields a directional
implication that a generic difficulty account does not. If
under-recognition reflects technological legibility rather than
difficulty, the two forms of complexity should be associated with Silent
Green in opposite directions. Moreover, if the resulting error is directional rather than generic classification noise, complexity should predict whether an administrative mistake takes the form of omission (Silent Green) rather than erroneous inclusion (False Green). The latter test directly connects patent characteristics to the sign of administrative measurement error.

\subsection{Four Competing Accounts of Administrative Error}
\label{sec:competing_accounts}

The preceding arguments yield four accounts of why Y02 errs,
arranged by the direction of the error and by where it originates
(Table~\ref{tab:accounts_design}). Two are strategic and concern what
applicants do. Greenwashing would generate False Green: applicants use
environmental language to attract the green tag. Greenhushing would
generate Silent Green: firms stay quiet about environmental achievements,
masking green function. Two concern how the label is produced under
bounded classification capacity, and each is indexed by a form of
technological complexity. Structural complexity, or combinatorial
diversity, offers many recognizable anchors, so that when placement is
wrong it is likely to be wrong in the direction of inclusion; we call this
over-matching. Reflection complexity, or capability scarcity, offers few,
so that a genuine climate function is likely to go unread; we call this
under-matching. The accounts are not mutually exclusive, and each is
likely to explain some individual cases. The empirical question is which explains the bias, the systematic component of the error, the part that varies with the technology rather than with the case.

Each account leaves a distinct signature, but not every signature
discriminates. Greenwashing predicts that False Green and explicit green
framing rise discretely when the green tag becomes salient, a prediction
that the introduction of Y02 tagging in 2010 and the launch of the CPC in
2013 allow us to test. Greenhushing predicts that genuinely green
inventions are described in technical rather than environmental
vocabulary; because non-strategic technical description predicts the
same, the account can be argued from cases but not directly tested. The
two capacity accounts predict opposite tilts in the direction of error, so
that complexity should predict not only whether Y02 errs but which way.
Section~\ref{sec:causes} tests these signatures row by row, the
strategic accounts in Section~\ref{sec:greenwashing_event}, the capacity
accounts in Section~\ref{sec:complexity_silent_green}, and
Table~\ref{tab:competing_accounts} reports the verdicts.

\begin{table}[H]
\centering
\caption[Two errors, four accounts: origins and predictions]{Two errors, four accounts: where each account originates and
what it predicts}
\label{tab:accounts_design}
\footnotesize
\setlength{\tabcolsep}{6pt}
\renewcommand{\arraystretch}{1.15}
\begin{tabularx}{\textwidth}{@{}>{\raggedright\arraybackslash}p{2.35cm} L L@{}}
\toprule
& {\color{fgorange}\textbf{Type~I}} $\cdot$ \textbf{False Green} $\cdot$ over-inclusion
& {\color{sgteal}\textbf{Type~II}} $\cdot$ \textbf{Silent Green} $\cdot$ omission \\
\midrule
\textbf{Strategic account}\newline\textit{what applicants do}
&
\acct{fgorange}{Greenwashing}\newline
Applicants use environmental language to attract the green tag.\newline
\textsc{Tested by} difference-in-differences in explicit green language
around the 2010 Y02 introduction and the 2013 CPC launch, and an
interrupted time series of the False Green rate
(Section~\ref{sec:greenwashing_event}).
&
\acct{sgteal}{Greenhushing}\newline
Firms stay quiet about environmental achievements, masking green
function.\newline
\textsc{Argued from} consensus-resolved cases whose climate function is
described in technical rather than green vocabulary
(Section~\ref{sec:greenwashing_event}).
\\[3pt]
& \predbar{Predicts a discrete break at either event}
& \predbar{Predicts technical, not green, vocabulary}
\\
\addlinespace[8pt]
\textbf{Bounded classification capacity}\newline
\textit{how the label is produced}
&
\acct{fgorange}{Structural complexity}\newline
\textbf{Combinatorial diversity.} High when knowledge exhibits diverse
network structure. Many recognizable anchors open routes for non-green
inventions into green categories (over-matching).\newline
\textsc{Tested by} logistic models of Silent Green and of error direction
(Section~\ref{sec:complexity_silent_green}).
&
\acct{sgteal}{Reflection complexity}\newline
\textbf{Capability scarcity.} High when a technology is tied to places
with scarce capabilities. The invention maps onto no familiar category, so
a real climate function goes unread (under-matching).\newline
\textsc{Tested by} the same models
(Section~\ref{sec:complexity_silent_green}).
\\[3pt]
& \predbar{Predicts errors tilted toward False Green}
& \predbar{Predicts errors tilted toward Silent Green}
\\
\bottomrule
\end{tabularx}

\vspace{0.4em}
\begin{minipage}{0.97\textwidth}
\footnotesize
\textit{Notes:} Rows distinguish where an error originates;
columns distinguish its direction. The accounts are not mutually
exclusive, and each can explain individual cases; the empirical question
is which carries the systematic component of the error. The bar under each
cell states the account's prediction; Table~\ref{tab:competing_accounts}
reports the corresponding verdicts. All tests are conducted within the
\num{517772} disagreement cases; errors that PatentSBERTa shares with Y02
generate no disagreement and lie outside this comparison. Every main
specification holds constant the atypicality of an invention's
technological combinations (Appendix~\ref{app:atypicality}).
\end{minipage}
\end{table}

\subsection{Model Error Correction and Transmission}
\label{sec:model_error_transmission}

The rapid growth of patent filings has increased the need for machine learning (ML) methods capable of processing large-scale, multi-label, and highly imbalanced patent data. Encoder-based models such as PatentSBERTa are particularly useful because they provide deterministic and high-throughput mappings from patent text to administrative classifications \parencite{Bekamiri2024PatentSBERTa}. However, when a model is trained to reproduce an administrative label, disagreement with the official classification does not necessarily indicate model failure. It may instead reveal a measurement error in the administrative label. Because PatentSBERTa is trained using administrative Y02 labels, its predictions are not independent of the administrative classification process. Systematic errors in Y02 may enter the training data as label noise and shape the model's decision boundary \parencite{FrenayVerleysen2014,Northcutt2021}. At the same time, PatentSBERTa may generate additional errors through class imbalance, threshold selection, limitations in textual representation, or insufficient coverage of emerging and atypical technologies.

We distinguish two principal relationships between administrative and model error: error correction and error transmission. Error correction occurs when PatentSBERTa challenges an incorrect administrative label and agrees with substantive adjudication:
\[
A_i \neq M_i,
\qquad
M_i = G_i
\]
where \(A_i\) denotes the administrative Y02 label, \(M_i\) the PatentSBERTa classification, and \(G_i\) the substantive adjudication of patent \(i\). Within the resolved disagreement pool, PatentSBERTa corrects \num{209849} administrative errors, comprising \num{180384} administrative Type~I errors and \num{29465} administrative Type~II errors.

Error transmission occurs when distortions associated with the administrative classification persist in, or become stronger through, the predictive model. Transmission may take two forms: bias preservation and bias amplification. Bias preservation occurs when the administrative classification and PatentSBERTa produce the same incorrect label:
\[
A_i = M_i \neq G_i
\]
In such cases, the model preserves an administrative error rather than correcting it. These shared errors remain hidden in disagreement-based validation because the two measurement systems generate no disagreement signal. The potential search space for preserved errors consists of all dual-measurement agreement cases, which constitute the vast majority of the corpus (Section~\ref{sec:results}). The potential search space for preserved errors contains \num{8557649} dual-measurement agreement cases, comprising \num{8216292} joint non-green classifications and \num{341357} joint green classifications. These agreement cases should not themselves be interpreted as preserved errors. Identifying preservation requires substantive adjudication of a representative sample drawn from this space. Bias amplification occurs when the predictive model introduces additional errors that reinforce and increase an existing administrative distortion \parencite{Zhao2017,WangRussakovsky2021}. Cases in which the construct-validity test supports the administrative label while PatentSBERTa is incorrect represent potential sources of amplification, because the model adds errors beyond those made by the administrative classification. They constitute evidence of amplification only when their technological distribution reinforces and increases an existing administrative pattern, such as the over-recognition of ICT technologies or the under-recognition of energy, transport, chemical, or industrial technologies. Amplification is therefore a comparative group-level effect and cannot be inferred automatically from an individual model error. Section~\ref{sec:conservative_by_design} shows that this
comparative condition is met: the same lexical cue that drives
over-inclusion by the administrative label inside Y02 also drives
over-inclusion by the model outside it.

In this study, dual-measurement disagreement is used as a diagnostic signal for potential classification error, consistent with methods that use model predictions to identify potentially mislabelled observations \parencite{Northcutt2021}. The subsequent evaluation and attribution of these disagreements draw on the algorithmic-auditing literature \parencite{Raji2020}. The purpose of the construct-validity test is therefore not merely to assess how accurately PatentSBERTa reproduces Y02, but to determine whether each disagreement is more consistent with an administrative error or a model error. PatentSBERTa functions as a scalable diagnostic instrument rather than as a replacement ground truth. Both administrative labels and model predictions are imperfect operationalizations of the underlying construct of substantive climate relevance \parencite{JacobsWallach2021,MishraGorana2021}. Independent substantive adjudication is therefore required to determine the likely source and direction of error. In the present study, this role is performed by the construct-validity test, a consensus-based LLM evaluation, while cases without sufficient agreement are retained as unresolved rather than assigned a definitive error source.

The empirical analysis directly measures error correction and, in
Section~\ref{sec:conservative_by_design}, documents bias amplification
within the full pool of dual-measurement disagreement cases. Bias preservation is not directly observed because preserved errors are located within the considerably larger agreement space. A comprehensive analysis of preservation would therefore require substantive tests of representative agreement cases to identify preserved errors. Disagreement-based validation thus uses predictive disagreement to locate possible measurement failures while recognizing that disagreement alone cannot identify shared errors.

\section{Methodology}\label{sec:methodology}

\begin{figure}[H]
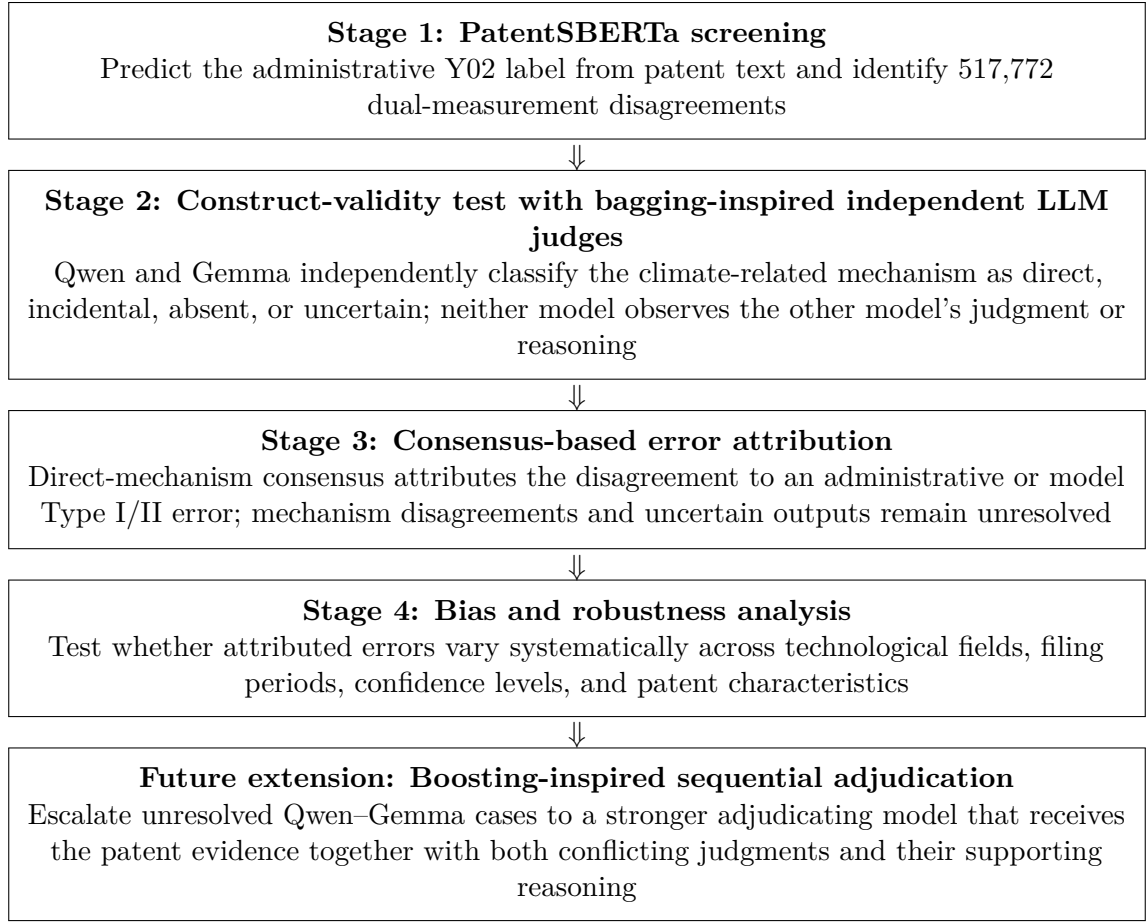

\centering
\setlength{\fboxsep}{7pt}

\begin{tabular}{c}

\fbox{\parbox{0.88\textwidth}{\centering
\textbf{Stage 1: PatentSBERTa screening}\\
Predict the administrative Y02 label from patent text and identify
\num{517772} dual-measurement disagreements}}
\\[0.8em]

$\Downarrow$
\\[-0.2em]

\fbox{\parbox{0.88\textwidth}{\centering
\textbf{Stage 2: Construct-validity test with bagging-inspired independent LLM judges}\\
Qwen and Gemma independently classify the climate-related mechanism as
direct, incidental, absent, or uncertain; neither model observes the
other model's judgment or reasoning}}
\\[0.8em]

$\Downarrow$
\\[-0.2em]

\fbox{\parbox{0.88\textwidth}{\centering
\textbf{Stage 3: Consensus-based error attribution}\\
Direct-mechanism consensus attributes the disagreement to an administrative
or model Type~I/II error; mechanism disagreements and uncertain outputs
remain unresolved}}
\\[0.8em]

$\Downarrow$
\\[-0.2em]

\fbox{\parbox{0.88\textwidth}{\centering
\textbf{Stage 4: Bias and robustness analysis}\\
Test whether attributed errors vary systematically across technological
fields, filing periods, confidence levels, and patent characteristics}}
\\[0.8em]

$\Downarrow$
\\[-0.2em]

\fbox{\parbox{0.88\textwidth}{\centering
\textbf{Future extension: Boosting-inspired sequential adjudication}\\
Escalate unresolved Qwen--Gemma cases to a stronger adjudicating model that
receives the patent evidence together with both conflicting judgments and
their supporting reasoning}}

\end{tabular}

\caption{Research workflow for detecting, adjudicating, and testing administrative
and model errors in Y02 green-patent classification. The construct-validity test uses a
bagging-inspired ensemble in which Qwen and Gemma evaluate each disagreement
independently. A future boosting-inspired extension would concentrate additional
adjudication on unresolved cases by providing a stronger model with the competing
first-stage judgments and their supporting evidence.}
\label{fig:researchworkflow}
\end{figure}

Figure~\ref{fig:researchworkflow} summarizes the research design. The workflow separates four tasks that are often conflated in conventional classification studies \parencite{DawidSkene1979,Northcutt2021,Raji2020}: learning the administrative label, detecting dual-measurement disagreement, adjudicating the likely source of resolved disagreement, and testing whether attributed errors are systematically distributed. PatentSBERTa first screens the full patent corpus and identifies cases in which its prediction differs from the administrative Y02 classification. These disagreement cases are then evaluated using a bagging-inspired independent LLM ensemble \parencite{Breiman1996}, which serves as the construct-validity test. Qwen and Gemma receive the same patent information, classification rubric, and direct-mechanism criterion, but evaluate each patent independently. Neither model observes the other model's prediction, explanation, or confidence assessment. Their judgments are subsequently aggregated through cross-model consensus. This parallel design reduces dependence on the idiosyncratic reasoning and error tendencies of any single LLM.

A case is substantively resolved only when Qwen and Gemma agree on whether the patent contains a direct climate-mitigation or adaptation mechanism. Consensus judgments are then used to attribute the disagreement to an administrative Type~I or Type~II error, or to a model Type~I or Type~II error. Cases involving mechanism disagreement or uncertain outputs are retained as unresolved rather than being forced into an error category. The resolved cases are subsequently used to test whether attributed errors vary systematically across technological fields, filing periods, confidence levels, and patent characteristics.

The treatment of unresolved cases also suggests a possible boosting-inspired sequential adjudication extension \parencite{FreundSchapire1997}. In such a procedure, only cases that remain unresolved after the independent first-stage test would be passed to a stronger adjudicating model. The adjudicator would receive the patent text together with the conflicting judgments, cited technical evidence, and reasoning produced by Qwen and Gemma, and would be asked to evaluate the competing interpretations. This approach is boosting-inspired because later evaluation is concentrated on difficult cases that the initial models failed to resolve, and the information generated at the earlier stage is used to guide the subsequent assessment. It nevertheless differs from classical statistical boosting because the models are not iteratively retrained, observations are not formally reweighted, and predictions are not combined through learned ensemble weights. To avoid anchoring or order effects, future implementation should randomize the presentation order of the competing judgments and require the adjudicating model to assess the underlying patent evidence rather than simply select between the earlier answers.

\subsection{Data}

Our empirical analysis is based on the full corpus of \num{9075421} USPTO granted patents issued between 1962, the earliest grant year represented in the assembled corpus, and 2024. For each patent, we combine patent text, CPC classifications, filing year, assignee information, citation-based indicators, technological-community measures, and geolocation variables. The administrative green label is defined using the CPC Y02 classification. A patent is coded as $A_i=1$ if it has been assigned at least one Y02 code, and as $A_i=0$ otherwise. In the full corpus, \num{592387} patents are administratively labelled as Y02 green patents, corresponding to approximately 6.5\% of all granted patents.

\begin{figure}[h]
    \centering
    \includegraphics[width=0.8\textwidth]{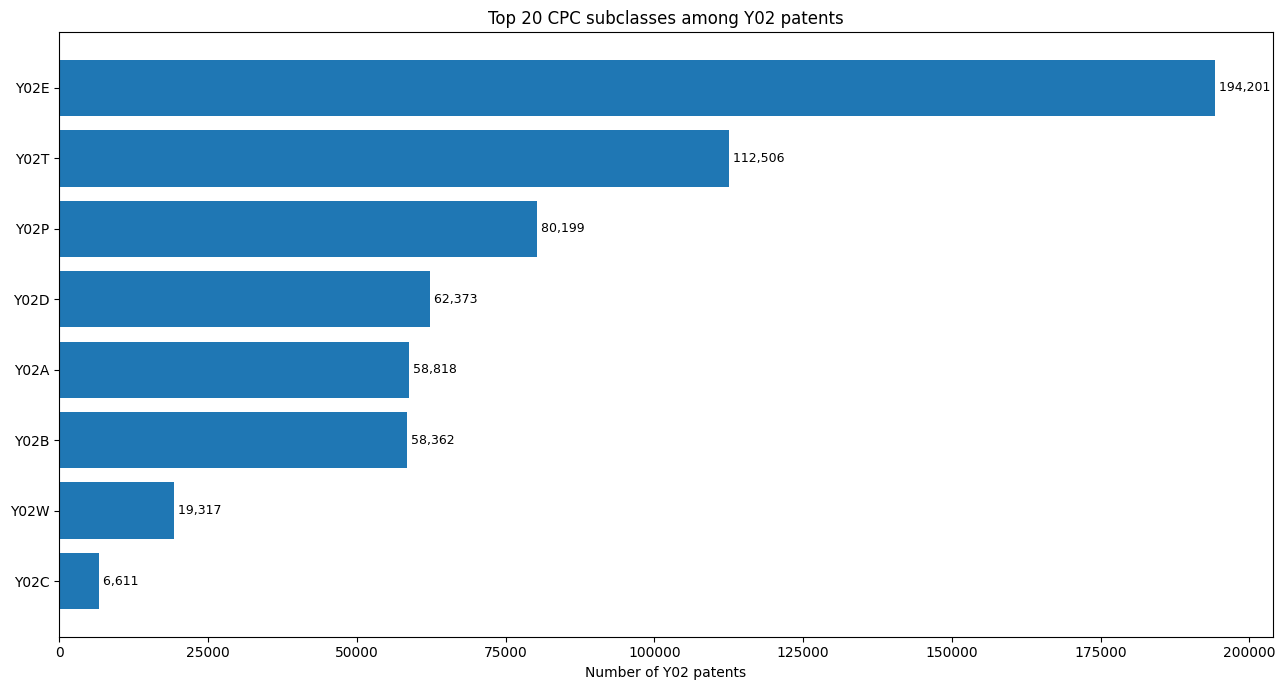}
    \caption{Distribution of the administrative CPC Y02 label}
    \label{fig:distribution_y02}
\end{figure}

The Y02 classification is widely used as a standardized patent-based indicator of climate-related technology. It covers technologies related to climate-change mitigation and adaptation, including renewable energy, energy efficiency, low-carbon transport, climate adaptation, carbon capture, and industrial decarbonization. As shown in Figure~\ref{fig:distribution_y02}, administratively labelled Y02 patents represent a relatively small share of the full patent universe, creating a highly imbalanced measurement setting.

\subsection{Dual-Measurement Disagreement as an Error Signal}

We fine-tuned PatentSBERTa \parencite{Bekamiri2024PatentSBERTa}, a BERT-based model pre-trained on patent data, and added a multilayer perceptron (MLP) classification head. In simple terms, this small neural-network layer converts the PatentSBERTa representation into a green/non-green probability used to predict the administrative Y02 label from patent text. The model's purpose is not to replace the Y02 taxonomy or directly define substantive greenness. It provides a scalable second measurement of the same underlying construct and thereby creates diagnostic tension with the administrative label. After fine-tuning, the classifier was deployed to the full corpus of \num{9075421} granted U.S. patents. Among these patents, \num{592387} were administratively labelled Y02. Using the validation-calibrated decision threshold of $\tau=0.88$, the model agreed with the administrative Y02 label for \num{341357} Y02 patents but classified \num{251030} administratively green patents as non-green. At the same time, it classified \num{266742} patents without an administrative Y02 label as potentially green. Table~\ref{tab:full_deployment_matrix} reports the deployment matrix.

\begin{table}[H]
\centering
\caption{PatentSBERTa deployment matrix on the full patent corpus}
\label{tab:full_deployment_matrix}
\begin{tabular}{lrrr}
\toprule
& \multicolumn{2}{c}{Model prediction} & \\
\cmidrule(lr){2-3}
Administrative label & non-green & green & Total \\
\midrule
non-Y02 & 8,216,292 & 266,742 & 8,483,034 \\
Y02     &   251,030 & 341,357 &   592,387 \\
\midrule
Total   & 8,467,322 & 608,099 & 9,075,421 \\
\bottomrule
\end{tabular}
\end{table}

If the administrative Y02 label is treated as ground truth, this deployment yields an accuracy of 0.943, precision of 0.561, recall of 0.576, and an $F1$ score of 0.569. These metrics measure agreement with Y02, not agreement with substantive climate relevance, and should therefore be interpreted cautiously. Let $M_i$ denote the model prediction. We define the disagreement signal as
\[
D_i=\mathbf{1}(A_i\neq M_i)
\]
The distinction between random error and systematic bias is central to the empirical design. Some disagreement is expected in any large-scale classification task and may reflect ordinary model noise. Our concern is whether the direction and frequency of consensus-attributed errors vary systematically across technological fields. We therefore use the model first to locate individual candidate errors and then use functional consensus to determine their source where possible. Patents with $A_i=1$ and $M_i=0$ form the downward disagreement pool. Before substantive validation, these cases may represent either administrative Type I errors (False Green) or model Type II errors (model false negatives). Patents with $A_i=0$ and $M_i=1$ form the upward disagreement pool. Before validation, these cases may represent either administrative Type II errors (Silent Green) or model Type I errors (model false positives). The procedure identifies \num{517772} total disagreement signals: \num{251030} downward and \num{266742} upward disagreements.

\subsection{The Construct-Validity Test: Independent LLM Judges and Error Attribution}
\label{sec:llm_attribution}

The complete disagreement pool was evaluated by two open-weight LLM judges: Qwen\footnote{The Qwen judge used the checkpoint \texttt{Qwen/Qwen3-32B}.} and Gemma\footnote{The Gemma judge used the quantized checkpoint \texttt{Gemma-4-31B-it-W4A16}.}. Each model returned a binary green judgment, a confidence level, a domain, a benefit type, and a short technical justification. The substantive criterion was stricter than the binary green-field criterion: an invention was treated as functionally green only when its primary technical function made a direct and credible contribution to climate change mitigation or adaptation. We call this the direct-mechanism criterion: the test requires a specific technical pathway linking the invention's primary function to climate mitigation or adaptation, rather than incidental or promotional environmental language. For model $m\in\{Q,G\}$, where $Q$ denotes Qwen and $G$ denotes Gemma, define
\[
L_{im}=
\begin{cases}
1, & \text{if benefit type is direct},\\
0, & \text{if benefit type is incidental or none},\\
\text{undefined}, & \text{if the mechanism output is uncertain or unparseable}.
\end{cases}
\]
A functional consensus judgment $L_i$ is assigned only when $L_{iQ}=L_{iG}$. If the two values differ, or if either value is undefined, the patent remains unresolved. Confidence is used to assess robustness but does not change the functional label. For downward disagreements ($A_i=1,M_i=0$), consensus $L_i=0$ identifies administrative Type I error, while consensus $L_i=1$ identifies model Type II error. For upward disagreements ($A_i=0,M_i=1$), consensus $L_i=1$ identifies administrative Type II error, while consensus $L_i=0$ identifies model Type I error. This procedure attributes \num{400186} disagreements and retains \num{117586} cases as unresolved. Figure~\ref{fig:error_as_signal} in Section~\ref{sec:results_who} reports the attribution outcomes.

The attribution rule sets a deliberately high bar. A disagreement
is attributed to the administrative label only when three
systems, PatentSBERTa, Qwen, and Gemma, all contradict Y02 in the same
direction; a single dissent turns the case into a model error or leaves it
unresolved. The design trades recall for defensibility, so that, for a
given direct-mechanism criterion, False Green and Silent Green are
conservative counts rather than maximal ones. Two further features work in
the same direction. Because PatentSBERTa is fine-tuned on Y02, it is more
likely to agree with the label than an independent classifier would be,
and any error it shares with the label generates no disagreement at all.
And to the extent that the judges encountered administrative
classifications during pre-training, that exposure would incline them
toward the label rather than against it.

Using generative models as a measurement instrument raises validity
questions of its own \parencite{deKok2025}. The test is designed to meet
them in four ways. The two judges are independent open-weight models with
published checkpoints, so the test can be rerun. Both apply the same fixed
rubric (Appendix~\ref{app:audit_prompt}), which asks what the invention's
primary function does rather than whether Y02 is right. Neither observes
the other's judgment, and only agreement counts. And disagreement between
the judges is retained as an explicit unresolved category rather than
broken by a tie rule. Agreement between independent judges establishes the
reliability of the test; establishing its validity requires comparison
with an independent human reference, which we discuss in
Section~\ref{sec:conclusion}.

The consensus-based corrected measure removes attributed administrative Type I errors from the Y02 count and adds attributed administrative Type II errors:
\[
\text{Consensus-corrected Green}
=
\underbrace{592387}_{\substack{\text{administrative}\\\text{Y02 count}}}
-
\underbrace{180384}_{\substack{\text{administrative Type I}\\\text{False Green}}}
+
\underbrace{29465}_{\substack{\text{administrative Type II}\\\text{Silent Green}}}
=
441468
\]
This equals 4.86\% of the full patent corpus, compared with 6.53\% under raw Y02. Because \num{50928} unresolved patents belong to the downward pool and \num{66658} belong to the upward pool, transparent extreme allocations produce a lower bound of \num{390540} and an upper bound of \num{508126}. The point estimate of \num{441468} should therefore be interpreted as a consensus-based correction, not as a final corpus-wide truth.

\subsection{Technological Complexity and the Direction of
Administrative Error}
\label{sec:complexity_method}

To test the capacity accounts, we ask whether administrative
omission, and the direction of administrative error, vary systematically
with the complexity of an invention's technological knowledge base. The
omission analysis is restricted to the consensus-corrected green portfolio. The dependent variable equals one for a Silent Green patent and zero for a patent retained as green under the consensus correction. Thus, the comparison group contains Y02 patents not attributed to administrative Type~I error; unresolved downward disagreements retain their original administrative status and are not treated as substantively adjudicated.

We measure complexity along two dimensions that capture distinct properties of the knowledge structure underlying a patent. The first is reflection complexity, based on the method-of-reflections tradition in technological complexity \parencite{BallandRigby2017}. The second is structural complexity, which captures the diversity of combinatorial network structures underlying technological development \parencite{Broekel2019}. The supplied CPC-level complexity scores are aggregated to patent-level means across the technologies assigned to each patent. Higher values therefore indicate patents whose assigned technologies score higher on the corresponding complexity dimension.

Complexity coverage is high in the consensus-corrected portfolio. Reflection and structural complexity are observed for \num{409612} of \num{412003} recognized corrected-green patents (99.42\%) and \num{27574} of \num{29465} Silent Green patents (93.58\%). All main specifications also
control for technological atypicality, the degree to which a patent's
rarest pair of CPC classes departs from contemporaneous co-classification
convention \parencite{Uzzi2013}, and its square, because unfamiliar
recombination is a separate source of classification difficulty
\parencite{Fleming2001}. Appendix~\ref{app:atypicality} details the
construction and reports atypicality's own association with omission.
Because atypicality is undefined for single-class patents, the common
estimation sample contains \num{347241} patents, including \num{19374} Silent Green patents, filed between 1963 and 2020;
Section~\ref{sec:complexity_silent_green} also reports estimates on the
maximum available sample without this control. All continuous predictors are standardized in that common estimation sample. Reflection complexity and atypicality are only weakly correlated ($r=-0.077$), while reflection and structural complexity correlate at $-0.110$.

Let $R_i$ denote standardized reflection complexity, $H_i$
standardized structural complexity, and $f(Z_i)=\beta_1 Z_i+\beta_2 Z_i^2$
the atypicality controls, where $Z_i$ is standardized maximum atypicality.
We estimate each complexity measure alone, then with $f(Z_i)$, and then
both jointly:
\[
\Pr(S_i=1)
=
\Lambda\!\left(
\alpha+\gamma_1 R_i+\gamma_2 H_i+f(Z_i)+\delta_t
\right),
\]
where $S_i$ indicates Silent Green status, $\delta_t$ represents
filing-year fixed effects, and $\Lambda(\cdot)$ is the logistic function.
Because the complexity measures need not capture the same technological property, we interpret their coefficients separately rather than combine them into a single index. The main specifications use
heteroskedasticity-robust standard errors. We also estimate nonlinear specifications and interactions between atypicality and reflection complexity as supplementary tests (Appendix~\ref{app:atypicality}).

\subsubsection{Does Complexity Predict the Direction of Administrative Error?}

The distinction between classification error and measurement bias implies a stronger test. If complexity simply makes classification more difficult, it could increase errors in either direction. If complexity systematically creates under-recognition, however, it should shift administrative errors toward Silent Green rather than False Green. We therefore restrict the sample to consensus-attributed administrative errors and define
\[
B_i=
\begin{cases}
1, & \text{if patent } i \text{ is Silent Green},\\
0, & \text{if patent } i \text{ is False Green}.
\end{cases}
\]
We estimate
\[
\Pr(B_i=1)
=
\Lambda\!\left(
\alpha+\gamma_1 R_i+\gamma_2 H_i+f(Z_i)+\delta_t
\right).
\]
A positive coefficient on reflection complexity implies that, conditional on an administrative error having occurred, greater reflection complexity tilts that error toward omission rather than erroneous inclusion. The full complete-case error-direction specification contains \num{157241} patents. We additionally report maximum-sample, post-2010, and 2005--2015 robustness specifications.

\section{What Types of Error Does Y02 Make?}\label{sec:results}

Sections~\ref{sec:results}, \ref{sec:idiosyncratic}, and~\ref{sec:causes}
answer the paper's three questions in turn. This section takes the first:
what types of error does Y02 make? The Error-as-Signal typology of
Section~\ref{sec:error_as_signal} allows four, two of the label and two of
the model, and the construct-validity test assigns each resolved
disagreement to one of them. In the vocabulary of
Section~\ref{sec:construct_validity}, the label's errors are failures of
substantive validity: inclusions of inventions that lack the relevant
property, and omissions of inventions that have it.

\subsection{Who Errs? Construct-Validity Failures and Model Errors}
\label{sec:results_who}

\begin{figure}[htbp]
\centering
\includegraphics[width=0.90\textwidth]{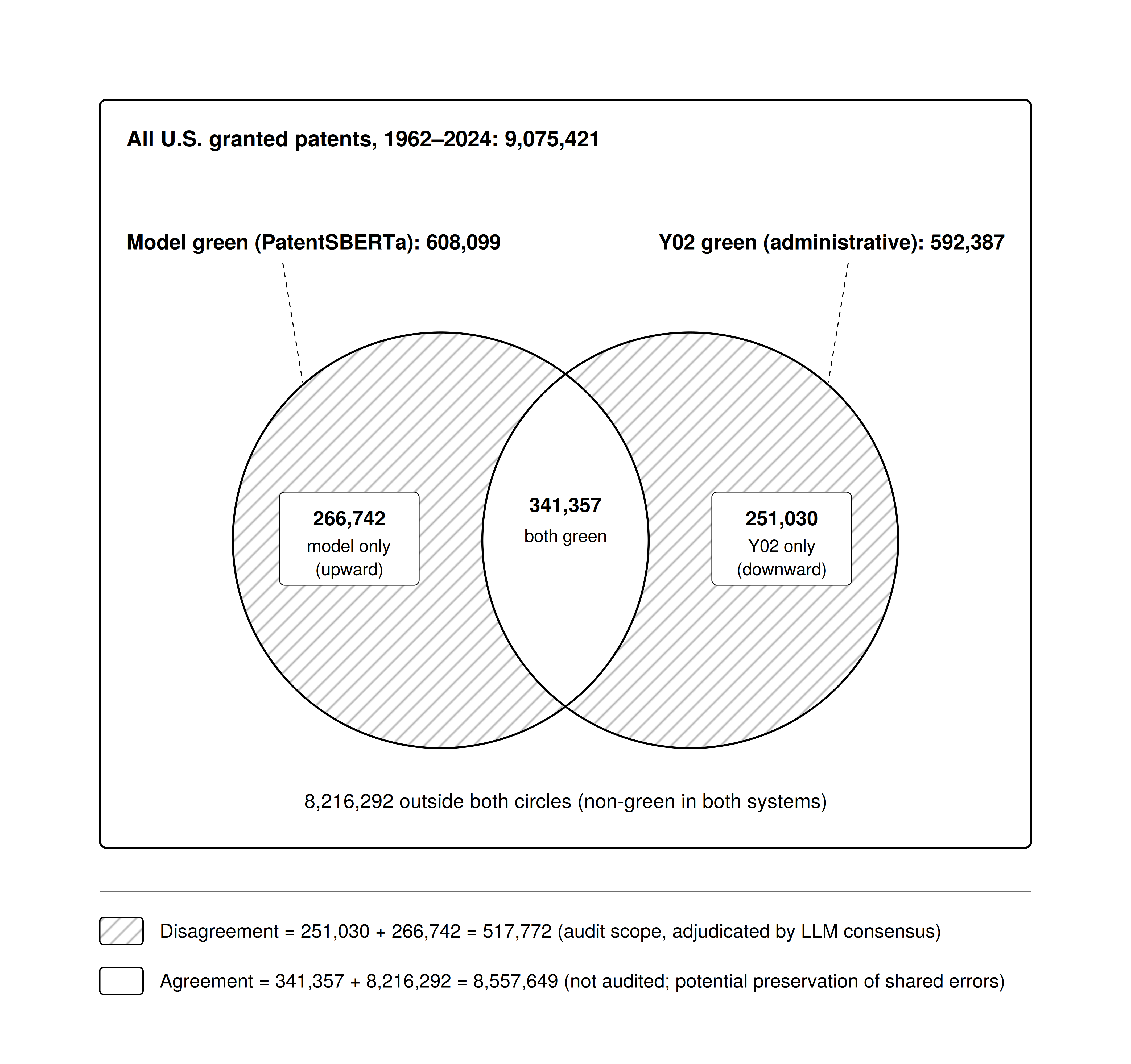}
\caption{Administrative Y02 and PatentSBERTa classifications across the full patent
corpus. The intersection contains \num{341357} patents classified as green by both
systems. The non-overlapping regions contain \num{517772} administrative--model
disagreements, comprising \num{251030} Y02-only green classifications and
\num{266742} model-only green classifications. The \num{8216292} patents outside
both circles are classified as non-green by both systems. Agreement cases are not substantively tested in the present study.}
\label{fig:venn_disagreement_scope}
\end{figure}

Figure~\ref{fig:venn_disagreement_scope} clarifies the scope of the construct-validity test. Among all \num{9075421} U.S. granted patents, \num{592387} are classified as green by the administrative Y02 system and \num{608099} are classified as green by PatentSBERTa. The two systems agree on \num{341357} patents as green and on \num{8216292} patents as non-green, producing \num{8557649} agreement cases in total. These agreement cases are not substantively tested in the present study and therefore constitute the potential search space for shared or preserved errors.

The test instead focuses on the \num{517772} administrative--model disagreement cases. These comprise \num{251030} downward disagreements, in which Y02 classifies the patent as green while PatentSBERTa classifies it as non-green, and \num{266742} upward disagreements, in which Y02 classifies the patent as non-green while PatentSBERTa classifies it as green. These non-overlapping regions define the empirical scope of the test, in which one measurement system challenges the other.

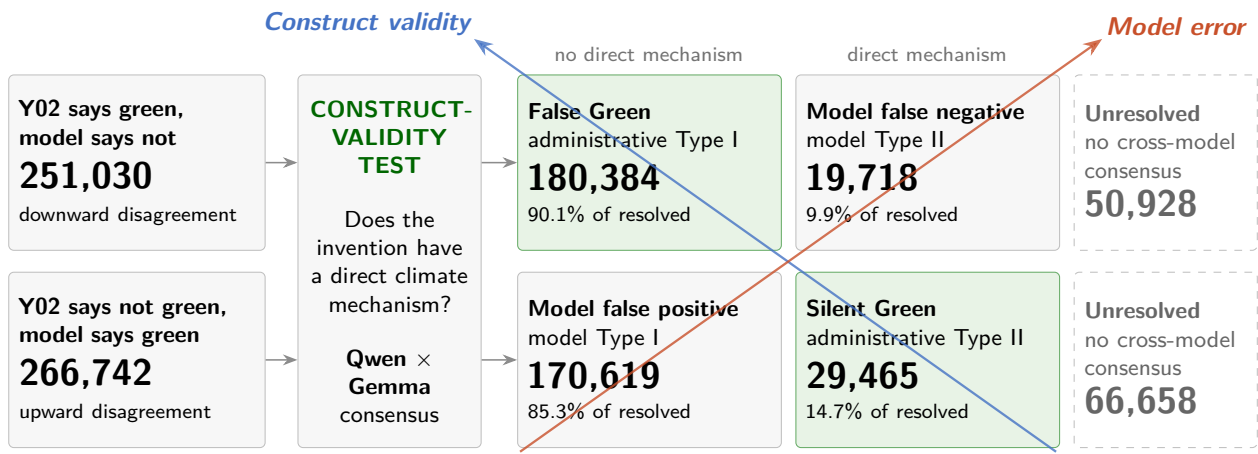
\begin{figure}[H]
\centering
\resizebox{\textwidth}{!}{%
\begin{tikzpicture}[
  x=1cm, y=1cm,
  font=\sffamily\footnotesize,
  every node/.style={inner sep=4pt},
  base/.style={draw=black!25, rounded corners=2pt, fill=black!3, align=left},
  cellbox/.style={base, minimum width=3.6cm, minimum height=2.4cm, text width=3.32cm},
  diag/.style={cellbox, fill=cellgreen, draw=green!45!black!60},
  unres/.style={base, dashed, fill=white, minimum width=2.5cm, minimum height=2.4cm, text width=2.22cm, text=black!60},
  flow/.style={-{Stealth[length=2mm]}, draw=black!45, line width=0.6pt}
]
\node[base, anchor=south west, minimum width=3.5cm, minimum height=2.4cm, text width=3.22cm] (down) at (0,2.7)
  {\textbf{Y02 says green,}\\ \textbf{model says not}\\[4pt] {\Large\bfseries 251,030}\\[2pt] {\scriptsize downward disagreement}};
\node[base, anchor=south west, minimum width=3.5cm, minimum height=2.4cm, text width=3.22cm] (up) at (0,0)
  {\textbf{Y02 says not green,}\\ \textbf{model says green}\\[4pt] {\Large\bfseries 266,742}\\[2pt] {\scriptsize upward disagreement}};
\node[base, anchor=south west, minimum width=2.5cm, minimum height=5.1cm, text width=2.22cm, align=center] (test) at (3.95,0)
  {{\color{green!40!black}\bfseries CONSTRUCT-}\\ {\color{green!40!black}\bfseries VALIDITY TEST}\\[10pt]
   Does the invention have a direct climate mechanism?\\[10pt]
   \textbf{Qwen $\times$ Gemma}\\ consensus};
\node[diag, anchor=south west] (fg) at (6.95,2.7)
  {\textbf{False Green}\\ administrative Type~I\\[4pt] {\Large\bfseries 180,384}\\[1pt] {\scriptsize 90.1\% of resolved}};
\node[cellbox, anchor=south west] (mfn) at (10.75,2.7)
  {\textbf{Model false negative}\\ model Type~II\\[4pt] {\Large\bfseries 19,718}\\[1pt] {\scriptsize 9.9\% of resolved}};
\node[unres, anchor=south west] (u1) at (14.55,2.7)
  {\textbf{Unresolved}\\ no cross-model consensus\\[4pt] {\Large\bfseries 50,928}};
\node[cellbox, anchor=south west] (mfp) at (6.95,0)
  {\textbf{Model false positive}\\ model Type~I\\[4pt] {\Large\bfseries 170,619}\\[1pt] {\scriptsize 85.3\% of resolved}};
\node[diag, anchor=south west] (sg) at (10.75,0)
  {\textbf{Silent Green}\\ administrative Type~II\\[4pt] {\Large\bfseries 29,465}\\[1pt] {\scriptsize 14.7\% of resolved}};
\node[unres, anchor=south west] (u2) at (14.55,0)
  {\textbf{Unresolved}\\ no cross-model consensus\\[4pt] {\Large\bfseries 66,658}};
\node[font=\sffamily\scriptsize, text=black!55, anchor=south] at (8.75,5.15) {no direct mechanism};
\node[font=\sffamily\scriptsize, text=black!55, anchor=south] at (12.55,5.15) {direct mechanism};
\draw[flow] (down.east) -- (down.east -| test.west);
\draw[flow] (up.east) -- (up.east -| test.west);
\draw[flow] (test.east |- fg.west) -- (fg.west);
\draw[flow] (test.east |- mfp.west) -- (mfp.west);
\draw[-{Stealth[length=2.6mm]}, draw=cvblue, line width=0.9pt, draw opacity=0.8]
  (14.31,-0.05) -- (6.36,5.6)
  node[above left, inner sep=1pt, text=cvblue, font=\sffamily\small\bfseries\itshape] {Construct validity};
\draw[-{Stealth[length=2.6mm]}, draw=fgorange, line width=0.9pt, draw opacity=0.8]
  (6.99,-0.05) -- (14.94,5.6)
  node[above right, inner sep=1pt, text=fgorange, font=\sffamily\small\bfseries\itshape] {Model error};
\end{tikzpicture}}
\caption[Error as signal: attribution by the construct-validity test]{Error as signal: attribution of the \num{517772} label--model
disagreements by the construct-validity test. Each disagreement is
ambiguous between an error of the administrative label and an error of
the model; the test, cross-model consensus on whether the invention's
primary function has a direct climate mechanism, decides which. Shaded
cells lie on the construct-validity diagonal: the test contradicts Y02,
and only these cells correct the indicator. Cells on the model-error
anti-diagonal correct the model, not the label, and the \num{117586}
cases without consensus are quarantined rather than forced into a
category. Percentages are shares of resolved cases within each
disagreement direction; across both directions, \num{209849} resolved
cases (52.4\%) are construct-validity failures of Y02 and \num{190337}
(47.6\%) are model errors. Screening: PatentSBERTa versus Y02 on
\num{9075421} USPTO granted patents, 1962--2024. Cross-model confidence
by outcome is reported in Table~\ref{tab:new_error_counts_confidence}.}
\label{fig:error_as_signal}
\end{figure}

Figure~\ref{fig:error_as_signal} shows how the disagreement cases are attributed through the construct-validity test. The test attributes \num{400186} of the \num{517772} disagreements and retains \num{117586} cases as unresolved. Among the \num{251030} downward disagreements, \num{180384} are administrative Type~I errors, \num{19718} are model Type~II errors, and \num{50928} remain unresolved. Among the \num{266742} upward disagreements, \num{29465} are administrative Type~II errors, \num{170619} are model Type~I errors, and \num{66658} remain unresolved. In the terminology of Section~\ref{sec:theory}, the test therefore documents \num{209849} instances of error correction, in which PatentSBERTa challenges an incorrect administrative label, and \num{190337} model errors that constitute potential sources of bias amplification.

The figure arranges the outcomes as the framework predicts. The
shaded cells, False Green and Silent Green, lie on the construct-validity
diagonal: the test contradicts Y02, and only these cells correct the
indicator. The cells on the model-error anti-diagonal correct the model,
not the label. The \num{117586} ambiguous cases are quarantined rather
than forced into a category, which is why the disagreement pool of
\num{517772} is a diagnostic rather than a verdict. In the vocabulary of
Section~\ref{sec:construct_validity}, \num{209849} of the \num{400186}
resolved cases, or 52.4\%, are failures of Y02's substantive validity.

Disagreement therefore identifies a candidate measurement problem, while Qwen--Gemma direct-mechanism consensus identifies the likely source of error only for resolved cases. Administrative Type~I errors correspond to False Green patents, where Y02 incorrectly includes a patent without a direct climate-relevant technical function. Administrative Type~II errors correspond to Silent Green patents, where Y02 omits a patent containing a direct climate-relevant technical function. Model Type~I and Type~II errors arise when the construct-validity test instead supports the original administrative classification. Table~\ref{tab:new_error_counts_confidence} reports these consensus outcomes together with the associated cross-model confidence levels.

\begin{table}[htbp]
\centering
\caption{Direct-mechanism consensus, confidence, and unresolved cases}
\label{tab:new_error_counts_confidence}
\small
\begin{tabular}{lrrrr}
\toprule
Outcome
& Count
& Share
& Both high confidence
& High-confidence rate \\
\midrule
Administrative Type~I & 180,384 & 34.84\% & 177,047 & 98.15\% \\
Administrative Type~II & 29,465 & 5.69\% & 24,415 & 82.86\% \\
Model Type~I & 170,619 & 32.95\% & 168,319 & 98.65\% \\
Model Type~II & 19,718 & 3.81\% & 16,429 & 83.32\% \\
Unresolved & 117,586 & 22.71\% & 84,333 & 71.72\% \\
\midrule
Total & 517,772 & 100.00\% & --- & --- \\
\bottomrule
\end{tabular}
\end{table}

Most unresolved cases reflect substantive disagreement about whether the climate mechanism is direct rather than low-confidence failure. Gemma identifies a direct mechanism while Qwen does not in \num{115779} cases, whereas the reverse occurs in \num{1752} cases. Only 55 cases contain at least one uncertain mechanism output. Moreover, \num{84333} unresolved cases receive high confidence from both models. High confidence therefore does not eliminate cross-model disagreement and should not be interpreted as calibrated accuracy. After recoding only the consensus-resolved administrative errors, the estimated green-patent count decreases from \num{592387} to \num{441468}. Because the \num{117586} unresolved disagreement cases are not reclassified, they retain their original administrative Y02 status: \num{50928} unresolved downward cases remain provisionally green, while \num{66658} unresolved upward cases remain provisionally non-green. Accordingly, \num{441468} should be interpreted as a consensus-adjusted count rather than as a complete substantive adjudication of the full corpus. For transparency, we also report the extreme-allocation sensitivity interval of \num{390540}--\num{508126}. These values are not alternative main estimates; they indicate how the consensus-adjusted count would change under the most extreme possible classifications of the unresolved cases.

\subsection{The Test Attributes More Cases Against the Model Than
Against the Label}
\label{sec:conservative_by_design}

Two of the four attributed outcomes are model errors. In
\num{170619} cases the test finds no direct climate mechanism in a patent
the screening model predicted to be green, and in \num{19718} cases it
finds one in a patent the model rejected. Together these \num{190337}
attributions resolve in favour of the administrative label and against the
instrument used to challenge it. This is a property of the design rather
than a residual: a case is attributed to administrative error only when
the screening model and both language-model judges contradict Y02 in the
same direction, and everything short of that unanimity, \num{190337}
model errors and \num{117586} unresolved cases, is excluded from the
correction. The reported \num{180384} False Green and \num{29465} Silent
Green patents are therefore conservative counts of administrative error
conditional on entry into the disagreement pool, not maximal ones.

The two disagreement directions behave very differently, and the
asymmetry is itself informative. Among resolved downward disagreements,
90.1\% are attributed to the administrative label (\num{180384} of
\num{200102}); among resolved upward disagreements, only 14.7\% are
(\num{29465} of \num{200084}). When PatentSBERTa withholds a green
prediction from a Y02-tagged patent it is usually vindicated; when it
volunteers one for an untagged patent it usually is not. Because the model
is fine-tuned on Y02, this is the error profile the label itself would be
expected to transmit: a system trained to recognize green language
over-predicts green wherever green language is present and a direct
mechanism is not.

Appendix~\ref{sec:lexical_error_structure} confirms the channel
directly. Digital-context vocabulary raises the odds that a downward
disagreement is administrative over-inclusion rather than model failure
(OR $=2.32$; 95\% CI 2.16--2.49) and lowers the odds that an upward
disagreement is administrative omission rather than model failure
(OR $=0.59$; 95\% CI 0.56--0.62). The same lexical cue therefore drives
over-inclusion by the administrative label inside Y02 and by the model
outside it. This is the comparative, group-level condition that
Section~\ref{sec:model_error_transmission} sets for bias amplification:
the model does not merely add errors, it adds errors that reinforce the
administrative pattern of over-recognizing digitally described
technologies. Silent Green is the count recovered against that headwind,
which is a further reason to read \num{29465} as a floor.

\subsection{Illustrative Mechanisms from the Full-Corpus Construct-Validity Test}
\label{sec:illustrative_mechanisms}

Table~\ref{tab:illustrative_error_mechanisms} presents five especially informative examples of each error type. The cases were purposively selected from the consensus-resolved cases of the test to reveal the substantive mechanisms behind the errors and the technological bias identified in the population-level analysis. They are illustrative rather than representative, and a broader catalogue of examples is reported in Appendix~\ref{app:illustrative_patent_cases}.

\begin{table}[H]
\centering
\caption{Illustrative False Green and Silent Green Patents from the Full-Corpus Construct-Validity Test}
\label{tab:illustrative_error_mechanisms}

\scriptsize
\setlength{\tabcolsep}{3pt}
\renewcommand{\arraystretch}{1.18}

\begin{tabularx}{\textwidth}{
    @{}
    >{\raggedright\arraybackslash}p{1.35cm}
    >{\raggedright\arraybackslash}X
    >{\centering\arraybackslash}p{0.55cm}
    >{\centering\arraybackslash}p{0.55cm}
    >{\centering\arraybackslash}p{0.75cm}
    >{\raggedright\arraybackslash}X
    @{}
}
\toprule
\textbf{Patent}
&
\textbf{Primary technical function}
&
\textbf{Y02}
&
\textbf{TF}
&
\textbf{LLMs}
&
\textbf{Mechanism and bias signal}
\\
\midrule

\multicolumn{6}{@{}l}{
\textbf{Panel A. Administrative Type I: False Green}
\quad
($Y02=1$, TF$=0$, LLM consensus$=0$)
}
\\
\addlinespace[2pt]

\#11417179
&
Automated retail checkout using image and voice tracking within a
``shopping environment''
&
1 & 0 & 0/0
&
\textbf{Semantic homonymy and ICT over-inclusion.}
The term ``environment'' denotes a retail or digital setting rather
than the natural environment.
\\
\addlinespace

\#9537918
&
Encrypted data sharing within a distributed secure ``environment''
&
1 & 0 & 0/0
&
\textbf{Generic computing language.}
A secure computing environment is classified as environmentally
relevant despite having no direct climate function.
\\
\addlinespace

\#7811444
&
Oxidation and upgrading of heavy-hydrocarbon asphaltenes for fuel or
petrochemical use
&
1 & 0 & 0/0
&
\textbf{Counter-directional taxonomic spillover.}
The invention supports fossil-resource processing rather than climate
mitigation.
\\
\addlinespace

\#11572899
&
Pressure-exchange system designed for hydraulic-fracturing operations
&
1 & 0 & 0/0
&
\textbf{Counter-directional application.}
The technology facilitates fossil-fuel extraction despite receiving an
administrative green label.
\\
\addlinespace

\#9993540
&
Immunotherapy for the treatment of brain cancer
&
1 & 0 & 0/0
&
\textbf{Adaptation-category spillover.}
A general medical invention is included without a specific causal
connection to climate adaptation.
\\

\midrule

\multicolumn{6}{@{}l}{
\textbf{Panel B. Administrative Type II: Silent Green}
\quad
($Y02=0$, TF$=1$, LLM consensus$=1$)
}
\\
\addlinespace[2pt]

\#7986539
&
Maximum-power-point tracking and power conversion for photovoltaic
electricity
&
0 & 1 & 1/1
&
\textbf{Power-electronics opacity.}
The renewable-energy contribution is embedded in feedback control,
voltage regulation, and conversion architecture.
\\
\addlinespace

\#6266975
&
Heat-pump and engine system for energy-efficient heating and cooling
&
0 & 1 & 1/1
&
\textbf{Embedded building efficiency.}
The climate mechanism is expressed through thermodynamic engineering
rather than explicit green terminology.
\\
\addlinespace

\#11525437
&
Mechanical storage and delivery of electricity using gravitational
potential energy
&
0 & 1 & 1/1
&
\textbf{Enabling-technology omission.}
A grid-scale storage function is described through lifting and
mechanical-system terminology.
\\
\addlinespace

\#11878950
&
Oxy-fuel cement-production process that concentrates carbon dioxide
for capture
&
0 & 1 & 1/1
&
\textbf{Industrial-process opacity.}
A direct mechanism for reducing hard-to-abate cement emissions is
embedded in specialized process-engineering language.
\\
\addlinespace

\#8877486
&
Microalgae photobioreactor for carbon-dioxide mitigation during
wastewater treatment
&
0 & 1 & 1/1
&
\textbf{Cross-domain under-coverage.}
The invention combines biological carbon capture and wastewater
treatment but remains outside the administrative green indicator.
\\

\bottomrule
\end{tabularx}

\vspace{2mm}
\begin{minipage}{0.98\textwidth}
\footnotesize
\textit{Notes:}
Y02 denotes the administrative classification and TF denotes the
PatentSBERTa prediction at the validation-calibrated threshold.
The LLM column reports the two independent direct-mechanism judgments:
1 indicates a direct climate-mitigation or adaptation mechanism and
0 indicates that no such direct mechanism was identified. All ten
examples received matching high-confidence judgments from both LLMs.
The cases were selected for explanatory clarity and are not estimates
of the relative frequency of the mechanisms.
\end{minipage}
\end{table}

\subsubsection{Administrative Type I Error: False Green}

The False Green examples reveal several forms of administrative over-inclusion. Patents \#11417179 and \#9537918 illustrate semantic homonymy in ICT. In both cases, the term ``environment'' describes a digital or commercial setting rather than the natural environment. These examples help explain the strong concentration of administrative Type I errors in computing and communications technologies. Patents \#7811444 and \#11572899 reveal a more consequential mechanism. Their primary functions support heavy-hydrocarbon processing and hydraulic fracturing, respectively. These are not merely inventions with weak or incidental environmental benefits; they facilitate fossil-resource production. Their inclusion, therefore, represents counter-directional taxonomic spillover. Patent \#9993540 illustrates a further form of category expansion, in which a general medical invention is associated with climate adaptation despite lacking a specific climate-related health mechanism.

False Green, therefore, cannot be reduced to a simple keyword error. It can arise from semantic ambiguity, technological-category proximity, overextension of adaptation categories, or insufficient consideration of the invention's primary technical function. As Section~\ref{sec:theory} notes, the first two examples
misapply the scheme, whereas the device-level power-management cases in
Appendix~\ref{app:illustrative_patent_cases} apply it correctly while its
scope exceeds the construct. Both inflate Y02-based measures; only the
first is a classification mistake in the scheme's own terms.

\subsubsection{Administrative Type II Error: Silent Green}

The Silent Green examples demonstrate the opposite measurement mechanism. Patent \#7986539 directly improves photovoltaic electricity conversion, Patent \#6266975 provides energy-efficient heating and cooling through heat-pump engineering, and Patent \#11525437 supplies mechanical electricity storage for grid integration. Nevertheless, their climate functions are expressed through power electronics, thermodynamics, and mechanical engineering rather than overt environmental vocabulary. The omissions extend beyond energy-system components. Patent \#11878950 embeds carbon capture within an oxy-fuel cement-production process, while Patent \#8877486 combines microalgae-based carbon mitigation with wastewater treatment. These examples illustrate why technologies crossing conventional classification boundaries, or contributing to decarbonization through specialized industrial functions, may remain administratively invisible.

\subsubsection{From Illustrative Errors to Measurement Bias}

The examples show what the errors look like, whereas the full-sample analysis establishes whether they constitute measurement bias. Dual-measurement disagreement first identifies a potentially problematic classification. Independent LLM consensus then attributes the disagreement according to the invention's primary technical function. Finally, comparisons across technological fields determine whether the attributed errors are systematically concentrated.

This combination reveals an opposing measurement structure: visible digital or environmentally suggestive language can generate over-inclusion, while technically embedded mechanisms in energy and industrial technologies can generate under-inclusion. The method therefore identifies not only individual errors but also how the administrative indicator redistributes measured green invention across the technological landscape.

The answer to the first question is therefore that Y02 makes both
types of error, and in volume: \num{180384} False Green inclusions and
\num{29465} Silent Green omissions, more than half of the resolved
disagreements. Whether these errors are idiosyncratic is the second
question.

\section{Are the Errors Idiosyncratic?}\label{sec:idiosyncratic}

An idiosyncratic error is one whose occurrence and direction are
unrelated to what the invention is: such errors add noise, threaten
reliability, and partly cancel in aggregate. A systematic error, a
bias, is one whose probability or direction depends on observable
characteristics of the invention, so that it threatens validity and does
not cancel (Section~\ref{sec:error_as_signal}). This section asks which
description fits Y02, first by relating the source of error to
technological field and patent language
(Section~\ref{sec:technological_bias}) and then by tracing what the errors
do to the measured landscape of green invention
(Section~\ref{sec:corrected_landscape}).

\subsection{Random Error or Systematic Bias? The Technological Structure of
Administrative Error}
\label{sec:technological_bias}

An individual False Green or Silent Green patent represents a classification error. Such errors constitute measurement bias when their probability or direction varies systematically across technological fields. We therefore examine error attribution separately within the two directions of PatentSBERTa--Y02 disagreement. Among resolved downward disagreements, we compare administrative Type I errors with model Type II errors ($N=\num{200102}$). Among resolved upward disagreements, we compare administrative Type II errors with model Type I errors ($N=\num{200084}$). These within-direction comparisons test whether error attribution varies across technologies among patents already selected by the same disagreement pattern.

\begin{table}[H]
\centering
\caption{Association of technological field and filing period with
error attribution in the resolved disagreement pools}
\label{tab:resolved_bias_tests}
\small
\setlength{\tabcolsep}{5pt}
\renewcommand{\arraystretch}{1.12}

\begin{tabularx}{\textwidth}{
    @{}
    >{\raggedright\arraybackslash}p{4.0cm}
    >{\raggedright\arraybackslash}X
    r
    r
    r
    r
    @{}
}
\toprule
\textbf{Resolved disagreement pool}
&
\textbf{Patent characteristic}
&
\textbf{$N$}
&
\textbf{$\chi^2$}
&
\textbf{df}
&
\textbf{Cram\'er's $V$}
\\
\midrule

Administrative Type I vs.\ Model Type II
&
Primary non-Y CPC section
&
200,102
&
13,143.73
&
8
&
0.256
\\

&
Primary non-Y CPC subclass
($n\geq500$; smaller subclasses pooled)
&
200,102
&
31,542.97
&
81
&
0.397
\\

&
Filing period
&
200,102
&
1,045.97
&
6
&
0.072
\\

\addlinespace

Administrative Type II vs.\ Model Type I
&
Primary non-Y CPC section
&
200,084
&
9,294.09
&
8
&
0.216
\\

&
Primary non-Y CPC subclass
($n\geq500$; smaller subclasses pooled)
&
200,084
&
32,125.25
&
92
&
0.401
\\

&
Filing period
&
200,084
&
618.30
&
6
&
0.056
\\

\bottomrule
\end{tabularx}

\vspace{0.35em}
\begin{minipage}{0.97\textwidth}
\footnotesize
\textit{Notes:}
All Pearson chi-square tests reject independence at $p<0.001$.
The primary non-Y CPC section represents the broad technological field
derived from each patent's primary conventional CPC classification.
The primary non-Y CPC subclass identifies the more detailed technology
within that field. Y02 classifications are excluded from both
technology measures. For the subclass tests, subclasses represented by
fewer than 500 patents in the relevant resolved-disagreement pool are
combined into a residual category. Filing periods are $\leq1994$,
1995--1999, 2000--2004, 2005--2009, 2010--2014, 2015--2019, and
2020 onward. Cram\'er's $V$ ranges from zero, indicating no association,
to one, indicating perfect association.
\end{minipage}
\end{table}

\begin{figure}[H]
    \centering
    \includegraphics[width=\textwidth]
    {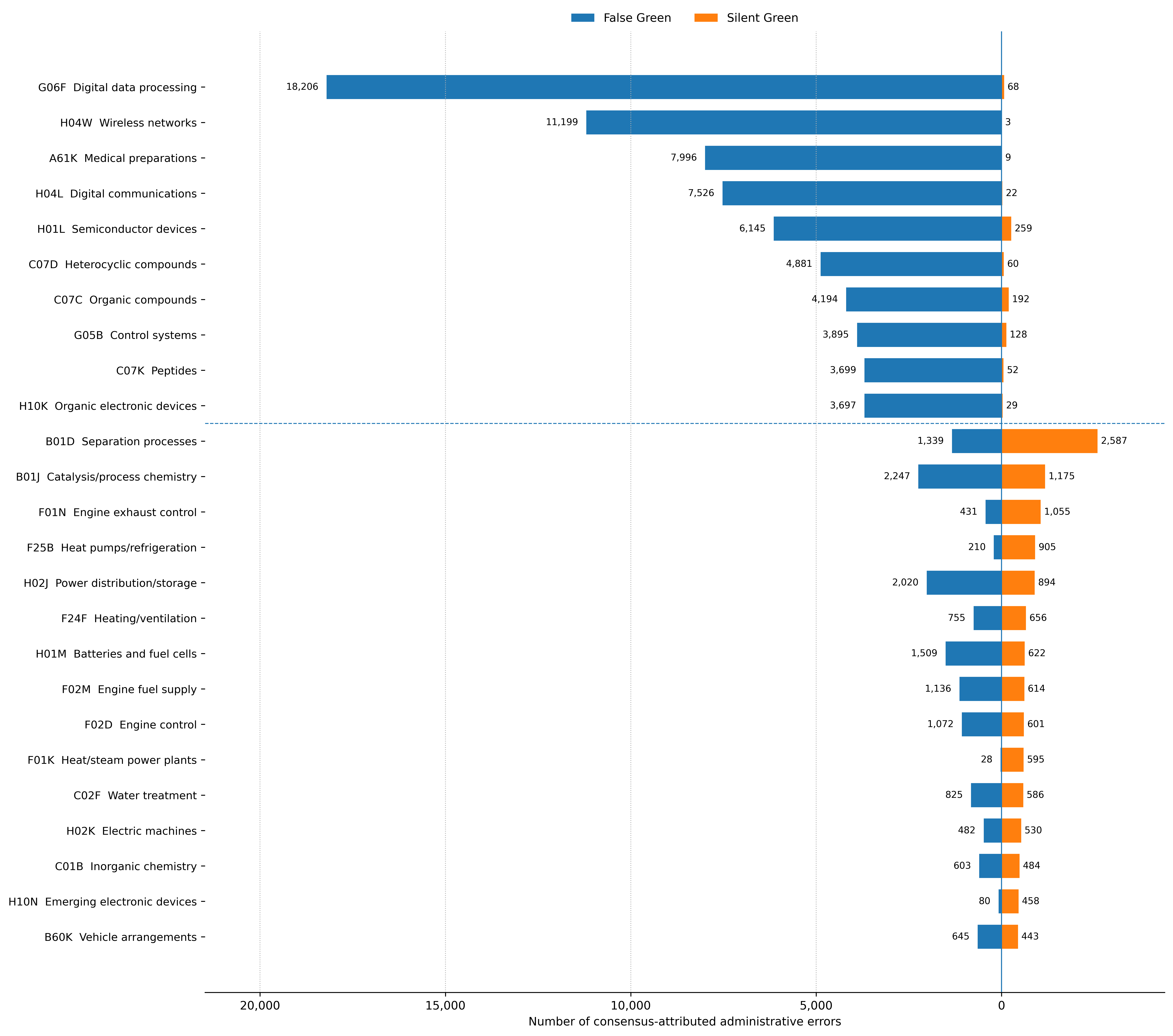}
    \caption{Technological concentration of
    consensus-attributed administrative errors.
    Leftward bars report False Green patents and rightward bars report
    Silent Green patents. The upper block contains the ten primary
    non-Y CPC subclasses with the largest False Green counts. The lower
    block contains the fifteen subclasses with the largest Silent Green
    counts. Because a displayed subclass can contain both types of
    error, both counts are reported for every subclass. Administrative
    over-inclusion is especially prominent in digital data processing,
    wireless networks, digital communications, semiconductor devices,
    and selected medical and chemical technologies. Administrative
    omission is concentrated in separation processes, catalysis,
    engine-exhaust control, heat pumps and refrigeration, power
    systems, batteries and fuel cells, water treatment, and other
    engineering technologies. Technology descriptions are abbreviated
    for readability. The figure describes the scale and technological
    location of the errors; the conditional tests in
    Table~\ref{tab:resolved_bias_tests} account for differences in field
    size.}
    \label{fig:primary_cpc_subclass_errors}
\end{figure}

Figure~\ref{fig:primary_cpc_subclass_errors} reports the technological location of the attributed administrative errors. The subclass-level results reveal two contrasting technological error profiles. False Green patents are concentrated in digital and communications technologies. Digital data processing (G06F) contains \num{18206} cases, followed by wireless networks (H04W; \num{11199}) and digital communications (H04L; \num{7526}). Semiconductor devices (H01L; \num{6145}) and control systems (G05B; \num{3895}) are also prominent. The concentration of \num{7996} cases in medical preparations (A61K), together with large numbers in organic chemistry and peptide subclasses, shows that administrative over-inclusion extends beyond ICT.

Silent Green patents display a different technological profile. The largest concentration occurs in separation processes (B01D; \num{2587}), followed by catalysis and process chemistry (B01J; \num{1175}), engine-exhaust control (F01N; \num{1055}), heat pumps and refrigeration (F25B; \num{905}), and electrical power distribution and storage (H02J; \num{894}). Further concentrations occur in heating and ventilation, batteries and fuel cells, engine fuel supply, engine control, heat and steam power plants, water treatment, electric machines, inorganic chemistry, and vehicle technologies. These technologies typically achieve climate benefits through physical, chemical, thermal, or systems-engineering mechanisms rather than through explicit environmental terminology.

The displayed counts show where the largest errors occur, but counts alone cannot establish bias because technological fields differ greatly in size. The conditional tests provide that evidence. Error attribution is substantially more strongly associated with detailed CPC subclasses than with broad CPC sections. Cram\'er's $V$ is approximately 0.40 in both disagreement directions at the subclass level, compared with 0.22--0.26 at the section level. Associations with filing period are much smaller ($V\leq0.072$). Thus, the source and direction of classification error depend primarily on the specific technology rather than on filing period or broad differences in field size.

The technological associations remain similar when the analysis is restricted to patents for which both independent LLMs report high confidence. Cram\'er's $V$ remains 0.392 for the administrative Type I subclass test and 0.408 for the administrative Type II subclass test; the corresponding section-level values are 0.246 and 0.208. The technological structure of the errors is therefore not driven by low-confidence adjudications. The sharply contrasting error profiles across independently defined non-Y CPC subclasses provide clear evidence that administrative misclassification is technology-specific rather than random. In the terms of Section~\ref{sec:construct_validity}, the error threatens validity, not merely reliability.

The same asymmetry appears in language. Adjusting for filing
period, primary non-Y CPC subclass, and eight competing lexical
categories, digital-context vocabulary raises the odds that a downward
disagreement is administrative over-inclusion rather than model failure by
a factor of 2.32 (95\% CI 2.16--2.49), while explicit adaptation
vocabulary raises the odds that an upward disagreement is administrative
omission rather than model failure by a factor of 14.76 (95\% CI
11.55--18.85). The first estimate is evidence of discriminant invalidity: the label partly responds to the vocabulary of a neighbouring construct, digital systems, rather than to climate function. Y02 reads environmental language well and technical
function poorly: it over-credits inventions that look green to a taxonomy
and overlooks those that merely are.
Appendix~\ref{sec:lexical_error_structure} reports all nine categories in
both directions.

\subsection{Where Correction Moves Green Invention}
\label{sec:corrected_landscape}

The systematic error structure produces a materially different map of green invention, the consequential-validity question of what follows from using Y02 as a measure. The administrative Y02 indicator identifies \num{592387} green patents, corresponding to 6.53\% of the $\sim$9 million granted patents. Removing the \num{180384} consensus-attributed False Green patents and adding the \num{29465} Silent Green patents produces a consensus-corrected estimate of \num{441468}. The correction therefore removes a net \num{150919} patents, reducing the measured green-patent population by 25.5\% and its corpus share from 6.53\% to 4.86\%. Unresolved downward disagreements retain their administrative green status, whereas unresolved upward disagreements remain administratively non-green. Table~\ref{tab:corrected_y02_landscape} reports the resulting family-level changes.

\begin{table}[H]
\centering
\caption{Changes in the measured green-patent landscape by Y02 family}
\label{tab:corrected_y02_landscape}
\small
\setlength{\tabcolsep}{4pt}
\renewcommand{\arraystretch}{1.12}

\begin{tabularx}{\textwidth}{
    @{}
    >{\raggedright\arraybackslash}X
    r
    r
    r
    r
    r
    @{}
}
\toprule
\textbf{Y02 family}
&
\textbf{Admin.}
&
\textbf{Corrected}
&
\textbf{Change}
&
\textbf{Admin. share}
&
\textbf{Corrected share}
\\
\midrule

Y02A: Adaptation
& 58,818 & 30,004 & -49.0\% & 9.93\% & 6.80\% \\

Y02B: Buildings
& 58,362 & 46,224 & -20.8\% & 9.85\% & 10.47\% \\

Y02C: Carbon capture
& 6,611 & 7,160 & +8.3\% & 1.12\% & 1.62\% \\

Y02D: ICT energy efficiency
& 62,373 & 20,237 & -67.6\% & 10.53\% & 4.58\% \\

Y02E: Energy
& 194,201 & 177,916 & -8.4\% & 32.78\% & 40.30\% \\

Y02P: Industrial production
& 80,199 & 44,017 & -45.1\% & 13.54\% & 9.97\% \\

Y02T: Transportation
& 112,506 & 98,908 & -12.1\% & 18.99\% & 22.40\% \\

Y02W: Waste and wastewater
& 19,317 & 17,002 & -12.0\% & 3.26\% & 3.85\% \\

\midrule
\textbf{Total}
&
\textbf{592,387}
&
\textbf{441,468}
&
\textbf{-25.5\%}
&
\textbf{100.00\%}
&
\textbf{100.00\%}
\\
\bottomrule
\end{tabularx}

\vspace{0.35em}
\begin{minipage}{0.97\textwidth}
\footnotesize
\textit{Notes:}
The corrected measure removes consensus-attributed False Green patents
and adds consensus-attributed Silent Green patents. Because Silent
Green patents have no observed administrative Y02 family, their
candidate family is assigned using their assessed technical domain or
primary non-Y CPC position. Family-level results are therefore
descriptive; the non-Y CPC tests in
Table~\ref{tab:resolved_bias_tests} provide the primary statistical
evidence of technological bias.
\end{minipage}
\end{table}

\begin{figure}[H]
    \centering
    \includegraphics[width=\textwidth]
    {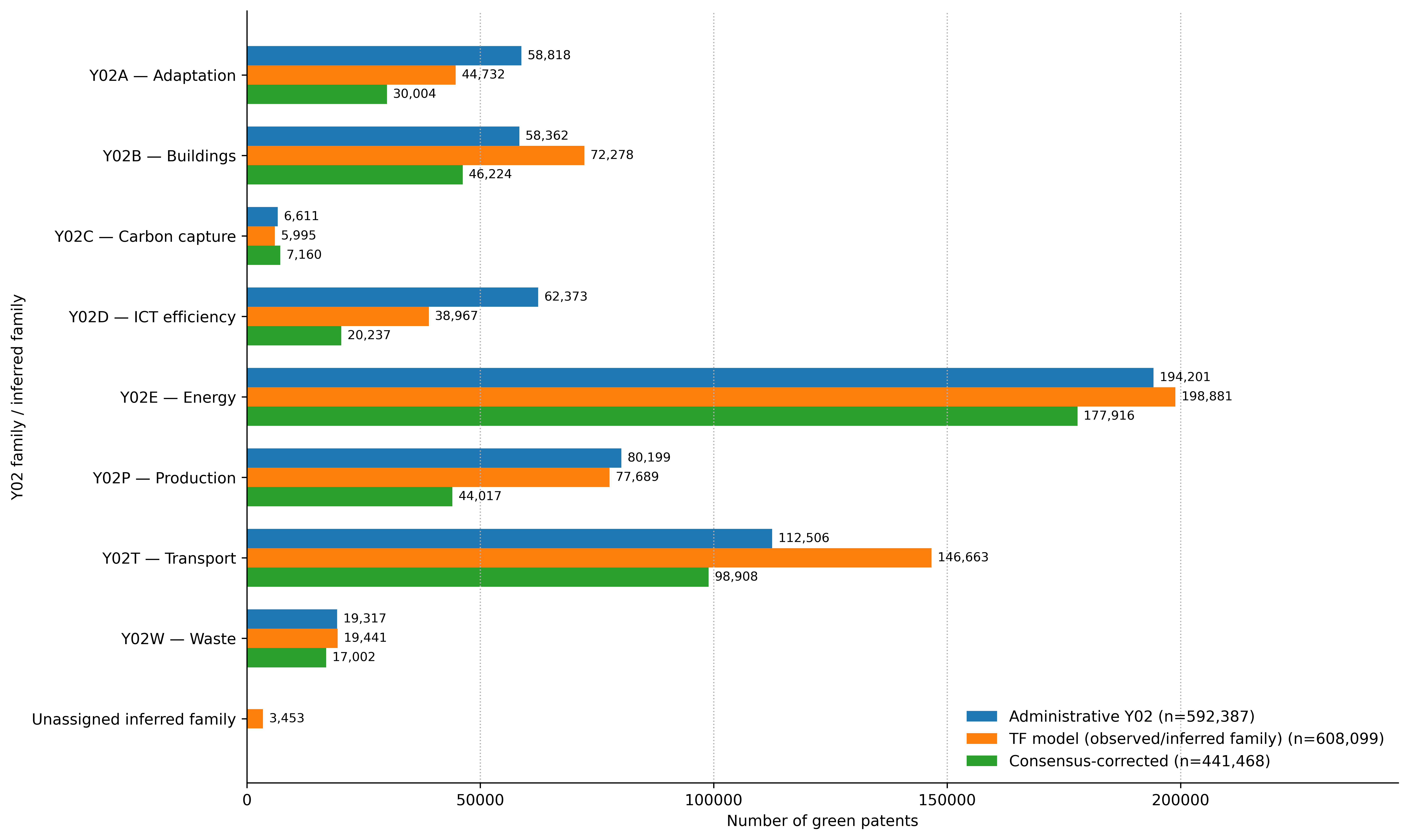}
    \caption{Green-patent counts under the administrative Y02
    classification, the TF model, and the consensus-corrected measure
    by Y02 family or inferred candidate family. The administrative
    measure contains \num{592387} patents, the TF model identifies
    \num{608099}, and the consensus-corrected measure contains
    \num{441468}. For TF-positive patents without an administrative Y02
    label, candidate-family assignment is based on the assessed domain or
    a primary non-Y CPC crosswalk. Candidate-family assignment affects
    the presentation of the TF distribution but does not cause a patent
    to enter the corrected measure; only cross-model consensus
    identifying a direct climate mechanism does so.}
    \label{fig:corrected_green_landscape}
\end{figure}

As Figure~\ref{fig:corrected_green_landscape} illustrates, correction changes both the scale and composition of measured green innovation. ICT energy efficiency declines from \num{62373} patents to \num{20237}, a reduction of 67.6\%, while Industrial production declines by 45.1\% and Adaptation by 49.0\%. By contrast, Energy declines by only 8.4\%, and Carbon capture records a small net increase after recovered Silent Green patents are added. The relative technological landscape consequently shifts. Energy rises from 32.8\% of administrative green patents to 40.3\% of the corrected population, and Transportation rises from 19.0\% to 22.4\%. ICT energy efficiency falls from 10.5\% to 4.6\%, while Industrial production falls from 13.5\% to 10.0\%. Thus, even where the broad ranking of large technology families remains stable, their measured importance changes substantially.

The decline of the industrial-production family (Y02P) does not
contradict the omission pattern; it shows both errors operating within one
domain. Y02P loses tagged patents without a direct climate mechanism,
including counter-directional fossil-resource processing (for example,
\#7811444 and \#10113118 in
Appendix~\ref{app:illustrative_patent_cases}), while separation,
catalysis, and emissions-control inventions omitted from Y02 enter the
corrected measure through their assessed domains
(Figure~\ref{fig:primary_cpc_subclass_errors}). The same industrial
landscape is thus over-credited where the tag follows category proximity
and under-credited where climate function is embedded in process
engineering.

Taken together, the non-Y CPC and corrected-landscape analyses establish two distinct aspects of measurement bias. The non-Y CPC tests show that the probability and direction of administrative error depend on the underlying technology. The corrected landscape shows the substantive consequence: the raw Y02 indicator overstates digitally visible and broadly framed technologies while under-recognizing climate mechanisms embedded in energy, transport, chemical, and industrial technologies. 

The answer to the second question is therefore no. The errors are
not idiosyncratic: whether a disagreement is an error of the label, and in
which direction, depends far more on the specific technology (Cram\'er's
$V\approx0.40$ across CPC subclasses) than on when the patent was filed
($V\leq0.072$), and correcting the errors changes the composition of
measured green invention, not merely its scale. They form a bias. The
third question is what causes it.

\section{What Are the Main Causes of the Errors?}\label{sec:causes}

Four accounts could explain why Y02 errs in these directions
(Section~\ref{sec:competing_accounts} and Table~\ref{tab:accounts_design}).
This section runs the comparison row by row: first the strategic accounts,
which locate the cause in what applicants do
(Section~\ref{sec:greenwashing_event}), then the classification-capacity
accounts, which locate it in how the label is produced
(Section~\ref{sec:complexity_silent_green}).
Section~\ref{sec:which_dominates} assembles the verdicts.

\subsection{The Strategic Accounts: Did Firms Game the Green
Patent System?}
\label{sec:greenwashing_event}

Table~\ref{tab:accounts_design} makes the strategic accounts
falsifiable: greenwashing predicts a discrete change when green
classification became salient and consequential. We test that prediction
at both candidate moments. Rather than the taxonomy failing to recognize inventions whose climate function is hard to read, firms may have learned to write patents in the language the classifier rewards, so that over-inclusion reflects strategic green framing rather than classification error. The two accounts imply different responses to the moments when green classification became salient and consequential, the introduction of Y02 tagging in 2010 and the launch of the CPC in 2013. Strategic framing makes two testable predictions: after these events, (i) False Green patents should adopt explicit green language at higher rates than genuinely green patents, and (ii) the prevalence of False Green should rise discretely. We test both.

To enable firm-level controls, the language test filters the \num{200102} resolved Y02 patents to include only those that can be successfully linked to a specific patenting firm and grant year. This requirement reduces the sample to \num{85541} patents (\num{79225} False Green and \num{6316} LLM-supported genuine green) from \num{3422} firms over 1964--2024. False Green status is defined here by the Qwen--Gemma consensus rather than by the screening model, and all specifications include CPC-section fixed effects with firm-clustered standard errors.\footnote{Formally, for each cutoff $c\in\{2010,2013\}$ we estimate $Y_{ift}=\alpha+\beta_1 FG_i+\beta_2(FG_i\times Post_{tc})+\mu_f+\delta_t+\theta_s+\varepsilon_{ift}$, where $FG_i$ identifies a consensus-attributed False Green patent, $Post_{tc}$ equals one for filing years at or after cutoff $c$, and $\mu_f$, $\delta_t$, and $\theta_s$ denote firm, filing-year, and CPC-section fixed effects. The filing-year effects absorb the main post indicator, so $\beta_2$ captures the differential change for False Green patents relative to LLM-supported Y02 patents. Standard errors are clustered by firm; the rare binary outcome uses a linear probability model to retain the high-dimensional fixed effects.} Explicit green language is rare throughout the corpus: only 312 of the \num{85541} patents (0.36\%) contain any explicit green phrase, so the test asks whether that rarity shifts differentially for False Green patents after each event. The phrase dictionary, the matching rule, and the two derived framing outcomes are detailed in Appendix~\ref{app:green_dictionary}.

\begin{table}[H]
\centering
\caption{Explicit green-language use by patent type and period, around the
2010 Y02 introduction and the 2013 CPC launch.}
\label{tab:greenwashing_descriptives}
\begin{tabular}{llrrr}
\toprule
Patent type & Period & $N$ & Any green language (\%) & Mean green rate \\
\midrule
\multicolumn{5}{l}{\textit{Panel A: cutoff 2010}}\\
LLM-supported genuine green & Before & \num{3089}  & 1.748 & 0.239 \\
LLM-supported genuine green & After  & \num{3227}  & 1.611 & 0.329 \\
False Green                 & Before & \num{34024} & 0.194 & 0.037 \\
False Green                 & After  & \num{45201} & 0.310 & 0.046 \\
\midrule
\multicolumn{5}{l}{\textit{Panel B: cutoff 2013}}\\
LLM-supported genuine green & Before & \num{3879}  & 1.908 & 0.290 \\
LLM-supported genuine green & After  & \num{2437}  & 1.313 & 0.277 \\
False Green                 & Before & \num{42878} & 0.238 & 0.042 \\
False Green                 & After  & \num{36347} & 0.286 & 0.042 \\
\bottomrule
\end{tabular}
\begin{minipage}{0.92\textwidth}
\vspace{4pt}
{\footnotesize \textit{Notes:} ``Any green language'' is the share of patents
containing at least one explicit green phrase; ``mean green rate'' is the
average intensity of green-phrase use. Sample: \num{85541} resolved Y02
patents with valid firm and grant year.}
\end{minipage}
\end{table}

The 2010 introduction of Y02 tagging produces no differential
change in either the probability of explicit green language ($0.373$
percentage points; SE $0.389$, $p=0.338$) or its intensity ($0.0065$;
SE $0.0101$, $p=0.518$). The 2013 CPC launch is followed by a small
relative increase in probability ($0.835$ percentage points; SE $0.402$,
$p=0.038$) and no significant change in intensity ($0.0204$; SE $0.0110$,
$p=0.065$). The estimate does not survive a Bonferroni correction across
the four related tests, which sets the threshold at $p<0.0125$. It is also
arithmetically a control-group effect: explicit green language among
genuinely green patents falls from 1.91\% to 1.31\% after 2013, while
among False Green patents it moves from 0.24\% to 0.29\%
(Table~\ref{tab:greenwashing_descriptives}). False Green patents did not
begin describing themselves as green. The comparison group stopped.

\begin{table}[H]
\centering
\caption{Difference-in-differences: the differential change in explicit green
language for False Green patents after each event
(False Green $\times$ post interaction).}
\label{tab:greenwashing_did}
\begin{tabular}{llrrr}
\toprule
Cutoff & Outcome & Interaction & SE & $p$ \\
\midrule
2010 & Any green language (pp) & 0.373  & 0.389  & 0.338 \\
2010 & Language intensity      & 0.0065 & 0.0101 & 0.518 \\
2013 & Any green language (pp) & 0.835  & 0.402  & 0.038 \\
2013 & Language intensity      & 0.0204 & 0.0110 & 0.065 \\
\bottomrule
\end{tabular}
\begin{minipage}{0.92\textwidth}
\vspace{4pt}
{\footnotesize \textit{Notes:} $N=\num{85541}$ patents across \num{3422} firms.
Estimates are the interaction between False Green status (Qwen--Gemma consensus)
and an indicator for grants after the cutoff, with firm, filing-year, and CPC-section fixed effects and firm-clustered standard errors. ``pp'' denotes percentage points.}
\end{minipage}
\end{table}

\begin{figure}[H]
    \centering
    \includegraphics[width=0.82\textwidth]{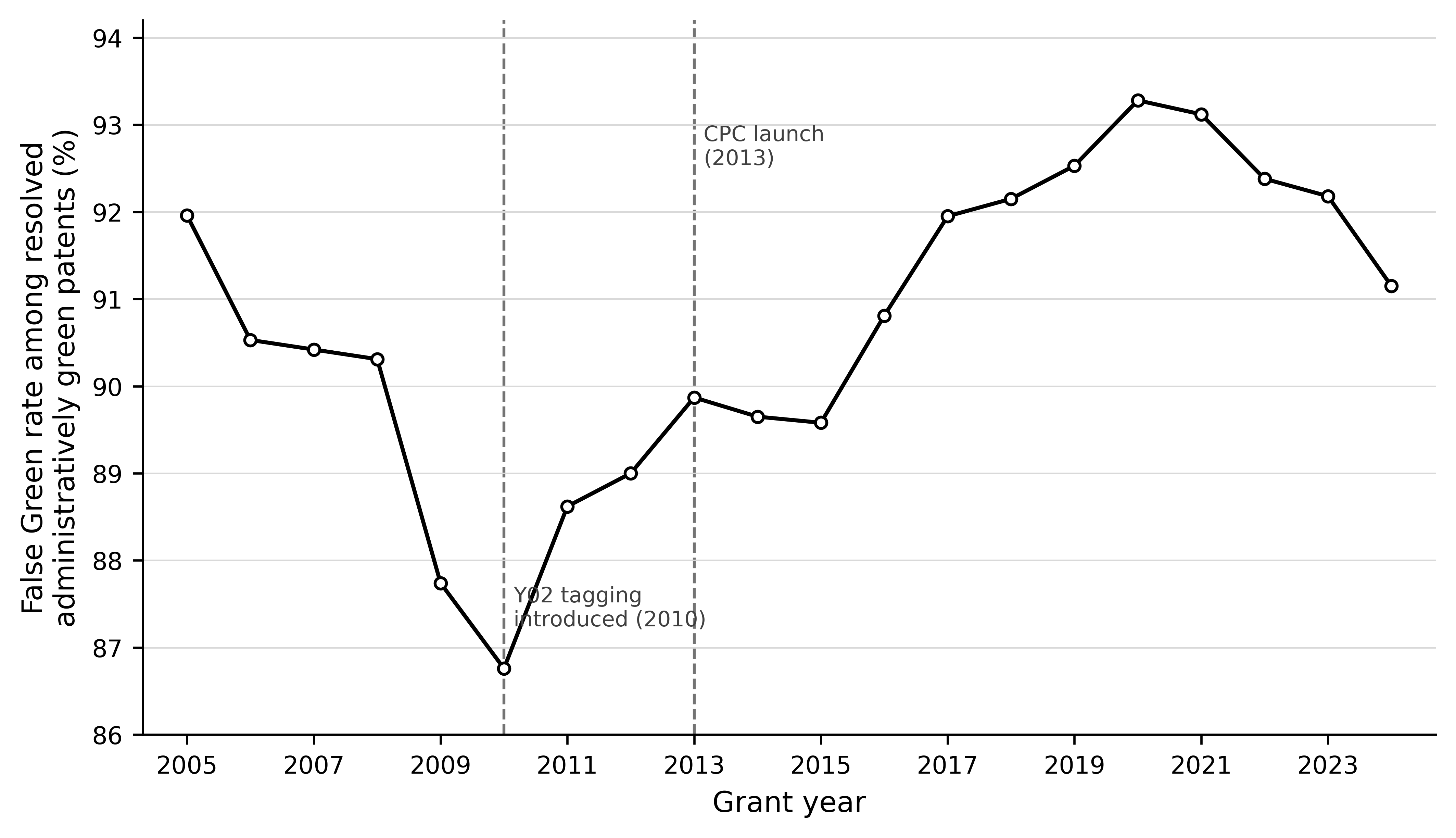}
    \caption{Annual False Green rate among resolved administratively green
    patents, 2005--2024. Dashed vertical lines mark the 2010 introduction of
    Y02 tagging and the 2013 CPC launch. The rate exhibits a dip around
    2009--2010 and a gradual upward drift, with no discrete break at either
    event.}
    \label{fig:false_green_rate_annual}
\end{figure}

The second test asks whether the prevalence of False Green patents spiked when green classifications were introduced. An interrupted-time-series analysis of all \num{200102} resolved administratively green patents (\num{180384} False Green and \num{19718} LLM-supported genuine green) reveals no discrete break at either event (Table~\ref{tab:false_green_rate_its}). For the 2010 Y02 introduction, the immediate level shift is actually downward (odds ratio $0.649$, $p<0.001$), with the annual False Green rate dipping from roughly 90\% in the mid-2000s to 86.76\% in 2010, before beginning a gradual upward drift (annual odds ratio $1.060$, trend $p<0.001$). Similarly, the 2013 CPC launch shows no significant immediate jump (odds ratio $0.879$, $p=0.144$), again followed by a gradual drift (annual odds ratio $1.058$, trend $p<0.001$). Both cutoffs therefore indicate a gradual post-event drift rather than a sudden, policy-induced response. As shown in the underlying annual series (Figure~\ref{fig:false_green_rate_annual}), the False Green rate fluctuates within a narrow 87--93\% band across two decades, drifting upward gradually with no visible discontinuity at either policy date.

\begin{table}[H]
\centering
\caption{Interrupted-time-series comparison of the False Green rate before and
after each event (Qwen--Gemma outcome).}
\label{tab:false_green_rate_its}
{\small
\begin{tabular}{lrrrrrccl}
\toprule
Cutoff & Pre $N$ & Post $N$ & Pre rate (\%) & Post rate (\%) &
Raw diff (pp) & Immed.\ level OR & Annual OR\\
\midrule
2010 & \num{80127}  & \num{119975} & 88.90 & 90.98 & 2.09 & $0.649^{***}$ & $1.060^{***}$  \\
2013 & \num{100467} & \num{99635}  & 88.76 & 91.54 & 2.78 & $0.879$       & $1.058^{***}$ \\
\bottomrule
\end{tabular}}
\begin{minipage}{0.98\textwidth}
\vspace{4pt}
{\footnotesize \textit{Notes:} All \num{200102} resolved administratively green
patents. ``Immed.\ level OR'' is the discontinuous shift at the cutoff;
``Annual OR'' is the post-cutoff annual trend. Immediate-level $p<0.001$ (2010)
and $p=0.144$ (2013); annual-trend $p<0.001$ and joint level/slope $p<0.001$ for
both cutoffs. $^{***}p<0.001$.}
\end{minipage}
\end{table}

Taken together, the greenwashing explanation receives at most weak and mixed support. The 2010 introduction of Y02 tagging produced no differential increase in green language and no discrete rise in misclassification. The 2013 CPC launch produced a small, one-margin, statistically significant relative increase in explicit green language, but no discrete jump in the False Green rate, and an effect that is economically small and partly an artifact of declining framing among genuinely green patents. Against the strong, consistent, technology-specific error structure documented in Section~\ref{sec:technological_bias}, we read the event evidence as leaving classification capacity, not applicant strategy, as the more plausible source of the systematic component, while acknowledging that a modest strategic contribution after 2013 cannot be excluded.

The greenhushing account cannot be tested in the same way, because
it predicts no salience break: a firm that stays quiet about a genuine
climate function can do so whether or not the tag is salient. The
consensus-resolved Silent Green cases are consistent with it. Their
climate function is typically described in power-electronics,
thermodynamic, or process-engineering vocabulary rather than
environmental language (Table~\ref{tab:illustrative_error_mechanisms}).
That signature, however, is exactly what non-strategic technical
description would also produce, so it does not discriminate between the
accounts. One implication can be checked. If omission reflected stable
firm-level disclosure strategies, its technological gradient would vary
across firms rather than within them; the atypicality gradient in
omission survives firm fixed effects (Appendix~\ref{app:atypicality}),
which rules out stable firm-level strategy as its source but leaves
technology-specific under-communication unidentified. We therefore treat
greenhushing as inferred, not directly tested. Neither strategic account
explains the technological patterning of the errors. The next subsection asks what does.

\subsection{The Capacity Accounts: Legibility, Not Difficulty}
\label{sec:complexity_silent_green}

With the strategic accounts unable to explain the technological patterning of the errors, we turn to the capacity accounts in the second row of Table~\ref{tab:accounts_design}. Descriptively, Silent Green patents have substantially greater reflection complexity than recognized corrected-green patents (means 53.61 versus 48.34), while their mean structural complexity is lower (9.49 versus 10.02). The two complexity measures therefore capture different aspects of technological structure rather than a common one-dimensional difficulty scale.

The standalone models already reveal the opposing pattern. With filing-year fixed effects, reflection complexity is positively associated with Silent Green ($\beta=0.485$, $SE=0.012$, OR $=1.62$, $p<0.001$), whereas structural complexity is negatively associated with Silent Green ($\beta=-0.476$, $SE=0.007$, OR $=0.62$, $p<0.001$). These coefficients change only modestly when technological atypicality is added, indicating that the two complexity dimensions are not proxies for atypical recombination.

Reflection complexity is strongly associated with Silent Green status. Controlling for technological atypicality, its quadratic term, and filing-year fixed effects, the coefficient on standardized reflection complexity is $0.473$ ($SE=0.012$, $p<0.001$), corresponding to an odds ratio of 1.61. Thus, a one-standard-deviation increase in reflection complexity is associated with approximately 61\% higher odds that a substantively green patent is administratively omitted. The atypicality controls, and the absence of an interaction between atypicality and reflection complexity, are reported in Appendix~\ref{app:atypicality}.

The magnitude is substantively meaningful. Average adjusted predictions increase from 3.26\% at one standard deviation below mean reflection complexity to 5.12\% at the mean, 7.95\% at one standard deviation above the mean, and 12.12\% at two standard deviations above the mean. The nonlinear reflection-complexity specification is also strongly supported ($\beta_R=0.613$, $\beta_{R^2}=0.338$, joint $p<0.001$), and the raw quartile pattern shows that the increase is especially concentrated in the upper tail: Silent Green rates are 4.94\%, 3.47\%, 3.63\%, and 10.28\% across successive reflection-complexity quartiles.

Structural complexity exhibits the opposite relationship. Conditional on atypicality, a one-standard-deviation increase in structural complexity is associated with lower odds of Silent Green ($\beta=-0.466$, OR $=0.63$, $p<0.001$). When both complexity measures and atypicality enter simultaneously, reflection complexity remains positively associated with omission ($\beta=0.372$, OR $=1.45$, $p<0.001$), while structural complexity remains negative ($\beta=-0.392$, OR $=0.68$, $p<0.001$). The result cautions against treating technological complexity as a single construct: the dimension of complexity matters for administrative recognition.

This is the complexity paradox. The two associations are near-mirror images. In the
specifications that add atypicality to a single complexity measure, the
reflection coefficient is $+0.473$ and the structural coefficient is
$-0.466$ (Table~\ref{tab:complexity_main}, columns 3 and 4); because odds
change multiplicatively, the protection conferred by structural complexity
is almost exactly as large as the penalty imposed by reflection
complexity. Entering both simultaneously attenuates each but leaves the
opposition intact ($+0.372$ and $-0.392$). Since the two measures
correlate at only $-0.110$, this is not one construct entering twice with
opposite signs.

The opposition identifies the operative variable as legibility
rather than difficulty. Reflection complexity is high when a technology
draws on capability bundles concentrated in few places; such an invention
offers the classifier no familiar template, and a real climate function
goes unread. We call this under-matching. Structural complexity is high
when a technology's knowledge base has a diverse combinatorial network
structure; such an invention offers many recognizable anchors, which makes
placement easy, and, when placement is wrong, makes it wrong in the
direction of inclusion. We call this over-matching. On this reading the
classification system does not simply fail more often on hard inventions.
It fails in a direction set by how many familiar handles the invention
offers.

\begin{table}[H]
\centering
\caption{Technological legibility, complexity, and the direction of administrative error}
\label{tab:complexity_main}
\small
\resizebox{\textwidth}{!}{%
\begin{tabular}{lcccccc}
\toprule
& \multicolumn{5}{c}{\textbf{Silent Green}} & \textbf{Error direction} \\
\cmidrule(lr){2-6}\cmidrule(lr){7-7}
& (1) & (2) & (3) & (4) & (5) & (6) \\
& Reflection only & Structural only & Reflection + atyp. & Structural + atyp. & Both + atyp. & Silent vs. False \\
\midrule
Reflection complexity
& $0.485^{***}$ & --- & $0.473^{***}$ & --- & $0.372^{***}$ & $0.897^{***}$ \\
& $(0.012)$ & --- & $(0.012)$ & --- & $(0.011)$ & $(0.011)$ \\
\quad Odds ratio & 1.62 & --- & 1.61 & --- & 1.45 & 2.45 \\
\addlinespace
Structural complexity
& --- & $-0.476^{***}$ & --- & $-0.466^{***}$ & $-0.392^{***}$ & $-0.150^{***}$ \\
& --- & $(0.007)$ & --- & $(0.007)$ & $(0.007)$ & $(0.008)$ \\
\quad Odds ratio & --- & 0.62 & --- & 0.63 & 0.68 & 0.86 \\
\midrule
Atypicality and its square & No & No & Yes & Yes & Yes & Yes \\
Filing-year FE & Yes & Yes & Yes & Yes & Yes & Yes \\
Patents & 347,241 & 347,241 & 347,241 & 347,241 & 347,241 & 157,241 \\
\bottomrule
\end{tabular}%
}

\vspace{0.4em}
\begin{minipage}{0.98\textwidth}
\footnotesize
\textit{Notes:}
Columns~(1)--(5) estimate the probability that a patent is Silent Green within the consensus-corrected green-patent portfolio. Columns~(1) and (2) include reflection complexity and structural complexity separately. Columns~(3) and (4) add technological atypicality and its quadratic term to the respective complexity measure. Column~(5) includes atypicality and both complexity measures simultaneously. Column~(6) restricts the sample to consensus-attributed administrative errors and codes Silent Green as one and False Green as zero. Continuous variables are standardized within their respective estimation samples. All specifications include filing-year fixed effects. Logistic coefficients are reported with heteroskedasticity-robust standard errors in parentheses. Odds ratios are $\exp(\beta)$; the percentages in
Table~\ref{tab:competing_accounts} are $100\times(\mathrm{OR}-1)$ from
columns~(3), (4), and (6). Coefficients on the atypicality controls are
reported in Table~\ref{tab:complexity_full}, and atypicality is defined in
Appendix~\ref{app:atypicality}. $^{*}p<0.05$; $^{**}p<0.01$; $^{***}p<0.001$.
\end{minipage}
\end{table}

Most importantly, reflection complexity predicts the direction of administrative error. Among administrative errors with observed complexity, Silent Green patents have mean reflection complexity of 54.33, compared with 40.00 for False Green patents. In the full error-direction model, which controls for structural complexity, technological atypicality, its quadratic term, and filing-year fixed effects, a one-standard-deviation increase in reflection complexity is associated with 2.45 times the odds that an administrative error is Silent Green rather than False Green ($\beta=0.897$, $SE=0.011$, $p<0.001$). Structural complexity again points in the opposite direction (OR $=0.86$, $p<0.001$); the atypicality controls are reported in Appendix~\ref{app:atypicality}.

This distinction is central to the measurement-bias argument. Reflection complexity does not merely identify patents for which classification is noisy. It systematically predicts whether an administrative mistake takes the form of under-recognition rather than over-recognition. The result therefore provides patent-level evidence of direction-specific measurement bias consistent with bounded classification capacity.

The association is robust to sample definition and time period. In the maximum available reflection-complexity sample, which does not require atypicality to be defined ($N=\num{437186}$), the estimated odds ratio is 1.71, compared with 1.62 for the corresponding standalone specification in the common estimation sample (Table~\ref{tab:complexity_main}, column~1). Restricting the analysis to patents filed from 2010 onward yields an odds ratio of 1.70, and a 2005--2015 window yields 1.64. These estimates are close to the main result and indicate that differential availability of atypicality scores and early-period classification history do not drive the complexity association. 

The opposing complexity associations also survive more demanding controls for technological and geographic composition. In the Silent Green specification with filing-year, CPC-subclass, and inventor-location fixed effects, reflection complexity remains positive and structural complexity negative; under two-way clustering by CPC subclass and inventor location, the corresponding odds ratios are 1.78 ($p=0.035$) and 0.53 ($p<0.001$). In the error-direction specification, the corresponding odds ratios are 2.91 for reflection complexity ($p<0.001$) and 0.76 for structural complexity ($p=0.012$). Because high-dimensional fixed-effect logit removes observations under separation, these estimates are interpreted as robustness evidence rather than as direct coefficient-magnitude comparisons with the baseline models. Full estimates are reported in
Appendix~\ref{app:complexity_examiner_robustness}, which also reports the
examination-sorting tests discussed in Section~\ref{sec:discussion}.

Both capacity accounts therefore leave their signatures, and in
opposite directions: structural complexity tilts administrative error
toward over-inclusion, and reflection complexity toward omission. Section~\ref{sec:which_dominates} assembles the four verdicts.

\subsection{Which Account Dominates?}
\label{sec:which_dominates}

Table~\ref{tab:competing_accounts} returns to the four accounts
with the evidence in hand. Neither strategic account explains the
technological patterning of the errors: greenwashing leaves no discrete
break at either salience event, and greenhushing, though consistent with
individual cases, is inferred rather than directly tested. Both capacity
accounts leave their signatures, and they point in opposite directions:
structural complexity toward over-inclusion, reflection complexity toward
omission. What remains is an indicator that reads environmental language
well and technical function poorly, over-crediting inventions that look
green to a taxonomy and overlooking those that merely are.
The answer to the third question is therefore that the main causes
of the errors lie in how the label is produced rather than in what
applicants do.

\begin{table}[H]
\centering
\caption[Two errors, four accounts: verdicts]{Two errors, four accounts: which one dominates?}
\label{tab:competing_accounts}
\footnotesize
\setlength{\tabcolsep}{6pt}
\renewcommand{\arraystretch}{1.15}
\begin{tabularx}{\textwidth}{@{}>{\raggedright\arraybackslash}p{2.35cm} L L@{}}
\toprule
& {\color{fgorange}\textbf{Type~I}} $\cdot$ \textbf{False Green} $\cdot$ over-inclusion
& {\color{sgteal}\textbf{Type~II}} $\cdot$ \textbf{Silent Green} $\cdot$ omission \\
\midrule
\textbf{Strategic account}\newline\textit{what applicants do}
&
\acct{fgorange}{Greenwashing}\newline
Applicants use environmental language to attract the green tag.\newline
\textsc{Tested by} difference-in-differences around the Y02 introduction
and the CPC launch, and green-language use.\newline
\textit{Evidence:} immediate-level OR of the False Green rate $0.65$ in
2010 (a fall) and $0.88$ in 2013 (n.s.); a $+0.8$\,pp relative framing
change after 2013 that is driven by the comparison group and does not
survive a Bonferroni correction.
&
\acct{sgteal}{Greenhushing}\newline
Firms stay quiet about environmental achievements, masking green
function.\newline
\textsc{Argued from} consensus-resolved cases described in technical
rather than green vocabulary.\newline
\textit{Evidence:} consistent with individual cases but equally produced
by non-strategic technical description; the within-firm atypicality
gradient rules out stable firm-level disclosure strategy
(Appendix~\ref{app:atypicality}).
\\[3pt]
& \predbar{No discrete break at either event}
& \predbar{Inferred, not directly tested}
\\
\addlinespace[8pt]
\textbf{Bounded classification capacity}\newline
\textit{how the label is produced}
&
\acct{fgorange}{Structural complexity}\newline
\textbf{Combinatorial diversity.} High when knowledge exhibits diverse
network structure.\newline
$-37\%$ lower odds of Silent Green per one-SD rise\newline
$-14\%$ lower odds that the error is omission
&
\acct{sgteal}{Reflection complexity}\newline
\textbf{Capability scarcity.} High when a technology is tied to places
with scarce capabilities.\newline
$+61\%$ odds of Silent Green per one-SD rise\newline
$+145\%$ odds that the error is omission rather than over-inclusion
\\[3pt]
& \verdictbar{fgorange}{False Green (over-inclusion)}
& \verdictbar{sgteal}{Silent Green (omission)}
\\
\bottomrule
\end{tabularx}

\vspace{0.4em}
\begin{minipage}{0.97\textwidth}
\footnotesize
\textit{Notes:} Neither strategic account explains the
technological patterning of the errors. What remains is an indicator that
reads environmental language well and technical function poorly.
Percentages are $100\times(\mathrm{OR}-1)$ from
Table~\ref{tab:complexity_main}: the Silent Green percentages from
columns~(3) and (4), which control for atypicality and filing year, and
the error-direction percentages from column~(6), which also controls for
the other complexity measure. Rows distinguish where an error originates
and columns its direction; the accounts are not mutually exclusive, and
the verdicts concern which carries the systematic component of the error.
Scope: all four accounts are assessed within the \num{517772}
disagreement cases. Because PatentSBERTa is fine-tuned on Y02, errors the
two systems share generate no disagreement signal and lie outside this
comparison. The event tests detect discrete breaks and cannot rule out
gradual strategic framing. Greenwashing and policy--practice decoupling:
\textcite{BromleyPowell2012,delmas2011drivers,marquis2016scrutiny,meyer1977institutionalized}.
Greenhushing and strategic silence:
\textcite{carlos2018strategic,font2017greenhushing,Hilton2025,MontgomeryRobertsonSummers2026}.
Bounded classification capacity:
\textcite{Simon1955,Simon1957,TverskyKahneman1974,JiangGoetz2025}.
\end{minipage}
\end{table}

\section{Discussion and Policy Implications}\label{sec:discussion}

Taken together, the three answers, Y02 makes two distinct types of error, those errors form a bias rather than noise, and their main cause lies in how the label is produced rather than in what applicants do, sharpen a debate that has been treated as a puzzle about firm
behavior. Measurement does not explain the ESG innovation disconnect away;
at the level of technologies, it hides part of it. The indicator's
classification capacity is unevenly distributed across the technological
landscape and is thinnest in the engineering-intensive fields in which
the firms that ESG ratings under-recognize do much of their green
inventing, so the patent instrument and the rating instrument err in the
same direction. This has a direct methodological implication: studies that use raw Y02 counts as a dependent or explanatory variable inherit a direction-specific bias that will attenuate or invert estimated relationships between environmental performance and green inventions, particularly for samples concentrated in engineering-intensive fields.

\subsection{From Classification Error to Measurement Bias}
\label{sec:classification_to_bias}

The findings demonstrate why individual classification errors and systematic measurement bias should not be treated as equivalent. A False Green, Silent Green, model false positive, or model false negative is an instance-level classification error. Such an error does not, by itself, establish a failure of construct reliability, which concerns whether similar inputs to a measurement model yield similar outputs \parencite{JacobsWallach2021}. Bias arises only when the probability or direction of those errors varies systematically across relevant groups. The present results instead provide evidence concerning construct validity, particularly content and substantive validity. These forms of validity concern whether an operationalization captures the substantive nature of the intended construct and incorporates the relevant, and only the relevant, properties \parencite{JacobsWallach2021}. False Green patents indicate that Y02 includes inventions without a direct climate mechanism, whereas Silent Green patents indicate that it omits inventions whose technical mechanisms are directly climate-relevant.

More importantly, these errors are not evenly distributed across technologies. Within the resolved disagreement pools, administrative Type I attribution varies substantially across CPC sections ($V=0.256$) and subclasses ($V=0.397$), while administrative Type II attribution exhibits similarly strong technological variation ($V=0.216$ and $V=0.401$, respectively, see Table~\ref{tab:resolved_bias_tests}). False Green errors are concentrated in ICT and communications technologies, whereas Silent Green errors are concentrated in engineering-intensive technologies such as energy conversion, emissions control, batteries, heat pumps, industrial separation, and water treatment. The problem is therefore not simply that Y02 contains classification errors, but that its operationalization of green invention exhibits technology-specific threats to construct validity: Y02 does not simply make mistakes, it makes different mistakes in different parts of the technological landscape.

The patent-level analyses provide a second and stronger form of evidence that under-recognition is systematic. They reveal two empirically distinct forms of technological complexity that act on legibility in opposite directions. The two correlate at only $-0.110$, and both retain explanatory power when entered jointly and when the atypicality of an invention's technological combinations is held constant (Appendix~\ref{app:atypicality}).

More importantly, reflection complexity predicts the direction rather than merely the incidence of administrative error. Conditional on belonging to the consensus-attributed administrative-error sample, a one-standard-deviation increase in reflection complexity is associated with 2.45 times the odds that the error is Silent Green rather than False Green. Thus, the classification system does not simply become less accurate for difficult-to-read inventions. Its errors become systematically asymmetric. Structural complexity moves in the opposite direction, underscoring that the relevant mechanism is not generic ``complexity'' but the particular technological structure captured by the reflection measure.

This result sharpens the bounded-capacity interpretation. Administrative taxonomies appear better able to recognize inventions whose climate relevance maps cleanly onto familiar and widely legible technological structures and less able to recover climate functions embedded in capability-scarce or difficult-to-map technologies. The finding does not establish the cognitive process used by individual classifiers, but it identifies an observable patent-level signature of direction-specific measurement bias. Appendix~\ref{app:atypicality} shows that under-recognition also rises, along an inverted U, when inventions depart from familiar recombinant templates.

The opposing signs help resolve the apparent complexity paradox. Reflection complexity captures technologies associated with scarce and sophisticated capability bundles; such sophistication can increase economic value while reducing administrative legibility when climate relevance is embedded in specialized knowledge that does not map cleanly onto familiar categories. Structural complexity captures a different feature of technological organization. When an invention exposes multiple recognizable technological relationships, those relationships may provide classificatory anchors that make the invention easier to place within an established taxonomy. In this sense, sophistication can reduce legibility while recognizable technological structure can increase it.

The same distinction offers a cautious interpretation of False Green. Conditional on an administrative error occurring, greater structural complexity tilts the error away from Silent Green and relatively toward False Green. One possible explanation is over-matching: multiple recognizable technological anchors may facilitate correct recognition in many cases but also create more routes through which a non-green invention can be mapped onto a green category. Capability-scarce technologies face the opposite risk of under-matching and administrative invisibility. The results therefore suggest two asymmetric matching failures, under-matching and over-matching, rather than a one-dimensional complexity penalty.

The examination-sorting tests point away from a simpler organizational explanation. Neither Art Unit fixed effects nor measured prior examiner specialization attenuate the negative structural-complexity association (Appendix~\ref{app:complexity_examiner_robustness},
Panels~B and~C). This does not prove the classificatory-anchor mechanism, because examiner assignment and patent difficulty may be jointly determined, but it makes preferential routing to specialized examination expertise an unlikely explanation for the main structural result.

The temporal-language evidence points the same way
(Section~\ref{sec:greenwashing_event}). Explicit green language is rare,
appearing in 0.36\% of the \num{85541} firm-linked patents; it shows no
differential rise among False Green patents after 2010, and after 2013
only a small relative increase that is driven by a decline in the
comparison group and does not survive correction for multiple testing.
The evidence therefore provides no robust support for the claim that
firms deliberately increased strategic green framing to exploit the
classification system.

This distinction also clarifies the objective of the construct-validity test. We do not assume that all error can or should be eliminated. Some residual error is unavoidable in any classification of millions of heterogeneous technical documents. The central objective is to reduce the systematic component of error that alters sectoral comparisons and substantive conclusions. Random measurement error primarily reduces precision and weakens construct reliability. By contrast, systematic over- or under-inclusion across technological groups indicates measurement bias and threatens construct validity \parencite{JacobsWallach2021}.

The Error-as-Signal framework changes the interpretation of model failure. When administrative labels are imperfect, disagreement with those labels can contain information about weaknesses in the underlying indicator. PatentSBERTa is therefore valuable not because it provides a new unquestioned truth, but because it generates a scalable competing measurement that identifies cases requiring deeper evaluation. The subsequent construct-validity test turns disagreement into attributed evidence about administrative and model failure.

\subsection{Implications for Validating Machine-Learning Measures}

The results reveal both the value and the limitations of dual-measurement disagreement as a validation method. Error correction becomes observable when PatentSBERTa challenges an incorrect administrative label and the construct-validity test supports the model prediction. By contrast, bias preservation remains hidden when the administrative system and the model produce the same incorrect classification, because no disagreement signal is generated. This limitation is particularly important when a model is trained on the administrative labels that it is subsequently used to validate. Future validation designs should therefore combine disagreement-based sampling with targeted substantive reviews of representative agreement cases, especially high-confidence agreements, emerging technologies, and domains underrepresented in the training data.

Bias amplification, by contrast, is observable within the
disagreement pool, and Section~\ref{sec:conservative_by_design} documents
it: the same digital-context vocabulary that raises the odds of
administrative over-inclusion inside Y02 also raises the odds of model
over-inclusion outside it. A model fine-tuned on an administrative label
does not merely inherit that label's noise; it reproduces the label's
directional sensitivity to surface language on cases the label never
touched. Validation of label-trained models should therefore test whether
model and label errors share a lexical or structural signature, not only whether the model reproduces the label accurately. The point extends the algorithmic-auditing literature \parencite{Raji2020}: where the reference labels are themselves administrative, validating a model against them is not the same as validating it against the construct.

The correction also shows why validation should examine differences between administrative categories rather than only aggregate accuracy. Consensus correction disproportionately reduces measured adaptation: the number of adaptation patents falls from \num{58818} to \num{30004} ($-49.0\%$), compared with a decline from \num{533569} to \num{411464} ($-22.9\%$) across mitigation-oriented Y02 families. Adaptation's share of measured green invention consequently decreases from 9.93\% to 6.80\%. This difference is driven by substantial over-inclusion in Y02A: \num{29347} administratively classified adaptation patents are identified as False Green, whereas only \num{533} omitted patents are added as Silent Green. Validation exercises should therefore test not only whether classification errors occur, but also whether their frequency and direction differ across substantive categories.

The bagging-inspired multi-model design separates scalable screening from substantive adjudication. PatentSBERTa screens the full patent corpus, while Qwen and Gemma independently evaluate whether each flagged invention contains a direct climate-mitigation or adaptation mechanism. Cross-model consensus permits scalable attribution of the \num{400186} resolved disagreements, whereas the \num{117586} mechanism disagreements form a transparent escalation set for evaluation by a stronger adjudicating model or by domain experts. This design reduces dependence on the reasoning and error tendencies of any single model while preserving uncertainty rather than forcing unresolved cases into definitive error categories.

\subsection{Implications for Managers, Investors, and
Classification Authorities}
\label{sec:managerial_implications}

The findings change three practical decisions. The first is when
to trust a Y02 tag. Users of green-patent data, ESG raters, asset
managers, and strategists benchmarking rivals' green innovation, should
not treat Y02 families as equally reliable. After correction, the ICT
energy-efficiency family (Y02D) retains 32\% of its administrative count
and the adaptation family (Y02A) 51\%, whereas energy (Y02E) retains 92\%
and transportation (Y02T) 88\%
(Table~\ref{tab:corrected_y02_landscape}). A Y02D count therefore
overstates assessed green invention roughly threefold, while a Y02E count
is within about ten percent. For ESG screens this matters beyond
accuracy: a screen that combines ratings with Y02-based green-patent
counts is likely to compound, rather than correct, the under-recognition
of the energy producers that \textcite{Cohen2026} find to be key green
innovators.

The second is when to trust the absence of a tag. Absence is
least informative exactly where climate function is hardest to read: in
separation processes, catalysis, exhaust control, heat pumps, and power
electronics (Figure~\ref{fig:primary_cpc_subclass_errors}), and more
generally among technologies with high reflection complexity. Firms whose
climate contribution sits in these fields are under-credited by
patent-based green metrics and cannot rely on those metrics to surface it;
documenting the climate mechanism of their patent portfolios in their own
disclosures, rather than their environmental aspirations, is the
corresponding remedy.

The third is where classification authorities should spend scarce
review capacity. Omission risk rises with reflection complexity and falls with structural complexity, whereas over-inclusion concentrates in ICT subclasses and in
the adaptation family; reviews targeted on these observable signatures
should recover more error per reviewed patent than uniform sampling. These
implications are timely: since 2 July 2026, ESG rating providers operating
in the European Union have been supervised by the European Securities and
Markets Authority under Regulation (EU) 2024/3005, which makes the
validity of rating inputs, including any built on green-patent counts, a
regulatory as well as a scholarly question. More generally, organizations
that train models on administrative labels should read disagreement as
information: in our setting, more than half of what standard evaluation
scores as model error is error in the label.

\subsection{Implications for Research and Policy}

For research, the immediate implication is that raw Y02 counts are not a neutral dependent variable. Analyses of induced innovation, green directed technical change, and the returns to environmental performance should either use corrected counts, bound their estimates using the sensitivity range reported here, or demonstrate robustness to the direction-specific error structure we document. For policy, indicators built on Y02, innovation scoreboards, green industrial-strategy dashboards, and sustainable-finance screens, currently over-reward sectors whose inventions are described in digital language and under-credit the mechanical, chemical, and process technologies that carry much of the burden of decarbonization. This interpretation extends \textcite{Cohen2026}, who document that traditional energy firms make substantial contributions to green innovation despite weak ESG recognition. Our results suggest that measurement bias in Y02 leads existing indicators to understate that disconnect rather than to produce it. This pattern complements \textcite{Cohen2026} and \textcite{Lan2025}, who document a disconnect between environmental recognition and green patenting. Our additional contribution is to identify a patent-measurement channel that conceals part of that disconnect: Y02 redistributes observed green invention systematically across technological fields rather than merely adding random noise. One force pulls the other way.
Some False Green patents support fossil-resource production, such as
heavy-hydrocarbon processing and hydraulic fracturing
(Table~\ref{tab:illustrative_error_mechanisms}), and removing them
withdraws green credit from energy producers themselves. Whether
correction widens the disconnect firm by firm therefore depends on the
mix of each firm's portfolio, which re-estimating the comparison of
\textcite{Cohen2026} with corrected counts would settle.
 Reallocating measurement attention is cheaper than reallocating capital wrongly.

The Error-as-Signal Framework is not specific to patents. Every administrative indicator that governs the net-zero transition, ESG ratings, the EU Taxonomy of sustainable activities, national R\&D and green-jobs statistics, and green-bond and green-fund labels, is a classification layered by finite human or algorithmic capacity onto a heterogeneous underlying reality, and every one can be tested by confronting its labels with an independent functional assessment and treating the disagreements as signal. Just as \textcite{Berg2022} decomposed ESG-rating divergence into scope, measurement, and weight, and \textcite{LernerSeru2022} catalogued the biases of patent-count data, we offer a portable diagnostic for the taxonomies of directed technical change. We therefore propose a research agenda: (i) apply the framework to ENV-TECH \parencite{HascicMigotto2015,Favot2023}, the IPC Green Inventory, and the EU Taxonomy to establish whether the bias toward information technology and against engineering-embedded mitigation is scheme-specific or general; (ii) extend the test to non-US patent offices to separate classification-capacity effects from CPC-diffusion effects; (iii) test whether correcting the indicator changes the estimated returns to green inventions, the measured direction of induced innovation, and the size of the ESG innovation disconnect at the firm level; and (iv) develop capacity-aware classification protocols that allocate examiner and algorithmic attention toward inventions exhibiting the observable signatures of under-recognition documented here, high reflection complexity and, secondarily, unconventional technological recombination.

\section{Conclusion and Limitations}\label{sec:conclusion}

We tested the construct validity of the most widely used green-patent indicator against the full corpus of \num{9075421} USPTO granted patents. Treating disagreement between the administrative Y02 label and an independent model as a diagnostic signal, and attributing errors only where two independent large language models agree on an invention's primary technical function, we identify \num{180384} False Green patents and \num{29465} Silent Green patents: Y02 makes both types of error. Correcting consensus-attributed administrative errors reduces the green-patent population by 25.5\%, from \num{592387} to \num{441468} patents, with sensitivity bounds of \num{390540} to \num{508126}.

Nor are the errors idiosyncratic. They are structured in both location and direction, which is what makes them a bias rather than noise. Over-inclusion concentrates in digital technologies, whereas omission concentrates in mechanical, chemical, energy, and industrial-process technologies. Reflection complexity predicts omission and structural complexity its absence, independently of the atypicality of an invention's technological combinations. Most importantly, among consensus-attributed administrative errors, a one-standard-deviation increase in reflection complexity is associated with 2.45 times the odds that the error is Silent Green rather than False Green. The technological characteristics of an invention therefore affect not simply whether Y02 is wrong, but the direction in which it is wrong.

Combined with the absence of a robust discontinuous increase in strategic green framing when green classification became salient, these findings answer the third question: they are more consistent with bounded classification capacity than with applicant gaming. The result bears on the ESG innovation disconnect, but not by explaining it away: inventions that are difficult for administrative taxonomies to recognize are disproportionately located in engineering-intensive domains central to decarbonization.

The instruments we use to measure the green transition are not neutral mirrors of innovation; they are lenses that magnify some technologies and dim others. When an indicator systematically over-credits the visible language of digital efficiency and under-recognizes the hard, unglamorous engineering through which much decarbonization happens, it does not merely mismeasure the transition, it misdirects it. Administrative indicators do not simply count innovation; they redistribute it. Validating these instruments is therefore not a technical afterthought but a first-order task of climate-innovation policy, and the Error-as-Signal Framework offers a practical way to begin.

The contribution is therefore not a more accurate count. It is a
direction. \textcite{Cohen2026} report that carbon-intensive incumbents
produce more, and higher-quality, green invention than their environmental
ratings would predict. We find that the instrument used to observe green
invention retains 92\% of the energy family and 88\% of transportation
while retaining only 32\% of ICT energy efficiency, lifting energy from
32.8\% to 40.3\% of the corrected population. Measured green invention
moves toward the engineering-embedded technologies whose innovation the
ESG literature finds unrecognized, and away from the digitally legible
technologies the raw indicator over-credits. Carbon capture, the branch in which they find energy producers to
be the leading innovators, is the only Y02 family that grows after
correction (by 8.3\%). At the level of technologies, correcting the
instrument therefore widens the gap they document rather than closing
it. The disconnect is not an artifact of what the instruments can see; it
is larger than what they show.

The study distinguishes individual error from systematic bias and shows that opposing errors can cancel numerically while continuing to distort the sectoral map of green inventions. Random or non-differential error mainly reduces precision; systematic error changes who appears to innovate, which technologies appear strategically important, and where policy and finance are directed. The findings reinforce recent work on patent classification and the ESG innovation disconnect. Text-based studies have shown that conventional green-patent measures can diverge substantially from narrower functional definitions \parencite{Santarlasci2023}. Our analysis extends that work by attributing resolved disagreement to administrative or model error and by showing that the consensus-attributed administrative errors follow distinct technological patterns. It also provides a patent-side measurement mechanism that may contribute to the limited recognition of industrial firms central to decarbonization under commonly used ESG and patent indicators \parencite{Cohen2026}; on the patent side, the bias compounds rather than offsets the under-recognition these firms receive from ESG ratings. The exploratory temporal-language test does not alter this measurement-based conclusion: it yields no robust evidence that firms deliberately increased green patent framing after either candidate Y02 cutoff.

Several limitations remain. First, disagreement-based validation cannot directly identify error convergence, in which Y02 and PatentSBERTa agree but are both wrong. Targeted sampling of agreement cases is therefore necessary to estimate residual shared error. Second, the consensus of the construct-validity test is a reference judgment rather than independently observed ground truth; agreement between independent judges establishes the reliability of the test, not its validity \parencite{deKok2025}. The \num{117586} unresolved cases, including \num{84333} high-confidence disagreements, show that model confidence is not equivalent to accuracy and that additional frontier-model, expert, or human adjudication is needed. Third, our benchmark is the construct that Y02 is used to
measure rather than the scheme's own inclusion rules
(Section~\ref{sec:theory}). Some attributed False Green cases therefore
reflect a scheme whose scope exceeds the construct rather than a tag
applied in error, and the two call for different remedies. Separating them
requires case-level coding of scheme conformity, which we leave to future
work. Fourth, the reported error rates are conditional on entry into the PatentSBERTa--Y02 disagreement pool and should not be interpreted as corpus-wide precision or recall for either system. Fifth, the substantive standard identifies direct and credible climate contributions but does not estimate realized environmental impact after commercialization. Sixth, the complexity analyses remain associational rather than causal. The opposing reflection- and structural-complexity associations survive detailed CPC-subclass and inventor-location fixed effects, two-way clustered inference, and a single-location sensitivity analysis, reducing concerns that broad technological or geographic composition drives the findings. Seventh, our evidence on the ESG
innovation disconnect is at the level of technologies: we show where
Y02's errors fall, not how they fall across firms with different ESG
ratings, and counter-directional False Green in fossil-production
technologies means that the firm-level effect cannot be read off the
technology-level one. Re-estimating the comparison of
\textcite{Cohen2026} with corrected counts would turn this implication
into a test. Future work should extend the framework to agreement-case tests, multilingual and older patent corpora, additional patent offices, and other administrative indicators used in ESG and innovation research.

The broader implication is that errors should not be viewed only as observations to remove. Once independently validated and examined across domains, they become evidence about how a measurement infrastructure sees, misses, and redistributes the phenomenon it is intended to represent. For green inventions, reducing systematic measurement bias is essential if policy and capital are to recognize the technologies most relevant to a net-zero transition.

\printbibliography

\appendix

\section{Technological Atypicality and Administrative Omission}
\label{app:atypicality}

This appendix reports the analysis of technological atypicality
that every main specification holds constant. Atypicality is a source of
classification difficulty distinct from the two forms of complexity
analysed in Section~\ref{sec:complexity_silent_green}: it concerns the
familiarity of an invention's technological combinations rather than the
structure of the knowledge base on which it draws. Its relationship with
omission follows an inverted U, it survives within-firm comparison, and it
does not interact with reflection complexity in the pooled sample.

\subsection{Rationale}

Technological atypicality provides an observable implication of the bounded-rationality account developed in Section~\ref{sec:theory}. Patents that recombine technologies in unfamiliar ways impose greater cognitive and taxonomic demands because their climate contribution may not fit established templates or familiar Y02 category mappings. Administrative omission should therefore become more likely as inventions move away from conventional technological configurations. This relationship need not be monotonic, however. Extremely atypical inventions may be unusually salient, may trigger closer scrutiny, or may contain distinctive specialist signals that facilitate recognition. The bounded-capacity argument consequently allows for an inverted-U relationship in which Silent Green is most likely among moderately-to-highly atypical inventions rather than among either conventional patents or the most extreme technological outliers. The empirical analysis tests this nonlinear implication while treating it as an association consistent with bounded classification capacity, not as direct evidence of the underlying cognitive mechanism.

Atypicality captures only one source of classification difficulty. An invention may combine familiar technologies yet remain difficult to classify because its climate-relevant function is embedded in a technologically complex structure; conversely, an atypical combination need not be complex. This distinction parallels research on recombinant technological search, which separates the familiarity of combinations from the uncertainty and difficulty generated by the underlying technological structure \parencite{Fleming2001}.

\subsection{Measure}
\label{sec:atypicality_method}

The sample and dependent variable are those of Section~\ref{sec:complexity_method}. We measure technological atypicality following the atypical-combinations approach of \textcite{Uzzi2013}, adapting it from journal pairs in scientific papers to technology-class pairs in patents. An invention is atypical to the extent that it combines technological components that rarely co-occur relative to what chance would predict in the contemporaneous patent landscape. Let $C_i$ denote the set of CPC classes assigned to patent $i$, and let $W(t_i)$ denote the cohort of patents in a time window around its filing year $t_i$. For each unordered class pair $\{j,k\}\subseteq C_i$, let $O_{jk}$ be the number of patents in $W(t_i)$ jointly assigned classes $j$ and $k$. We compare this observed co-assignment frequency with a null distribution obtained from $R$ degree-preserving randomizations of the patent--class incidence matrix within the window, holding fixed both the number of classes per patent and the total frequency of each class \parencite{Uzzi2013}. Writing $\mu_{jk}$ and $\sigma_{jk}$ for the mean and standard deviation of $O_{jk}$ across randomizations, the pairwise conventionality score is the $z$-score
\[
z_{jk} \;=\; \frac{O_{jk}-\mu_{jk}}{\sigma_{jk}} .
\]
A large positive $z_{jk}$ denotes a conventional pairing that co-occurs more often than chance, whereas a negative $z_{jk}$ denotes an atypical, rarer-than-chance pairing. We orient the score so that higher values indicate greater atypicality and summarize each patent by its single most atypical combination:
\[
\tilde{A}_i \;=\; \max_{\{j,k\}\subseteq C_i}\!\bigl(-z_{jk}\bigr)
\;=\; -\!\min_{\{j,k\}\subseteq C_i} z_{jk}.
\]
We refer to $\tilde{A}_i$ as the maximum atypicality of patent $i$: the degree to which its rarest technological combination departs from contemporaneous convention, so that higher values indicate a more unusual technological configuration. The measure is standardized within the complete-case estimation sample,
\[
Z_i \;=\; \frac{\tilde{A}_i-\bar{\tilde{A}}}{s_{\tilde{A}}},
\]
so that one unit corresponds to one standard deviation. The complete-case sample contains \num{349898} corrected-green patents, including \num{19747} Silent Green patents. The reduction from the full corrected-green portfolio of \num{441468} patents is due to unavailable maximum-atypicality scores, for instance, single-class patents, for which no within-patent combination exists, rather than to missing filing years.

\subsection{Estimation}

Because bounded classification capacity may generate a nonlinear relationship, we estimate a logistic model containing both linear and quadratic atypicality terms:
\[
\Pr(S_i=1\mid Z_i,t)
=
\Lambda\!\left(
\alpha+\beta_1 Z_i+\beta_2 Z_i^2+\delta_t
\right),
\]
where $S_i$ indicates Silent Green status, $Z_i$ denotes standardized maximum atypicality, $\delta_t$ represents filing-year fixed effects, and $\Lambda(\cdot)$ is the logistic function. An inverted-U relationship corresponds to $\beta_1>0$ and $\beta_2<0$, with an interior maximum at $Z^{\ast}=-\beta_1/(2\beta_2)$. The main specification uses heteroskedasticity-robust standard errors.

As a robustness test, we estimate a logistic model with both filing-year and firm fixed effects, clustering standard errors by firm. This specification compares patents within the same firm while controlling for common filing-year shocks. Firms without within-firm variation in Silent Green status do not identify the fixed-effect relationship and are excluded, leaving \num{136030} patents, \num{6909} Silent Green patents, and 734 firms. This specification also bears on the greenhushing account set out in Table~\ref{tab:accounts_design}: stable
firm-level disclosure strategy is absorbed by the firm fixed effect, so a
relationship that survives within firms cannot be produced by it.

\subsection{Results}
\label{sec:atypicality_silent_green}

Silent Green exhibits a pronounced nonlinear relationship with technological atypicality. The observed rate rises from 2.12\% in the lowest atypicality quartile to 9.04\% in the third quartile, before declining to 4.93\% in the highest. Patents in the third quartile are therefore over four times as likely to be Silent Green as those in the lowest quartile. This inverted-U pattern indicates that administrative omission is most prevalent among moderately-to-highly atypical technologies, rather than among conventional patents or extreme outliers.

\begin{table}[H]
\centering
\caption{Silent Green rates across technological atypicality}
\label{tab:atypicality_rates}
\begin{tabular}{@{}lrrrr@{}}
\toprule
\multicolumn{5}{@{}l}{\textit{Panel A. Observed rates by atypicality quartile}}\\
\midrule
Atypicality group & Patents & Silent Green & Rate & Rel.\ to Q1 \\
\midrule
Q1: Lowest  & 87,475 & 1,852 & 2.12\% & 1.00 \\
Q2          & 87,474 & 5,671 & 6.48\% & 3.06 \\
Q3          & 87,474 & 7,912 & 9.04\% & 4.26 \\
Q4: Highest & 87,475 & 4,312 & 4.93\% & 2.33 \\
\addlinespace
\multicolumn{5}{@{}l}{\textit{Panel B. Filing-year-adjusted predicted rates}}\\
\midrule
Atypicality level & Std.\ value & Adjusted rate & Rel.\ to mean & Per 1,000 \\
\midrule
Low atypicality       & $-1.000$ & 4.19\% & 0.643 & 42 \\
Mean atypicality      & $0.000$  & 6.52\% & 1.000 & 65 \\
Estimated peak        & $0.638$  & 7.04\% & 1.081 & 70 \\
High atypicality      & $1.000$  & 6.87\% & 1.054 & 69 \\
Very high atypicality & $2.000$  & 4.93\% & 0.757 & 49 \\
\bottomrule
\end{tabular}

\vspace{0.5em}
\begin{minipage}{0.95\textwidth}
\footnotesize
\textit{Notes:} Panel~A reports unadjusted Silent Green rates within equal-sized quartiles of the complete-case sample; ``Rel.\ to Q1'' is each quartile's rate divided by the Q1 rate. Panel~B reports average predicted probabilities from the logistic model with filing-year fixed effects (Table~\ref{tab:atypicality_regressions}, column~1), evaluated at the indicated standardized atypicality value while retaining the observed filing-year distribution; ``Rel.\ to mean'' indexes each predicted rate to the mean-atypicality prediction. Because atypicality is right-skewed, the fitted peak at $Z=0.638$ falls within the upper part of the Q3 range, so the model peak and the highest observed quartile rate coincide in location while differing in level, the latter being unadjusted and pooled over a range of $Z$. All rates refer to the probability of Silent Green within the consensus-corrected green estimation sample, not within the full patent corpus.
\end{minipage}
\end{table}

Multivariate analyses accounting for filing year confirm this nonlinear relationship (linear $\beta = 0.262$, $p < 0.001$; quadratic $\beta = -0.205$, $p < 0.001$), with the implied within-sample maximum occurring at $0.638$ standard deviations above mean atypicality. The effect is substantively meaningful: the predicted probability of Silent Green rises from 4.19\% (one standard deviation below the mean) to 7.04\% at the peak. This 68\% relative increase corresponds to approximately 28 additional omitted patents per 1,000 otherwise comparable inventions in the corrected-green portfolio. Crucially, the relationship persists when comparing patents within the same firm and filing year. In a firm- and year-fixed-effects specification, the nonlinear association remains robust (linear $\beta = 0.277$, $p < 0.001$; quadratic $\beta = -0.146$, $p < 0.001$), peaking at $0.945$ standard deviations above the mean. Stable firm-level differences in patenting strategy therefore cannot explain the omission gap. This result bears on the greenhushing account specified in
Table~\ref{tab:accounts_design}. Because the firm fixed effect absorbs any
time-invariant firm-level disclosure strategy, a greenhushing account
operating through such a strategy cannot generate the within-firm
atypicality gradient observed here. Technology-specific
under-communication, a firm that hushes about some inventions and not
others, is not separately identified and remains possible.

\begin{table}[H]
\centering
\caption{Logistic estimates of Silent Green and technological atypicality}
\label{tab:atypicality_regressions}
\begin{tabular}{@{}lcc@{}}
\toprule
& \textbf{(1)} & \textbf{(2)} \\

\midrule
Atypicality ($Z$)           & $0.262^{***}$  & $0.277^{***}$  \\
                            & $(0.011)$      & $(0.027)$      \\
\addlinespace
Atypicality$^2$ ($Z^2$)     & $-0.205^{***}$ & $-0.146^{***}$ \\
                            & $(0.009)$      & $(0.019)$      \\
\midrule
Implied peak (SD)           & $0.638$   & $0.945$   \\
Filing-year FE              & Yes       & Yes       \\
Firm FE                     & No        & Yes       \\
Patents                     & 349,898   & 136,030   \\
Silent Green patents        & 19,747    & 6,909     \\
Firms                       & ---       & 734       \\
\bottomrule
\end{tabular}

\vspace{0.5em}
\begin{minipage}{0.85\textwidth}
\footnotesize
\textit{Notes:} The dependent variable equals one for Silent Green patents and zero for patents retained as green under the consensus correction. Atypicality is standardized within the complete-case sample (mean $0$, SD $1$). The table reports logistic coefficients with standard errors in parentheses. Column~(1) includes filing-year fixed effects with heteroskedasticity-robust standard errors; column~(2) includes filing-year and firm (GVKEY) fixed effects with standard errors clustered by firm. Because each model contains a quadratic term, the implied within-sample maximum occurs at $-\beta_{1}/2\beta_{2}$. Significance: $^{*}\,p<0.05$; $^{**}\,p<0.01$; $^{***}\,p<0.001$.
\end{minipage}
\end{table}

Together, these results are consistent with bounded classification capacity. Routine administrative tagging works well for familiar technological configurations, but omission rates rise when climate relevance is embedded in less familiar combinations. The subsequent decline in omissions among the most atypical patents should be interpreted cautiously; it may reflect greater administrative scrutiny for extreme outliers, though our analysis establishes only the nonlinear association rather than the underlying mechanism.

\subsection{Atypicality in the Complexity Specifications}
\label{app:atypicality_controls}

Table~\ref{tab:complexity_full} reports the complete
specifications summarized in Table~\ref{tab:complexity_main}, including
the atypicality controls. Atypicality remains independently predictive of omission when reflection complexity is included ($\beta=0.283$, $p<0.001$; $\beta_{Z^2}=-0.207$, $p<0.001$). The interaction between atypicality and reflection complexity is not significant in the pooled sample ($\beta=0.004$, $p=0.820$), indicating that the two dimensions operate primarily as independent correlates of omission rather than as a single multiplicative mechanism. The interaction becomes positive in the post-2010 sample; we therefore treat complementarity between atypicality and complexity as a secondary robustness result rather than the central mechanism. In the error-direction model, atypicality also tilts errors toward Silent Green (OR $=1.84$, $p<0.001$; quadratic OR $=0.80$, $p<0.001$).

\begin{table}[H]
\centering
\caption{Full specifications behind Table~\ref{tab:complexity_main}, including the atypicality controls}
\label{tab:complexity_full}
\small
\resizebox{\textwidth}{!}{%
\begin{tabular}{lcccccc}
\toprule
& \multicolumn{5}{c}{\textbf{Silent Green}} & \textbf{Error direction} \\
\cmidrule(lr){2-6}\cmidrule(lr){7-7}
& (1) & (2) & (3) & (4) & (5) & (6) \\
& Reflection only & Structural only & Reflection + atyp. & Structural + atyp. & Both + atyp. & Silent vs. False \\
\midrule

Atypicality ($Z$)
& ---
& ---
& $0.283^{***}$
& $0.232^{***}$
& $0.249^{***}$
& $0.611^{***}$ \\
& ---
& ---
& $(0.012)$
& $(0.012)$
& $(0.012)$
& $(0.014)$ \\

Atypicality$^2$
& ---
& ---
& $-0.207^{***}$
& $-0.195^{***}$
& $-0.187^{***}$
& $-0.218^{***}$ \\
& ---
& ---
& $(0.009)$
& $(0.009)$
& $(0.009)$
& $(0.009)$ \\

Reflection complexity
& $0.485^{***}$
& ---
& $0.473^{***}$
& ---
& $0.372^{***}$
& $0.897^{***}$ \\
& $(0.012)$
& ---
& $(0.012)$
& ---
& $(0.011)$
& $(0.011)$ \\

Structural complexity
& ---
& $-0.476^{***}$
& ---
& $-0.466^{***}$
& $-0.392^{***}$
& $-0.150^{***}$ \\
& ---
& $(0.007)$
& ---
& $(0.007)$
& $(0.007)$
& $(0.008)$ \\

\midrule
Filing-year FE
& Yes & Yes & Yes & Yes & Yes & Yes \\

Patents
& 347,241
& 347,241
& 347,241
& 347,241
& 347,241
& 157,241 \\

\bottomrule
\end{tabular}%
}

\vspace{0.4em}
\begin{minipage}{0.98\textwidth}
\footnotesize
\textit{Notes:}
Columns~(1)--(5) estimate the probability that a patent is Silent Green within the consensus-corrected green-patent portfolio. Columns~(1) and (2) include reflection complexity and structural complexity separately. Columns~(3) and (4) add technological atypicality and its quadratic term to the respective complexity measure. Column~(5) includes atypicality and both complexity measures simultaneously. Column~(6) restricts the sample to consensus-attributed administrative errors and codes Silent Green as one and False Green as zero. Continuous variables are standardized within their respective estimation samples. All specifications include filing-year fixed effects. Logistic coefficients are reported with heteroskedasticity-robust standard errors in parentheses. $^{*}p<0.05$; $^{**}p<0.01$; $^{***}p<0.001$.
\end{minipage}
\end{table}

\section{Robustness to Technological Composition and Examination
Sorting}
\label{app:complexity_examiner_robustness}

Table~\ref{tab:complexity_examiner_robustness} summarizes the most demanding robustness and examination-sorting tests. Panel~A reports high-dimensional fixed-effect logit specifications for Silent Green and error direction. Panel~B reports linear-probability specifications used to assess Art Unit and examiner-specialization channels on closely matched samples. The examiner-specialization measure is the share of an examiner's strictly prior applications in the focal USPC class, conditional on at least 20 prior applications and controlling for overall prior examiner experience. Panel~C reverses the dependent variable and asks whether patent characteristics predict assignment to a more specialized examiner in the first place; they do not, which is what makes the Panel~B result interpretable.

\begin{table}[H]
\centering
\caption{Robustness to technological composition and examination sorting}
\label{tab:complexity_examiner_robustness}
\small
\resizebox{\textwidth}{!}{%
\begin{tabular}{llcccc}
\toprule
Panel & Specification & Reflection & Structural & Examiner specialization & $N$ \\
\midrule

\multicolumn{6}{l}{\textit{Panel A. High-dimensional fixed-effect logit}} \\

A & Silent Green: Year + CPC + Location FE; two-way cluster CPC + Location
& $0.574^{*}$ & $-0.641^{***}$ & --- & 276,069 \\
& & $(0.273)$ & $(0.123)$ & --- & \\

A & Error direction: Year + CPC + Location FE; two-way cluster CPC + Location
& $1.068^{***}$ & $-0.274^{*}$ & --- & 117,132 \\
& & $(0.227)$ & $(0.109)$ & --- & \\

\addlinespace

\multicolumn{6}{l}{\textit{Panel B. Examination-sorting linear-probability tests}} \\

B & Silent Green: Year + CPC + Location FE
& $0.034^{***}$ & $-0.042^{***}$ & --- & 329,457 \\
& & $(0.001)$ & $(0.001)$ & --- & \\

B & + Art Unit FE
& $0.039^{***}$ & $-0.044^{***}$ & --- & 329,403 \\
& & $(0.001)$ & $(0.001)$ & --- & \\

B & + Art Unit FE + examiner specialization; HC1
& $0.038^{***}$ & $-0.043^{***}$ & $0.003^{***}$ & 319,038 \\
& & $(0.001)$ & $(0.001)$ & $(0.000)$ & \\

B & + Art Unit FE + examiner specialization; cluster CPC + Examiner
& $0.038^{**}$ & $-0.043^{***}$ & $0.003^{**}$ & 319,038 \\
& & $(0.014)$ & $(0.010)$ & $(0.001)$ & \\

\addlinespace

\multicolumn{6}{l}{\textit{Panel C. Assignment-to-specialization test}} \\

C & Outcome: examiner specialization; Year + CPC + Location + Art Unit FE; cluster Examiner + CPC
& $0.013$ & $0.022^{\dagger}$ & --- & 319,038 \\
& & $(0.015)$ & $(0.012)$ & --- & \\

\bottomrule
\end{tabular}%
}

\vspace{0.4em}

\begin{minipage}{0.98\textwidth}
\footnotesize
\textit{Notes:}
Panel~A reports logit coefficients; Silent Green equals one for administratively
omitted patents within the consensus-corrected green portfolio, while the
error-direction outcome equals one for Silent Green and zero for False Green
among consensus-attributed administrative errors.
Panel~B reports linear-probability coefficients.
Panel~C reverses the dependent variable and asks whether patent characteristics
predict assignment to an examiner with greater prior same-USPC-class
specialization.
The examiner-specialization outcome and patent complexity measures are
standardized in the mechanism sample.
High-dimensional logit samples are reduced by fixed-effect separation, so
Panel~A is used as a robustness test rather than for direct coefficient-magnitude
comparisons with the baseline models.
$^{\dagger}p<0.10$;
$^{*}p<0.05$;
$^{**}p<0.01$;
$^{***}p<0.001$.
\end{minipage}

\end{table}

\section{Lexical Fingerprints of Administrative and Model Errors}
\label{sec:lexical_error_structure}

This appendix examines whether the language used in patent titles and abstracts distinguishes errors attributed to the administrative Y02 classification from errors attributed to PatentSBERTa. The analysis does not treat raw disagreement between Y02 and PatentSBERTa as evidence that either system is incorrect. Instead, the dependent variable is constructed from the final Qwen--Gemma direct-mechanism consensus, which identifies the likely source of error among resolved disagreement cases.

Nine theory-informed lexical categories capture digital context, power management, generic environmental wording, efficiency, explicit climate mechanisms, recycling and waste, energy conversion and storage, industrial engineering, and climate adaptation. The lexical categories used in this error-attribution analysis differ deliberately from the explicit green-framing dictionary reported in Table~\ref{tab:green_dictionary}. The present categories form a contrastive lexical taxonomy designed to distinguish consensus-attributed administrative errors from PatentSBERTa errors. They therefore include both climate-mechanism language and potentially ambiguous or technology-specific language, such as digital context, power management, and industrial engineering. By contrast, Table~\ref{tab:green_dictionary} measures whether applicants explicitly present an invention as environmentally beneficial and is therefore restricted to recognizable environmental claims. The two dictionaries operationalize different constructs: lexical error attribution and explicit environmental framing, respectively. Separate logistic models are estimated within the two disagreement directions. All nine lexical indicators are included simultaneously, together with controls for filing period and primary non-Y CPC subclass.

\begin{figure}[H]
\centering

\textbf{a}\par
\includegraphics[width=0.72\textwidth]
{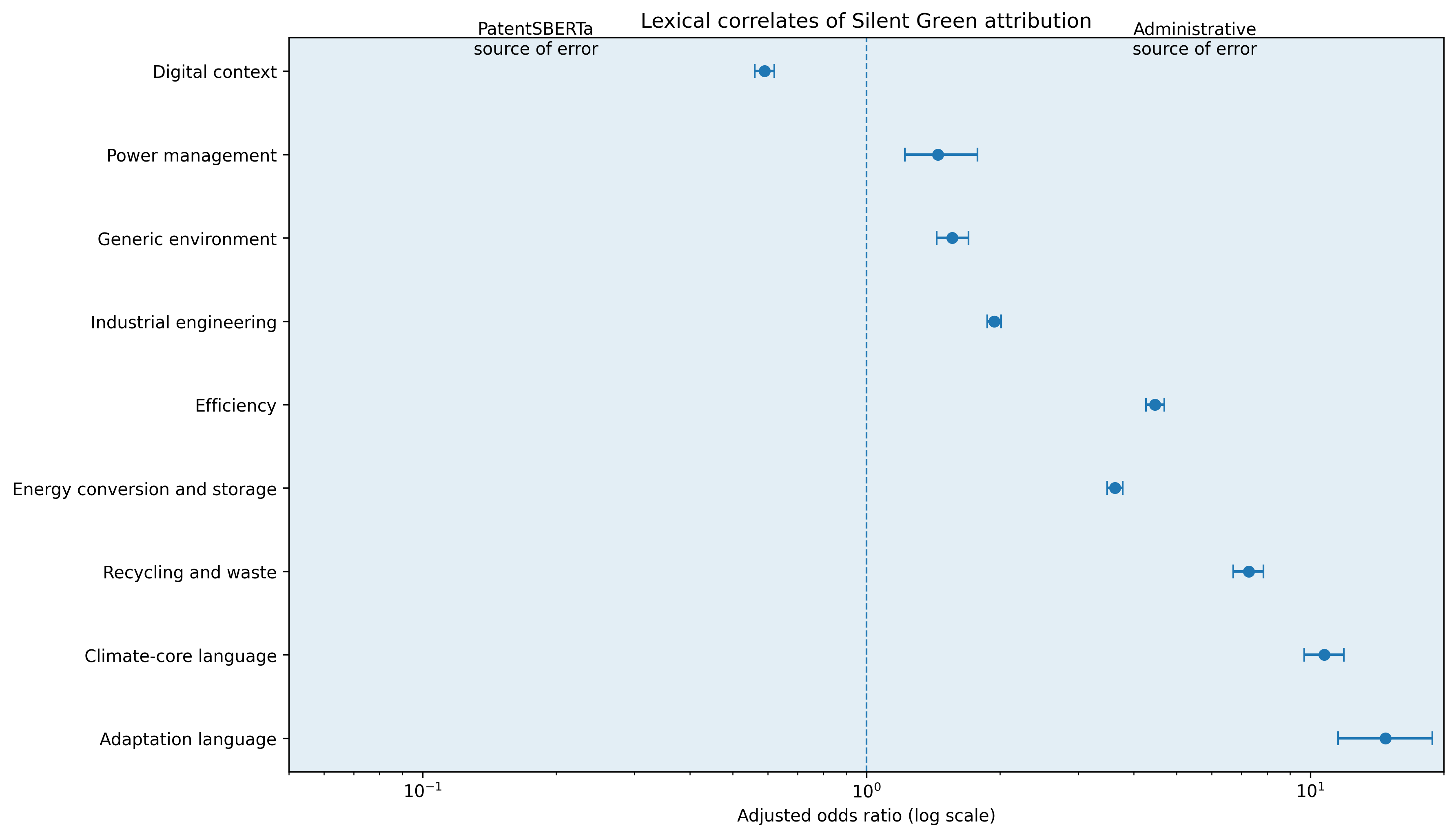}

\vspace{0.8em}

\textbf{b}\par
\includegraphics[width=0.72\textwidth]
{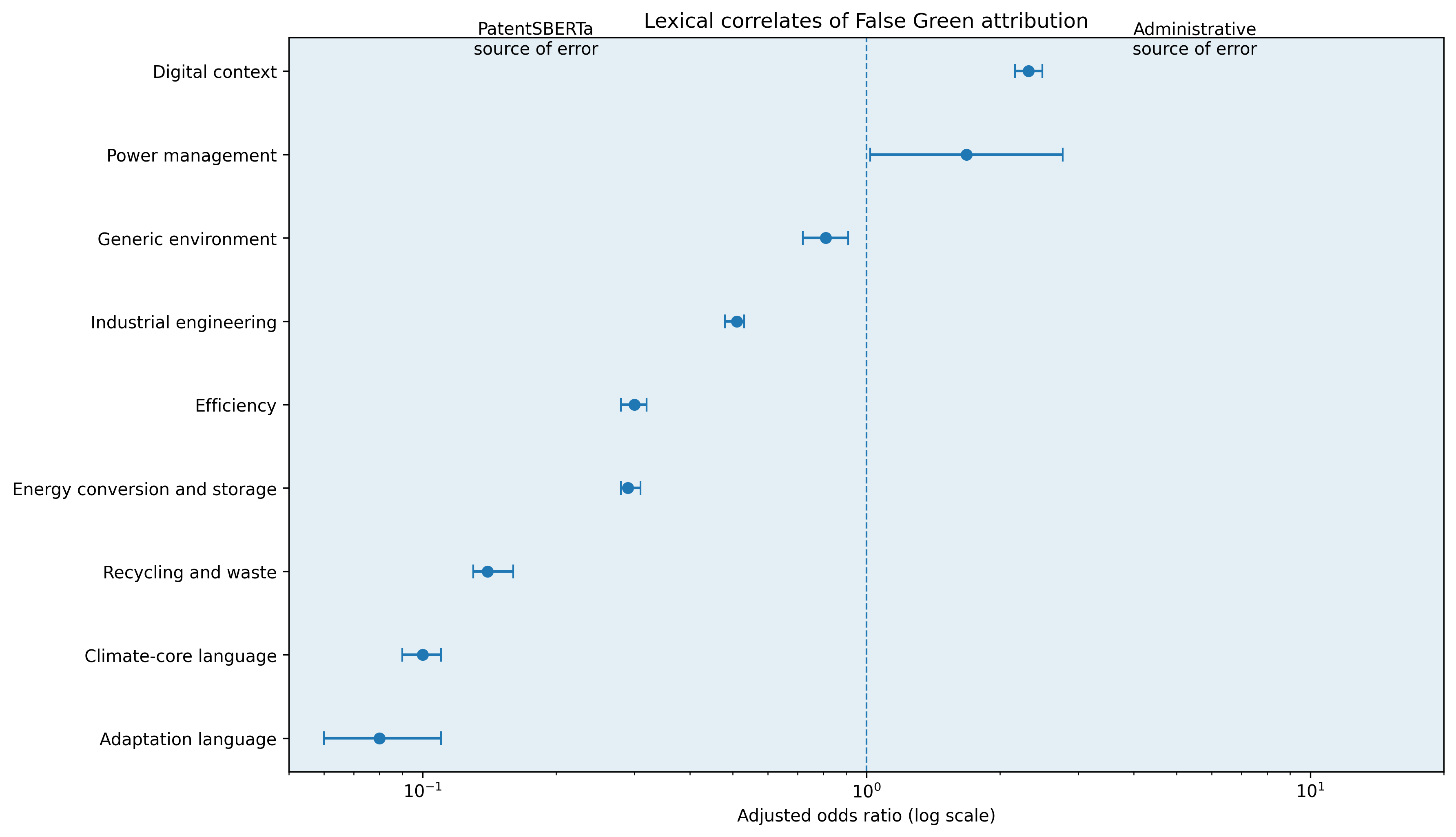}

\caption{Lexical correlates of consensus-attributed
classification error.
\textbf{a}, Adjusted odds ratios for administrative Type~I error
(False Green) rather than model Type~II error among resolved downward
disagreements.
\textbf{b}, Adjusted odds ratios for administrative Type~II error
(Silent Green) rather than model Type~I error among resolved upward
disagreements. In both panels, odds ratios greater than one indicate
that the lexical category is associated with a higher probability that
the administrative classification is the attributed source of error.
Odds ratios below one indicate a higher probability that PatentSBERTa
is the attributed source of error. Points show adjusted odds ratios and
horizontal lines show 95\% confidence intervals. The vertical dashed
line denotes the null value of one. Models include all nine lexical
indicators simultaneously and control for filing period and pooled
primary non-Y CPC subclass. Lexical indicators are derived from the
original patent title and abstract rather than from LLM-generated
explanations. Associations are conditional on entry into the
PatentSBERTa--Y02 disagreement pool and cross-model direct-mechanism
consensus.}
\label{fig:lexical_error_correlates}
\end{figure}

Among resolved downward disagreements, the dependent variable equals one when the LLM consensus attributes the disagreement to an administrative Type I error, or False Green, and zero when it attributes the disagreement to a PatentSBERTa Type II error, or model false negative. Digital-context language is strongly associated with administrative-error attribution. After adjustment for filing period, technological subclass, and the other lexical categories, patents containing digital or computing language have more than twice the odds of being identified as False Green rather than as a PatentSBERTa false negative (adjusted odds ratio, OR = 2.32; 95\% confidence interval, CI = 2.16--2.49; Figure~\ref{fig:lexical_error_correlates}a). For illustration, if the probability of False Green attribution were 20\% among otherwise similar patents without digital language, the corresponding probability would be approximately 37\% among patents containing such language. Power-management language shows a similar, although less precisely estimated, association (OR = 1.68; 95\% CI = 1.02--2.77).

The opposite pattern appears when patents contain language describing a specific climate or engineering mechanism. Adaptation language is associated with approximately 92\% lower odds of False Green attribution (OR = 0.08; 95\% CI = 0.06--0.11), while climate-core language is associated with approximately 90\% lower odds (OR = 0.10; 95\% CI = 0.09--0.11). Recycling and waste language is associated with approximately 86\% lower odds (OR = 0.14; 95\% CI = 0.13--0.16). Energy-conversion and storage language (OR = 0.29; 95\% CI = 0.28--0.31), efficiency language (OR = 0.30; 95\% CI = 0.28--0.32), and industrial-engineering language (OR = 0.51; 95\% CI = 0.48--0.53) show the same general pattern. Thus, when an administratively Y02-labelled patent contains specific climate or engineering content, the resolved disagreement is more likely to reflect an incorrect rejection by PatentSBERTa than an incorrect administrative inclusion.

Among resolved upward disagreements, the dependent variable equals one when the LLM consensus attributes the disagreement to an administrative Type II error, or Silent Green, and zero when it attributes the disagreement to a PatentSBERTa Type I error, or model false positive. Adaptation and climate-related language strongly distinguish Silent Green patents from PatentSBERTa false positives (Figure~\ref{fig:lexical_error_correlates}b). Patents containing adaptation language have almost fifteen times the odds of Silent Green attribution (OR = 14.76; 95\% CI = 11.55--18.85). Climate-core language is associated with more than ten times the odds (OR = 10.74; 95\% CI = 9.70--11.90), while recycling and waste language is associated with more than seven times the odds (OR = 7.26; 95\% CI = 6.71--7.85).

Strong positive associations are also observed for efficiency language (OR = 4.47; 95\% CI = 4.26--4.69), energy-conversion and storage language (OR = 3.63; 95\% CI = 3.49--3.78), and industrial-engineering language (OR = 1.94; 95\% CI = 1.87--2.01). These results show that patents omitted from Y02 are more likely to be confirmed as Silent Green when their climate contribution is expressed through adaptation, energy, recycling, efficiency, or industrial-engineering mechanisms.

Digital-context language again shows the reverse association. Among patents outside Y02 that PatentSBERTa classifies as green, digital language is associated with 41\% lower odds of Silent Green attribution (OR = 0.59; 95\% CI = 0.56--0.62). Such cases are therefore more likely to represent PatentSBERTa false positives than genuine administrative omissions. Read together with the downward-pool estimate of
OR $=2.32$, this is the two-sided signature of bias transmission discussed
in Section~\ref{sec:conservative_by_design}: one lexical cue raises the
probability of over-inclusion by the administrative label inside Y02 and
by the model outside it. Generic environmental wording shows only a modest positive association with Silent Green attribution (OR = 1.56; 95\% CI = 1.44--1.70) and is negatively associated with False Green attribution (OR = 0.81; 95\% CI = 0.72--0.91). The findings therefore do not support the broad claim that generic environmental wording alone causes administrative over-inclusion.

Taken together, the two analyses reveal a clear lexical asymmetry. Digital-system language is associated with administrative over-inclusion among Y02-labelled patents and with PatentSBERTa over-inclusion among patents outside Y02. By contrast, specific language concerning adaptation, emissions, recycling, energy systems, efficiency, and industrial engineering is associated with valid Y02 inclusion in the downward pool and administrative omission in the upward pool. The likely source and direction of classification error therefore depend not only on technological field but also on how the invention's climate-related function is expressed in patent text.

These associations are descriptive rather than causal. They are conditional on entry into the PatentSBERTa--Y02 disagreement pool and on Qwen--Gemma direct-mechanism consensus. They should therefore be interpreted as evidence about the lexical structure of consensus-attributed classification errors, rather than as corpus-wide effects of individual words.

\section{Judge Prompt and Decision Rubric for the Construct-Validity Test}
\label{app:audit_prompt}

This appendix summarizes the prompt and decision rubric used by the two judges in the construct-validity test. Its purpose is to make the adjudication logic more transparent and reproducible. The goal of the test was not to reproduce the administrative Y02 tag, but to determine whether a patent's primary technical function makes a direct and credible contribution to climate-change mitigation or adaptation.

\subsection{Inputs Provided to the Agents}

For each patent, the reviewing agents were given access to the following materials, subject to availability in the patent record:

\begin{itemize}[leftmargin=*]
    \item Patent title
    \item Patent abstract
    \item Patent claims
    \item Full patent description
    \item Relevant CPC Y02 taxonomy definitions
\end{itemize}

These materials were used to ensure that the decision was based on the invention's technical substance rather than on administrative labeling alone.

\subsection{Judge Prompt}

The judge prompt is reproduced below:

\begin{quote}
Read the patent materials and identify the invention's primary technical function. Then determine whether that primary function makes a direct and verifiable contribution to climate-change mitigation or climate adaptation. In your assessment, distinguish direct technical contributions from incidental, symbolic, or weakly related environmental language. Provide: (1) a brief summary of the invention's core technical function, (2) the causal mechanism, if any, linking that function to mitigation or adaptation, (3) an explanation of why the contribution is direct or not direct, and (4) a final label: ``True Green'' or ``Non-Green.''
\end{quote}

\subsection{Decision Rubric}

To improve consistency across cases, the agents applied the following decision rubric:

\begin{enumerate}[leftmargin=*]
    \item \textbf{Identify the core technical function:}
    Determine what the invention primarily does in technical terms, rather than relying on broad or promotional language.

    \item \textbf{Assess direct climate relevance:}
    Evaluate whether the invention's primary function directly contributes to at least one of the following:
    \begin{itemize}[leftmargin=*]
        \item reduction of greenhouse-gas emissions,
        \item improvement in energy or material efficiency with a clear decarbonization pathway,
        \item enabling renewable energy, electrification, emissions control, recycling, or other recognized mitigation technologies,
        \item strengthening climate adaptation capacity (e.g., resilience to flooding, heat, drought, or related climate risks).
    \end{itemize}

    \item \textbf{Distinguish direct from incidental effects.}
    Do not classify a patent as True Green merely because it contains broad terms such as environment, efficiency, resource conservation, or power management. Such language was treated as insufficient unless the patent's primary technical function clearly and directly supported mitigation or adaptation.

    \item \textbf{Prioritize technical function over wording:}
    If the patent used highly technical engineering or chemical language, the review focused on whether the described mechanism substantively enabled decarbonization, even if explicit environmental terminology was absent.

    \item \textbf{Assign the final label:}
    A patent was labeled True Green only when the primary technical function had a direct and credible mitigation or adaptation role. Otherwise, it was labeled Non-Green.
\end{enumerate}

\subsection{Escalation of Unresolved Cases}

The two language-model agents produced independent labels and written justifications. Agreement on a direct mechanism was retained as the consensus-based decision. Cases in which the models disagreed about directness were not automatically resolved. In the present large-scale results, \num{117586} such cases remain an explicit escalation set. Future adjudication should review the patent text and both
model rationales using the same rubric, with technically consequential
cases prioritized for domain-expert assessment. A stratified
human-coded validation sample, blind to the Y02 label, the model
prediction, and both model judgments, is the appropriate next step; it
would also permit case-level coding of scheme conformity
(Section~\ref{sec:theory}), which the present design does not separate
from construct conformity.

%

\section{Held-Out Validation of Non-Random Disagreement}
\label{sec:heldout_validation}

The consensus analysis directly tests the technological distribution of attributed administrative and model errors within the complete disagreement pool. A held-out test-set analysis provides an independent robustness check by asking whether PatentSBERTa--Y02 agreement and disagreement are associated with technological domain before LLM attribution. It therefore evaluates the structure of the screening signal rather than determining the substantive truth or error source of individual patents. Table~\ref{tab:main_omnibus_tests} reports the omnibus association tests.

\begin{table}[H]
\centering
\caption{Held-out evidence of non-random model--administrative
disagreement}
\label{tab:main_omnibus_tests}
\small

\begin{tabular}{lrr}
\toprule
\textbf{Specification}
&
\textbf{$N$}
&
\textbf{Cram\'er's $V$}
\\
\midrule

Primary non-Y CPC sections
& 298,459 & 0.157 \\

Primary non-Y CPC subclasses (58 categories)
& 298,459 & 0.341 \\

Primary Y02 family
& 22,568 & 0.222 \\

\bottomrule
\end{tabular}

\vspace{0.35em}
\begin{minipage}{0.95\textwidth}
\footnotesize
\textit{Note:}
All Pearson chi-square tests reject the null of independence at
$p<0.001$. Full test statistics are reported in the replication
repository accompanying this article.
\end{minipage}
\end{table}

The held-out results independently confirm that disagreement is technologically structured. The strongest association occurs at the primary non-Y CPC subclass level ($V=0.341$), indicating that disagreement is concentrated in specific technologies rather than being explained only by broad sector differences. Batteries, fuel cells, power systems, engine control, emissions-control technologies, heat pumps, separation processes, and industrial technologies are among the most disagreement-prone domains. The results also reveal substantial heterogeneity within Y02. Greater agreement among administratively labelled patents in a particular family should not be interpreted as complete measurement validity, because genuinely green inventions may remain outside Y02 altogether. Reliability conditional on receiving a Y02 label does not guarantee coverage of the underlying construct.

Taken together, the consensus and held-out analyses provide complementary evidence. The consensus analysis attributes the source and direction of error for \num{400186} disagreements and demonstrates strong conditional technological bias. The held-out analysis independently shows that the original model--administrative disagreement signal is associated with technological domain. The remaining \num{117586} mechanism disagreements are retained as unresolved rather than being forced into an attributed error category.

\section{Explicit Green-Language Dictionary}
\label{app:green_dictionary}

The strategic-framing test in Section~\ref{sec:greenwashing_event} measures explicit environmental framing by matching each patent's title and abstract field against a predefined dictionary of green phrases. Matching is case-insensitive and word-boundary anchored, and captures common morphological variants (singular and plural forms, and both \textit{-ize} and \textit{-ise} spellings of \textit{decarbonize}). Table~\ref{tab:green_dictionary} lists the dictionary by theme. Two outcomes are derived: a binary indicator equal to one if a patent contains at least one phrase, and framing intensity, the number of matched phrases per 1{,}000 words, log-transformed as $\log(1+x)$ in the regressions. In the estimation sample, 312 of \num{85541} patents (0.36\%) contain at least one phrase. The energy phrases refer specifically to \textit{energy} efficiency and energy saving; the bare term \textit{efficiency}, which frequently appears in non-climate contexts, is deliberately excluded, consistent with the judges' rubric (Appendix~\ref{app:audit_prompt}). The exact regular-expression patterns are provided in the replication repository. This explicit green-framing dictionary is distinct from the broader error-attribution lexical taxonomy used in Appendix~\ref{sec:lexical_error_structure}.

\begin{table}[H]
\centering
\caption{Explicit green-language dictionary, by theme}
\label{tab:green_dictionary}
\small
\renewcommand{\arraystretch}{1.25}
\begin{tabularx}{\textwidth}{@{}>{\raggedright\arraybackslash}p{0.30\textwidth} >{\raggedright\arraybackslash}X@{}}
\toprule
\textbf{Theme} & \textbf{Phrases} \\
\midrule
Climate change & climate change; global warming; climate mitigation; climate adaptation \\
Greenhouse gases & greenhouse gas(es); greenhouse effect; carbon footprint; carbon dioxide emissions; CO\textsubscript{2} emissions \\
Carbon reduction \& decarbonization & carbon neutral; carbon neutrality; decarbonize/decarbonise (and variants); low-carbon; zero-carbon; carbon emissions; carbon capture; carbon sequestration \\
Net zero & net-zero \\
Emissions reduction & emission(s) reduction; reduce emissions; fossil-fuel reduction \\
Renewable \& clean energy & renewable energy; clean energy; green energy; green technology \\
Energy efficiency & energy saving; energy-efficient; energy efficiency \\
Sustainability & sustainable; sustainability \\
Pollution control & pollution reduction; pollution control; environmental pollution \\
Environmental (general) & environmental benefit; environmentally friendly; environmentally beneficial; eco-friendly; environmental impact; environmental protection; cleaner production \\
\bottomrule
\end{tabularx}
\end{table}

\clearpage

\section{Broader Catalogue of Consensus-Resolved Administrative Errors}
\label{app:illustrative_patent_cases}

This appendix provides a broader catalogue of patents selected from the full-corpus construct-validity test to illustrate the substantive mechanisms underlying administrative Type I and Type II errors. The examples are purposively chosen for explanatory clarity and technological diversity; they are not a random or representative sample and should not be used to estimate the relative frequency of individual mechanisms.

Every patent reported below is consensus-resolved. For False Green cases, the administrative indicator classifies the patent as green ($Y02=1$), the TF model classifies it as non-green ($TF=0$), and both independent LLMs conclude that no direct climate-mitigation or adaptation mechanism is present ($LLMs=0/0$). For Silent Green cases, the administrative indicator omits the patent ($Y02=0$), the TF model identifies it as green ($TF=1$), and both LLMs identify a direct climate mechanism ($LLMs=1/1$). All listed examples received high-confidence judgments from both LLMs.

\begin{landscape}

\footnotesize
\setlength{\LTleft}{0pt}
\setlength{\LTright}{0pt}
\renewcommand{\arraystretch}{1.15}
\setlength{\tabcolsep}{3pt}

\begin{longtable}{
    @{}
    >{\raggedright\arraybackslash}p{1.35cm}
    >{\centering\arraybackslash}p{1.15cm}
    >{\centering\arraybackslash}p{1.15cm}
    >{\centering\arraybackslash}p{0.65cm}
    >{\centering\arraybackslash}p{0.65cm}
    >{\centering\arraybackslash}p{0.85cm}
    >{\raggedright\arraybackslash}p{5.4cm}
    >{\raggedright\arraybackslash}p{5.6cm}
    @{}
}

\caption{Broader Catalogue of Illustrative False Green and Silent Green
Patents from the Full-Corpus Construct-Validity Test}
\label{tab:appendix_error_catalogue}
\\

\toprule
\textbf{Patent}
&
\textbf{Primary CPC}
&
\textbf{Admin. Y02}
&
\textbf{Y02}
&
\textbf{TF}
&
\textbf{LLMs}
&
\textbf{Primary technical function}
&
\textbf{Administrative-error mechanism}
\\
\midrule
\endfirsthead

\multicolumn{8}{c}{
\textit{Table~\ref{tab:appendix_error_catalogue} continued}
}
\\
\toprule
\textbf{Patent}
&
\textbf{Primary CPC}
&
\textbf{Admin. Y02}
&
\textbf{Y02}
&
\textbf{TF}
&
\textbf{LLMs}
&
\textbf{Primary technical function}
&
\textbf{Administrative-error mechanism}
\\
\midrule
\endhead

\midrule
\multicolumn{8}{r}{\textit{Continued on next page}}
\\
\endfoot

\bottomrule
\endlastfoot

\multicolumn{8}{@{}l}{
\textbf{Panel A. Administrative Type I Errors: False Green}
\quad
($Y02=1$, $TF=0$, $LLMs=0/0$)
}
\\
\addlinespace[2pt]

\#11417179
&
G07G
&
Y02D
&
1
&
0
&
0/0
&
Image- and voice-based tracking of customers and products in a
cashierless retail ``shopping environment.''
&
\textbf{Semantic homonymy.}
The term ``environment'' refers to a commercial or digital setting,
not to the natural environment or a climate-related technical function.
\\
\addlinespace

\#9537918
&
G06F
&
Y02D
&
1
&
0
&
0/0
&
Client-side encryption and file sharing within a distributed secure
computing environment.
&
\textbf{Semantic homonymy in ICT.}
A secure digital environment is administratively associated with
environmental technology despite the absence of a direct climate
mechanism.
\\
\addlinespace

\#10057074
&
H04L
&
Y02D
&
1
&
0
&
0/0
&
Method and apparatus for supplying electrical power through a
Power-over-Ethernet communication network.
&
\textbf{Scheme scope exceeding the construct.}
The invention manages power within communications infrastructure but
does not establish a direct mitigation or adaptation pathway.
\\
\addlinespace

\#10139895
&
G06F
&
Y02D
&
1
&
0
&
0/0
&
Maintaining sufficient device power to preserve a fundamental
computing function.
&
\textbf{Scheme scope exceeding the construct.}
Ordinary device-level power management is included as climate
technology without evidence of a substantive decarbonization function.
\\
\addlinespace

\#10181139
&
G06Q
&
Y02P
&
1
&
0
&
0/0
&
Automated monitoring and management of workspace use through digital
business-process methods.
&
\textbf{Business-method spillover.}
Resource or workspace management is classified near sustainable
production despite lacking a direct climate mechanism.
\\
\addlinespace

\#7811444
&
C10G
&
Y02P
&
1
&
0
&
0/0
&
Oxidation and upgrading of asphaltenes obtained from heavy-hydrocarbon
sources for subsequent fuel or petrochemical use.
&
\textbf{Counter-directional taxonomic spillover.}
The invention facilitates fossil-resource processing rather than
climate mitigation.
\\
\addlinespace

\#10113118
&
C10G
&
Y02P
&
1
&
0
&
0/0
&
Process and apparatus for producing hydrocarbon products from a
hydrocarbon feedstock.
&
\textbf{Counter-directional process inclusion.}
The primary output is additional hydrocarbon production, with no
identified direct decarbonization mechanism.
\\
\addlinespace

\#11572899
&
F04F
&
Y02A
&
1
&
0
&
0/0
&
Pressure exchanger designed for use in hydraulic-fracturing
operations.
&
\textbf{Counter-directional application.}
The invention supports fossil-fuel extraction despite being included
within a climate-adaptation classification.
\\
\addlinespace

\#9993540
&
C07K
&
Y02A
&
1
&
0
&
0/0
&
Immunotherapy involving pharmaceutical compositions for the treatment
of brain and neuronal tumours.
&
\textbf{Adaptation-category spillover.}
A general medical invention is included without a specific causal
connection to a climate-related health risk.
\\
\addlinespace

\#9147960
&
H01R
&
Y02T
&
1
&
0
&
0/0
&
Electrical connector with a housing and cover designed to remain
operable when components become frozen.
&
\textbf{Technological-domain proximity.}
Resistance to freezing is treated as climate relevance even though the
invention has no direct transport-decarbonization or adaptation
mechanism.
\\

\midrule

\multicolumn{8}{@{}l}{
\textbf{Panel B. Administrative Type II Errors: Silent Green}
\quad
($Y02=0$, $TF=1$, $LLMs=1/1$)
}
\\
\addlinespace[2pt]

\#7986539
&
G05F
&
--
&
0
&
1
&
1/1
&
Maximum-power-point tracking and power conversion that improve the
extraction and delivery of photovoltaic electricity.
&
\textbf{Power-electronics opacity.}
The renewable-energy contribution is expressed through feedback loops,
voltage control, and conversion architecture rather than explicit
climate terminology.
\\
\addlinespace

\#6266975
&
F24F
&
--
&
0
&
1
&
1/1
&
Combined heat-pump and engine system for energy-efficient heating and
cooling.
&
\textbf{Thermodynamic opacity.}
The mitigation function is embedded in heat-transfer and mechanical
engineering rather than overt environmental framing.
\\
\addlinespace

\#11525437
&
B66C
&
--
&
0
&
1
&
1/1
&
Mechanical storage and delivery of electricity using elevated masses
and gravitational potential energy.
&
\textbf{Enabling-technology omission.}
A grid-relevant storage mechanism is described primarily through
lifting equipment and mechanical-system terminology.
\\
\addlinespace

\#11878950
&
C04B
&
--
&
0
&
1
&
1/1
&
Oxy-fuel cement-production process that produces a concentrated
carbon-dioxide stream suitable for capture and storage.
&
\textbf{Industrial-process opacity.}
A direct hard-to-abate-sector mitigation mechanism is embedded in
specialized cement and combustion engineering.
\\
\addlinespace

\#8877486
&
C12M
&
--
&
0
&
1
&
1/1
&
Microalgae photobioreactor that uses carbon dioxide during wastewater
treatment and biomass production.
&
\textbf{Cross-domain under-coverage.}
The invention combines biological carbon mitigation and wastewater
treatment across conventional technological boundaries.
\\
\addlinespace

\#10301977
&
F01K
&
--
&
0
&
1
&
1/1
&
Conversion of waste heat from gas-processing operations into usable
electrical power.
&
\textbf{Waste-heat recovery opacity.}
The climate contribution appears through thermodynamic energy
recovery rather than explicit environmental language.
\\
\addlinespace

\#10008737
&
H01M
&
--
&
0
&
1
&
1/1
&
Lithium-ion battery design that improves electrochemical efficiency
and energy-storage performance.
&
\textbf{Component-level omission.}
The enabling contribution to electricity storage is embedded in
battery materials and electrochemical design.
\\
\addlinespace

\#10247448
&
F25B
&
--
&
0
&
1
&
1/1
&
Refrigeration or heat-cycle system using a refrigerant with a low
global-warming potential.
&
\textbf{Materials-based mitigation opacity.}
The direct climate benefit is contained in refrigerant chemistry and
system design rather than in overt green terminology.
\\
\addlinespace

\#10370277
&
C02F
&
--
&
0
&
1
&
1/1
&
Anaerobic treatment of wastewater sludge that produces biogas and
recovers renewable energy.
&
\textbf{Joint waste--energy omission.}
The mitigation pathway crosses wastewater treatment and renewable
energy, making it less visible within a single administrative category.
\\
\addlinespace

\#11884602
&
C04B
&
--
&
0
&
1
&
1/1
&
Carbon-dioxide mineralization process that converts carbon into stable
mineral products.
&
\textbf{Chemistry-intensive carbon-removal opacity.}
Permanent carbon sequestration is expressed through mineral reactions
and materials-processing terminology.
\\

\end{longtable}

\noindent
\textit{Notes:}
Y02 is the binary administrative classification, and TF is the
PatentSBERTa classification at the validation-calibrated threshold.
The administrative Y02 column reports the assigned Y02 family for
False Green patents; Silent Green patents have no administrative Y02
classification. The LLM column reports the independent
direct-mechanism judgments in the order Qwen/Gemma. A value of 1 denotes
a direct climate-mitigation or adaptation mechanism, whereas 0 denotes
that such a mechanism is absent. All examples received high-confidence
judgments from both models. The descriptions summarize the patents'
primary technical functions and the mechanism revealed by the test;
they do not estimate realized environmental impact after
commercialization.

\end{landscape}

\end{document}